\documentclass[printer]{aa}
\usepackage[pagebackref,colorlinks,citecolor=blue,urlcolor=blue,linkcolor=blue]{hyperref}
\usepackage[varg]{txfonts}
\usepackage{graphicx}
\usepackage{natbib}
\usepackage{amssymb}
\usepackage{amsmath}
\usepackage{tabularx}
\usepackage{hhline}
\usepackage{makecell}
\usepackage{placeins}
\usepackage{tikz}
\usepackage{subcaption}
\usepackage{lipsum}
\usetikzlibrary{arrows,positioning,shapes} 
\usepackage{xspace}
\usepackage{mathtools}
\usepackage{{booktabs}}

\usepackage{siunitx}

\newcommand{\SD}{StarDICE\xspace}

\newcommand{\Qsolar}{Q_{\mathrm{solar}}}
\newcommand{\Qsolarmes}{Q_{\mathrm{solar}}^{\mathrm{mes}}}
\newcommand{\Qsolarcal}{Q_{\mathrm{solar}}^{\mathrm{cal}}}

\newcommand{\Qphot}{Q_{\mathrm{phot}}}
\newcommand{\Qphotmes}{Q_{\mathrm{phot}}^{\mathrm{mes}}}
\newcommand{\Qphotcal}{Q_{\mathrm{phot}}^{\mathrm{cal}}}

\newcommand{\Qccd}{Q_{\mathrm{ccd}}}
\newcommand{\Qccdcal}{Q_{\mathrm{ccd}}^{\mathrm{cal}}}
\newcommand{\Qccdmes}{Q_{\mathrm{ccd}}^{\mathrm{mes}}}

\newcommand{\Qspectro}{Q_{\mathrm{spectro}}}

\newcommand{\Qspectromain}{Q_{\mathrm{spectro}}^{\mathrm{main}}}

\newcommand{\Ssolarstat}{\sigma_{\mathrm{solar}}^{\mathrm{stat}}}
\newcommand{\Sphotstat}{\sigma_{\mathrm{phot}}^{\mathrm{stat}}}

\newcommand{\Rcbp}{R_{\mathrm{CBP}}}
\newcommand{\Rtel}{R_{\mathrm{tel}}}

\newcommand{\Esolar}{\epsilon_{\mathrm{SC}}}
\newcommand{\Espectro}{\epsilon_{\mathrm{spectro}}}

\newcommand{\Ephot}{\epsilon_{\mathrm{phot}}}

\newcommand{\spinhole}{\SI{75}{\micro\meter}\xspace}
\newcommand{\bpinhole}{\SI{5}{\milli\meter}\xspace}

\newcommand{\contcomp}{{}_{\mathrm{, \, comp}}}  

\newcommand{\Kghost}{K_{G_1/G_0}(\lambda)}
\newcommand{\Kpinholes}{K_\mathrm{\spinhole/\bpinhole}(\lambda)}

\newcommand{\Rwindow}{R_{\mathrm{win}}(\lambda)\xspace}
\newcommand{\Rccd}{R_{\mathrm{ccd}}(\lambda)\xspace}

\newcommand{\Kghostfit}{K_{G/A}}
\newcommand{\Kghostfitfirst}{K_{G_1/A}}

\newcommand{\up}[1]{\textsuperscript{#1}}

\usepackage[outputdir={fig},debug]{dot2texi}
\usepackage{xspace}

\newcolumntype{M}[1]{>{\centering\arraybackslash}m{#1}}

\graphicspath{{fig/}}


\title{StarDICE III: Characterization of the photometric instrument with a Collimated Beam Projector}

\author{
Thierry Souverin\inst{1}
 \and Jérémy Neveu\inst{1,2}
 \and Marc Betoule\inst{1}
 \and Sébastien Bongard\inst{1}
 \and Christopher W. Stubbs\inst{3}
 \and Elana Urbach\inst{3}
 \and Sasha Brownsberger\inst{3}
 \and Pierre Éric Blanc\inst{4}
 \and Johann Cohen-Tanugi\inst{6,7}
 \and Sylvie Dagoret-Campagne\inst{2}
 \and Fabrice Feinstein\inst{5}
 \and Delphine Hardin\inst{1}
 \and Claire Juramy\inst{1}
 \and Laurent Le Guillou\inst{1}
 \and Auguste Le Van Suu\inst{4}
 \and Marc Moniez\inst{2}
 \and Éric Nuss \textsuperscript{\textdagger}\inst{6}
 \and Bertrand Plez\inst{6}
 \and Nicolas Regnault\inst{1}
 \and Eduardo Sepulveda\inst{1}
 \and Kélian Sommer\inst{6}
 \and the LSST Dark Energy Science Collaboration\inst{}
}
\institute{
LPNHE, CNRS/IN2P3 \& Sorbonne Université, 4 place Jussieu, 75005 Paris, France
 \and Universit\'e Paris-Saclay, CNRS, IJCLab, 91405, Orsay, France
 \and Department of Astronomy, Harvard University, 60 Garden St., Cambridge, MA 02138, USA
 \and Université d’Aix-Marseille \& CNRS, Observatoire de Haute-Provence, 04870 Saint Michel l’Observatoire, France
 \and Aix Marseille Univ, CNRS/IN2P3, CPPM, Marseille, France
 \and LUPM, Université Montpellier \& CNRS, F-34095 Montpellier, France
 \and LPC, Université Clermont Auvergne, CNRS, F-63000 Clermont-Ferrand, France
}

\abstract{The measurement of Type Ia supernovae magnitudes provides cosmological distances, which can be used to constrain dark energy parameters. Yet, current and upcoming large photometric surveys require a substantial improvement in the calibration precision of their photometry to reduce systematic uncertainties in cosmological constraints.}{The StarDICE experiment is designed to establish accurate broadband flux references for these surveys, aiming for sub-percent precision in magnitude measurements. This requires a precise measurement of the filter bandpasses of both the StarDICE and survey instruments with sub-nanometer accuracy. To that end, we have developed the Collimated Beam Projector (CBP), an optical device capable of calibrating the throughput of an astronomical telescope and of its filters.}{The CBP is built from a tunable laser source and a reversed telescope to emit a parallel monochromatic light beam that is continuously monitored in flux and wavelength. The CBP output light flux is measured using a large area photodiode, previously calibrated relative to a NIST photodiode. We then derive the \SD{} telescope throughput and filter transmissions from the CBP measurements, anchoring it to the absolute calibration provided by the NIST.}{After carefully analyzing the systematic uncertainties, we have achieved sub-nanometer accuracy in determining filter central wavelengths, measured each filter transmission with a precision of \textasciitilde 0.5\% per \SI{1}{nm} bin, and detected out-of-band leakages at the \num{e-4} relative value. Furthermore, we have synthesized the equivalent transmission for full pupil illumination from four sample positions in the StarDICE telescope mirror, with an accuracy of approximately \SI{0.2}{nm} for central wavelengths and \SI{7}{mmag} for broadband fluxes.}{We have demonstrated our ability to characterize a telescope throughput down to the millimagnitude, and paved the way for future developments, such as a portable CBP version for in-situ transmission monitoring.}

\authorrunning{T. Souverin et al.}
\titlerunning{Characterisation of the photometric instrument with a Collimated Beam Projector}

\begin{document}

\authorrunning{T. Souverin et al.}
\titlerunning{Characterisation of the photometric instrument with a Collimated Beam Projector}

\maketitle

%


\section{Introduction}

The calibration of optical wide-field surveys needs to reach new levels of precision to meet the requirements of type Ia supernovae (SNe Ia) cosmology. SNe Ia are standard candles, a class of objects with predictable luminosity used as probes to characterize dark energy in the late Universe. We can infer dark energy properties by measuring the luminosity distance of SNe Ia at different redshifts. This luminosity distance is obtained by measuring the maximum amplitude of the SN Ia light curve, which is observed within different optical bands depending on its redshift. Errors in the relative flux calibration between the different bands have a knock-on effect on systematic errors in the Hubble diagram, which are then propagated to dark energy parameters constraints.

Current photometric surveys like the Dark Energy Survey (DES) \citep{Brout_2019} or Subaru Hyper Suprime-Cam (HSC) \citep{hsc_2019} observed hundreds of SNe Ia. The Zwicky Transient Facility (ZTF) \citep{ztf_2022} should total up to about \num{10000} spectroscopically confirmed SNe Ia, a consequent increase compared to previous surveys. Joint analysis has been performed to benefit from the different existing surveys, reducing the statistical uncertainty on the measurement of cosmological parameters (\citealt{Betoule_2014,Scolnic_2018,Brout_2022,rubin2023union}). Additionally, we expect the SNe Ia catalog to reach a new order of magnitude within the next decade thanks to the Legacy Survey of Space and Time (LSST) undertaken by the Vera Rubin Observatory \citep{lsst}, which is expected to observe between 120,000 and 170,000 SNe Ia up to redshifts $z \sim 0.3$ \citep{lsst_2022}. With this tremendous increase, the statistical uncertainty on cosmological parameters will consequently diminish, leaving the photometric calibration as one primary source of systematic uncertainty. Therefore, the photometric calibration needs to reach sub-percent precision to benefit from the incoming statistics of the present and future surveys.

SNe Ia survey bandpasses are calibrated relatively to the CALSPEC catalog of spectrophotometric standard stars (\cite{Bohlin_2020}), which relies on the radiative transmission model of white dwarf atmospheres (\cite{Narayan_2019}). To transfer this calibration to the photometric survey, it is necessary to consider two additional components: (i) the terrestrial atmospheric transmission and (ii) the survey filter transmissions as a function of wavelength, with particular attention given to the bandpasses edges wavelength position. The former can be inferred by airmass regression with observations of CALSPEC reference stars or by slitless spectrophotometric analysis with a dedicated telescope such as the Rubin Observatory's auxiliary telescope (AuxTel), as detailed in \cite{Neveu_2024}. The latter needs beforehand precise measurements of the bandpasses throughput. Multiple strategies have been developed to provide this measurement on different surveys. The most common approach consists of using calibrated sensors such as the ones supplied by the National Institute of Standards and Technologies (NIST) \citep{houston2008detectors} to monitor a light source used to illuminate a telescope to measure its throughput and filter transmissions. Several approaches involve diffusion on a flat-field screen toward the instrument \citep{stubbs2006,marshall2013}. Other designs have been developed, like \cite{Lombardo_2017}, which involves integrating spheres and parabolic mirrors to redirect the light in a parallel beam.

The StarDICE experiment (\citealt{Betoule_2023}) proposes a metrology chain from laboratory flux references toward the measurement of standard star spectra. Several steps are needed to calibrate increasingly sensitive detectors and finally transfer this calibration to on-sky sources. The StarDICE \SI{40}{\centi\meter} diameter telescope is calibrated with a stable light source positioned far enough (\textasciitilde {\SI{100}{\meter}) to appear as a pointlike source and provides in-situ calibration of the instrument. Composed of LEDs, the calibrated light will emit broadband flux, which will be used to monitor the $ugrizy$ filters of the \SD{} telescope at the millimagnitude level. Beforehand, it is necessary to have a laboratory measurement of the filter transmission at high wavelength resolution to interpret the broadband LED and star measurements. In this context, a Collimated Beam Projector (hereafter CBP) has been developed to accurately measure the StarDICE telescope response $\Rtel(\lambda)$, including the optics, the CCD camera quantum efficiency, and the filter transmission. 

The CBP was initially designed for the LSST telescope (\citealt{ingraham2016}), and prototypes have been used to measure the throughput of the CTIO \SI{0.9}{\meter} telescope and the DECam wide field imager, respectively, in \cite{coughlin2018} and \cite{coughlin2016}. Compared to flat-field illumination devices, the CBP can project monochromatic light in a collimated beam, which provides parallel monochromatic illumination over a portion of the primary mirror. A first prototype of the CBP for StarDICE has been developed in \cite{Mondrik_2023} as a proof-of-concept and measured the instrument throughput with a precision of \textasciitilde 3\% for wavelengths between \SI{400}{\nano\meter} and \SI{800}{\nano\meter}, and a wavelength calibration estimated at \textasciitilde \SI{0.2}{\nano\meter}. 

This paper details the enhanced version of the StarDICE CBP, now equipped with a tunable laser as a monochromatic source that injects light at the focal point of a Ritchey-Chrétien telescope mounted backward. The output light of the CBP is collimated and illuminates the StarDICE telescope pupil to measure its throughput. Combining the calibration effort provided by \cite{houston2008detectors} and \cite{solarcell} for the detectors and all the lessons learned from previous prototypes, we aim at measuring the StarDICE telescope throughput and $ugrizy$ filter transmissions at the sub-percent level. The other goal is to confirm the wavelength calibration accuracy of \SI{0.2}{\nano\meter}. This study has two major intents: (i) to provide a first measurement of the StarDICE filter transmission to contribute to the metrology chain of the experiment, and (ii) to serve as a pathfinder for the future measurement of the LSST telescope with its dedicated version of the CBP.

The following sections of the paper are structured as follows: Section~\ref{sec:setup} presents the laboratory setup of the experiment and gives an overview of the campaign of measurements. Section~\ref{sec:rcbp} details the measurements of the CBP optics response. Section~\ref{sec:rsd} presents the obtained measurements of StarDICE telescope throughput and filter transmissions. The StarDICE full-pupil synthesizing methodology and main results are presented in Section~\ref{sec:pupil_stitching} and discussed in Section~\ref{sec:discussion}.


\section{Laboratory setup}
\label{sec:setup}

Our setup consists of three primary components: the \SD telescope, a large area photodiode employed as a calibration reference, and the CBP
that illuminates one or the other. The following section will describe each component and detail its operations.

Compared to the original CBP design outlined in \cite{Mondrik_2023}, a significant enhancement is using a solar cell to calibrate and monitor the CBP optical transmission between the optical sphere and the telescope output. Other noteworthy improvements include (1) replacing the laser light source from NIST with an EKSPLA NT 252, which offers a slightly more favorable wavelength power cutoff, (2) implementing a filtering system to prevent laser harmonics from being injected into the light beam and (3) synchronizing the photocurrent readings with laser light emission, facilitating the extraction of the signal, (4) replacing the Hasselblad camera by a Ritchey-Chretien telescope as collimating optics to remove all chromatic effects and enlarge the illumination region.	

\subsection{\SD}
\label{sec:stardice}

The \SD photometric instrument consists of a Newton telescope with a primary mirror of \SI{40}{\centi\meter} diameter (16'') and \SI{1.6}{\meter} focal length ($f/D = 4$). The focal plane hosts an Andor Ikon-M DU934P-BEX2-DD camera equipped with a deep depleted, back-illuminated CCD sensor (E2V DU934P). The active area of the sensors is $\SI{13.3}{\milli\meter}\times\SI{13.3}{\milli\meter}$ divided in $1024\times 1024$ square pixel of \SI{13}{\micro\meter} side. In this baseline setup, the pixel resolution is \SI{1.68}{\arcsec} and the field of view $\SI{28.6}{\arcmin}\times\SI{28.6}{\arcmin}$.
"ugrizy"
A 9-slots \SI{28.5}{\milli\meter} filter wheel positioned in front of the camera features 6 interference filters in the $ugrizy$ photometric system, a Star Analyser 200 diffraction grating, and a \SI{0.2}{\milli\meter} pinhole. The remaining slot is left empty. Aside from the optional filters, the only other glass part in the light path is the non-coated fused-silica window of the CCD cell\footnote{The manufacturer code for this window is WN35FS(BB-VV-NR)W}. A \SI{0.5}{\degree} wedge affects the window's two sides.

The z-position of the camera-filter wheel assembly is adjustable over \SI{9}{\centi\meter}, allowing to focus from distances as close as \SI{35}{m} up to infinity. The secondary mirror, with a diameter of 11cm, is oversized to ensure the fully-illuminated plane extends over the sensor with a comfortable margin in all optical configurations.

In operation, the camera sensor is thermoelectrically cooled down to a temperature of \SI{-70}{\celsius}, delivering a median dark current of \SI{0.15}{e^-/s}, neglected for the $\sim 1s$ exposures considered in this study. When operated in the lab, the telescope was mounted on a custom altazimuth mount to enable easy alignment with the CBP.

\subsection{Collimated Beam Projector}
\label{sec:cbp}

The Collimated Beam Projector (CBP) general setup requires the following components: a tunable monochromatic light source and an optic device able to recreate a parallel beam from a point source. In our case, the light source is an Ekspla NT252 tunable laser, using a Q-switched pump laser at \SI{1064}{\nano\meter} and non-linear crystals to produce powerful monochromatic pulses from 335 to \SI{2600}{\nano\meter}. The pulse duration is fixed and lies between 1 and \SI{4}{\nano\second} with an energy of \SI{1.1}{\milli\joule} in the near-infrared. The energy can be decreased at will by a factor of 2 using the tuning of the Q-switch (namely QSW in the following), degrading the resonant cavity's quality factor. Pulses are shot with a fixed frequency of \SI{1}{\kilo\hertz} and can be shot in two modes. The first one is called the "continuous mode", since it shoots pulses continuously with a \SI{1}{\kilo\hertz} frequency, while the second one is called "burst mode", sending packets of pulses by bursts. Each burst is composed of 1 to 1000 pulses, meaning the duration of a burst is restrained between \SI{1}{\milli\second} and \SI{1}{\second}. The pulses have a maximum linewidth below \SI{10}{\per\cm}, which can be converted into a maximum spectroscopic width below \SI{0.4}{\nano\meter} around \SI{600}{\nano\meter}. This is an upper limit quoted by the manufacturer, but measurements taken using a similar laser show bandwidths going from \SI{0.08}{\nm} to \SI{0.48}{\nm} in the 350 to \SI{1100}{\nm} range \citep{woodward2018}. The standard deviation of pulse energy is around 2.5\%. This laser is composed of three different operational configurations, from 335 to \SI{669}{\nano\meter}, from 670 to \SI{1064}{\nano\meter} and above \SI{1064}{\nano\meter}, which results in three different regimes of power and light contamination in the CBP. This laser was chosen for its wide wavelength range and the high pulse energy, as the source power is a crucial criterion for CBP calibration with the solar cell, our reference calibration photodiode (see Section~\ref{sec:solarcell}). 

The laser output is polluted by light from the pump laser of other resonances in the system, both spatially and chromatically. The spatial pollution is stopped with a diaphragm. On the other hand, a filter wheel that contains three different broad bandpass filters purifies the laser light from pump photons or another parasite signal. In particular, we use a red-pass filter RazorEdge LP03-532RU-25 to filter out the \SI{532}{\nano\meter} pump photons in the regime 645 to \SI{1074}{\nano\meter}, and a infrared-pass filter RazorEdge LP02-1064RU-25 above \SI{1074}{\nano\meter} to filter $<\SI{1064}{\nano\meter}$ photons appearing in this regime. We also use a blue-pass filter BrightLine Multiphoton FF01-680/SP-25 in the regime 530 to \SI{645}{\nano\meter} to remove contamination not detected in the spectrograph and which we were not able to identify. A black metallic box encloses the entire optical stage to minimize light scattering in the room.

Being filtered, the light is focused and injected into an optical fiber Thorlabs MHP910L02 with a wide core diameter of \SI{910}{\micro\meter}, which is plugged into an IS200-4 $\Phi$2'' integrating sphere from Thorlabs, which has an internal \SI{50}{\mm} diameter and is composed of 1 input and 4 output ports. The integrating sphere dilutes the flux, breaks the laser light coherence, and eliminates any spatial dependence to achieve a uniform surface brightness. Two monitoring instruments are plugged into the sphere. First, a silicon photodiode Thorlabs SM05PD3A with high efficiency in the UV is mounted on an output port orthogonal to the laser input port to control the laser power stability. It is mounted behind a pinhole to reduce the photon flux and ensure it works in its linear regime. It is read by a Keithley 6514 electrometer at a rate of \SI{50}{\hertz}. Additionally, an OceanOptics QE65000 fiber spectrograph is plugged into another output to monitor the true laser wavelength and spectral purity. One port is left free to plug a calibration lamp to calibrate the spectrograph when needed. Finally, a slider with pinholes of different diameters is mounted on the output port connected to the CBP optics. Figure~\ref{fig:sphere} shows a schematic of the integrating sphere and the instruments plugged into it.

We used three different pinholes, of diameter \SI{75}{\micro\meter}, \SI{2}{\milli\meter} (respectively P75HK and P2000HK from Thorlabs) and \SI{5}{\milli\meter} (homemade). The \SI{5}{\mm} pinhole is the largest possible given the \SD field of view and gives the maximum flux. The other pinholes are used for systematic checks and filter edge analysis. The pinhole slider is attached to the ocular of a 154/1370 Ritchey-Chrétien Omegon telescope to position the pinhole at the focal point of the optics. The light injected in the telescope mounted backward will project a parallel and collimated beam. An iris diaphragm is positioned \SI{16}{mm} after the pinhole and adjusted to cut light that would otherwise miss the telescope's secondary mirror. A small amount of light scatters on the iris blades, best observed in telescope images of the CBP taken with the \SI{2}{mm} pinhole configuration. In these images, it forms a faint ring with a radius of approximately \SI{340}{pixels}, distinct from the main spot. When comparing the annular photometry of the ring to that of the main spot, the fraction of scattered light is usually smaller than \num{6e-4} between 400 and \SI{900}{nm}.

The assembly is mounted on a robotic Celestron NexStar Evolution 6 altazimuth mount. As a last step, we cropped the output of the Ritchey-Chrétien Omegon with a mask shaped as a quarter of a disk corresponding to one quadrant of its secondary mirror spider so that the actual beam shape is only a fourth of the complete aperture. This ensures that the entire beam can fit inside the reference solar cell footprint.

\begin{figure}
    \centering
    \includegraphics[width=1\columnwidth]{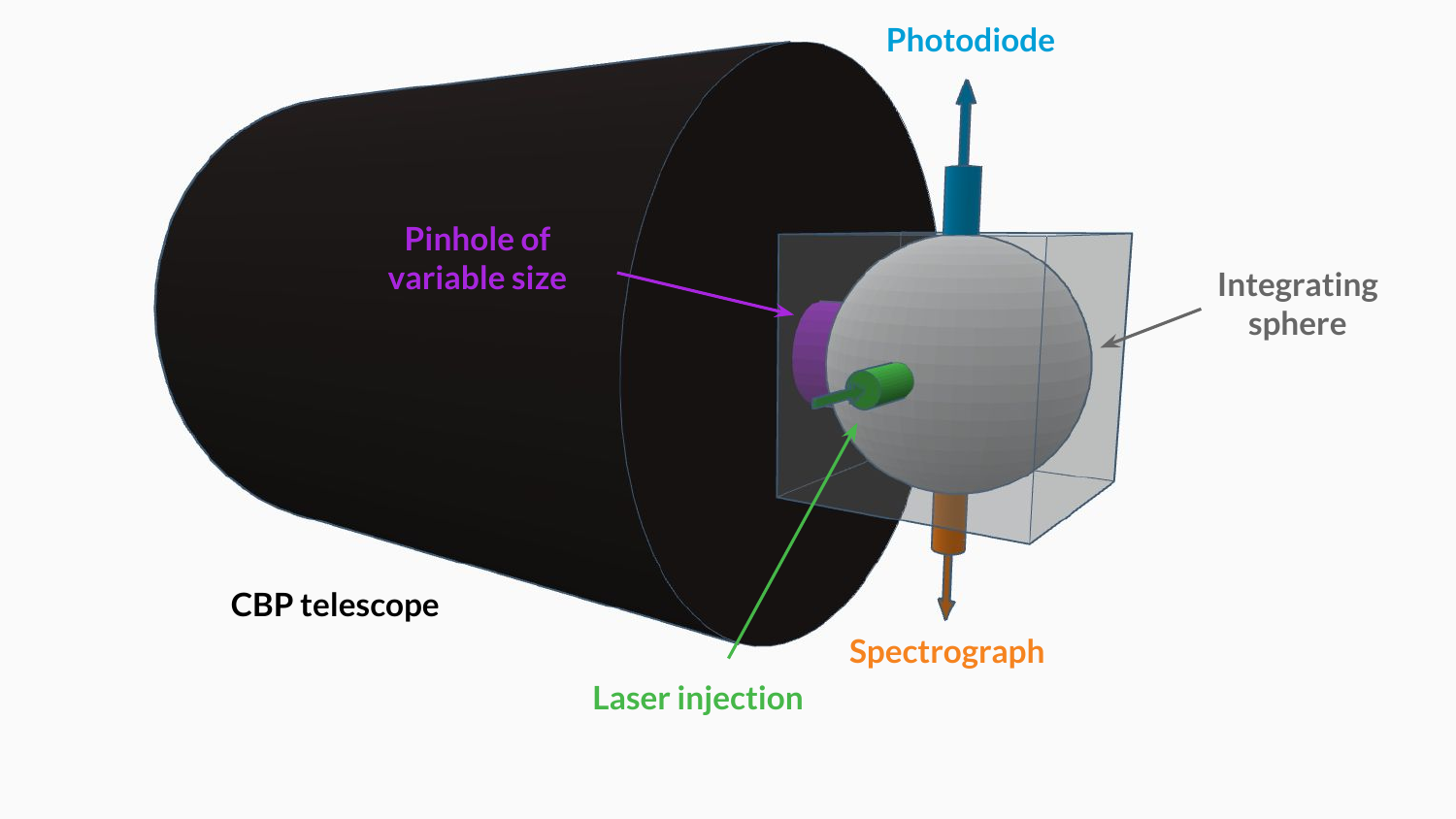}
    \caption{Schematic of the integrating sphere}
    \label{fig:sphere}
\end{figure}

\begin{figure*}[ht]
\centering
\includegraphics[width=\textwidth]{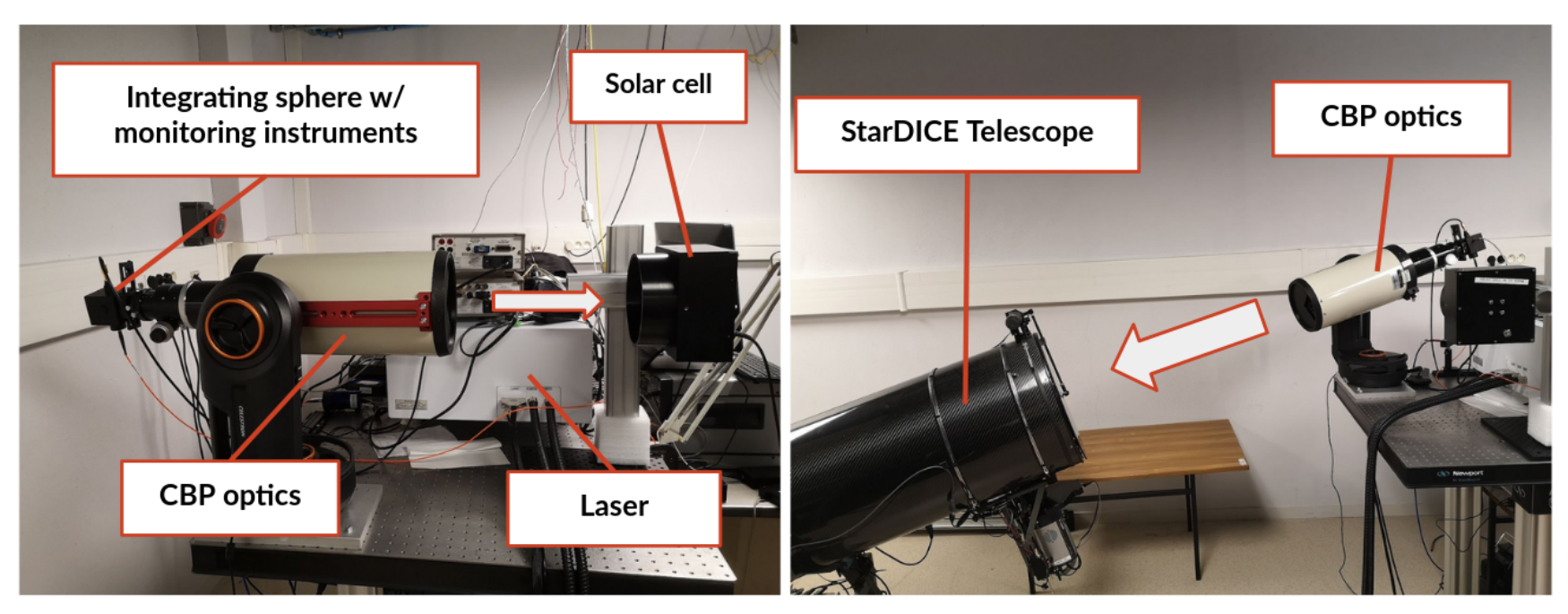}
\caption{Pictures of the different CBP setups. Left: CBP setup when shooting in the solar cell. Right: CBP setup when shooting in the \SD telescope.}
\label{fig:cbp_setup}
\end{figure*}

\subsection{Solar cell description}
\label{sec:solarcell}

We use a C60 solar cell of 3$^{\mathrm{rd}}$ generation from Sunpower as our calibration reference. Its sensitive area forms a square with \SI{12.5}{\centi\meter} side, setting the maximum size of the CBP beam, which can be accurately calibrated. This solar cell is set on a two-axis mount, one that allows a movement in the direction of the optical axis of the CBP telescope and another vertical axis, which allows the adjustment of the height of the solar cell. It is placed at \SI{16}{\cm} approximately from the telescope aperture. This solar cell is connected via a coaxial cable to a Keysight B2987A electrometer, the same that was used for its calibration. The charges are measured at a rate of \SI{500}{\hertz}. In Figure~\ref{fig:cbp_setup} right, we show a picture of the setup when the CBP is aiming at the \SD telescope described in section~\ref{sec:stardice}.

\subsubsection{Quantum efficiency measurement}
 
The quantum efficiency (QE) as a function of wavelength for the solar cell was
measured relative to a NIST-calibrated photodiode by using a monochromator as a
light source and using an electrometer to measure the current of the solar cell
and of the photodiode at each wavelength. The monochromator wavelength was
calibrated with a spectrograph relative to a mercury calibration source. Details
of the setup are discussed in ~\cite{solarcell}. The quantum efficiency of the
solar cell used for these measurements is shown in Fig.~\ref{fig:sc_qe}. Five
measurements were taken per source wavelength with a \SI{1}{\nm} step. The
smoothed average per wavelength is reported in Fig.~\ref{fig:sc_qe}. The cited
uncertainty arises from the RMS of the current measurements. The glitch at
\textasciitilde\SI{550}{\nm} caused by inserting a 550 nm long pass
filter to cut off second-order light contamination is masked by a linear
interpolation.

\begin{figure}[!h]
\centering
\includegraphics[width=\columnwidth]{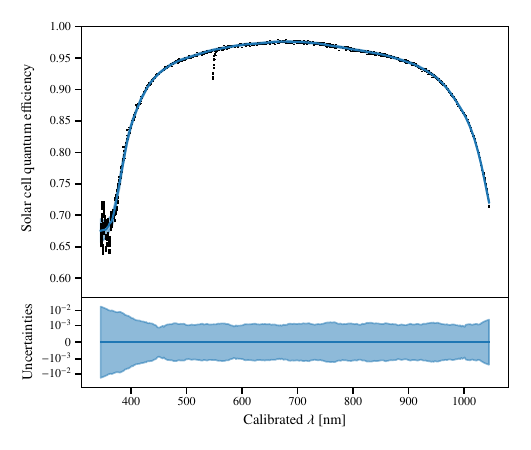}
\caption{Solar cell quantum efficiency with respect to wavelength (top) and error bar sizes (bottom).}
\label{fig:sc_qe}
\end{figure}

The QE of the solar cell was measured within a temperature range from \SI{32}{\degreeCelsius} to \SI{39}{\degreeCelsius}. The QE increased slightly at longer wavelengths as the temperature increased (see Fig.~\ref{fig:SC_temp}). This trend is consistent with the temperature dependence of silicon's QE \citep{Green_2008}. At 1050 nm, the QE changes by more than 0.1 percent per degree.
\begin{figure}[!h]
\centering
\includegraphics[width=\columnwidth]{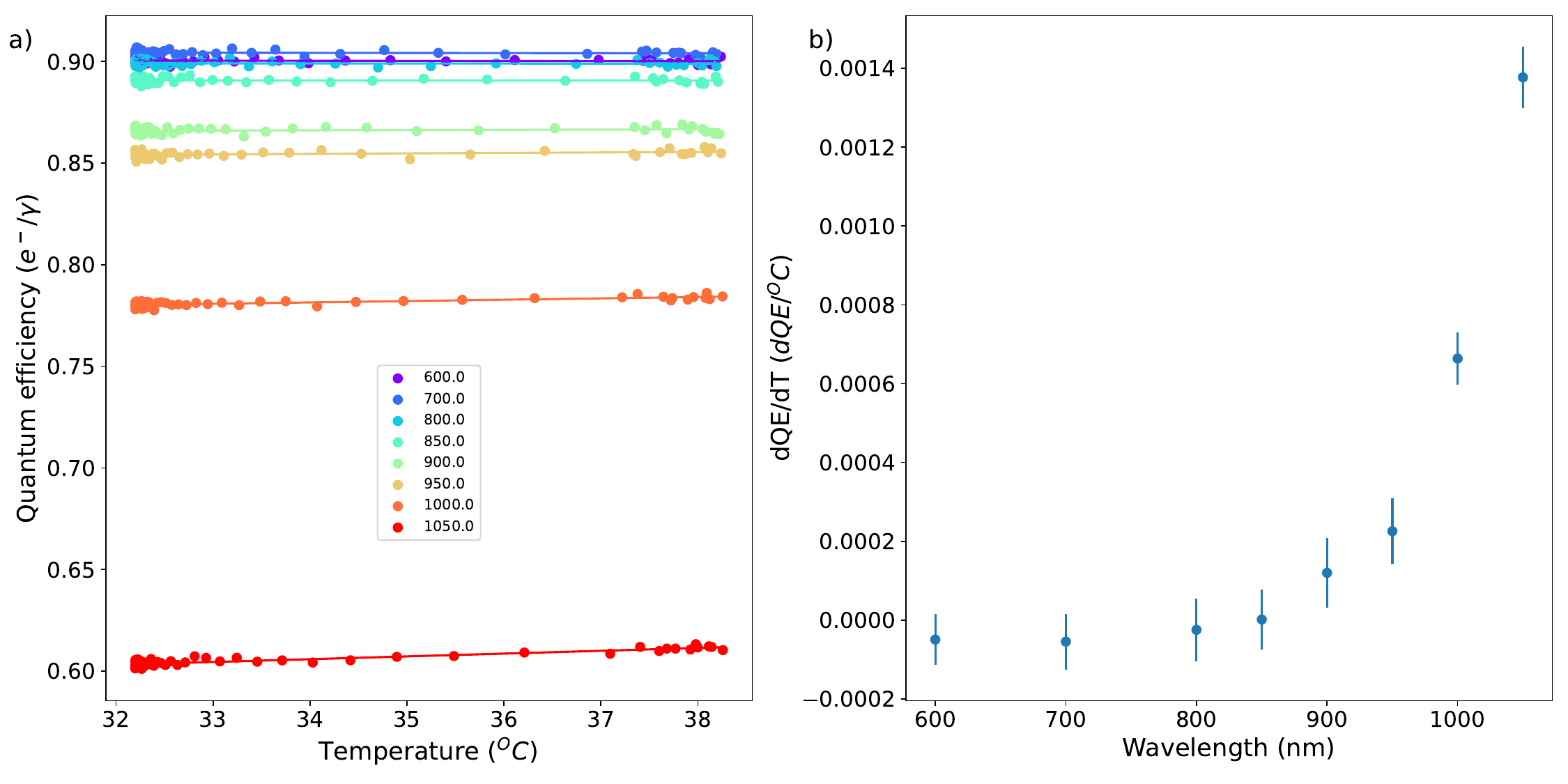}
\caption{(a) Solar cell quantum efficiency vs temperature for wavelengths ranging from 600 to 1050 nm. (b) Fitted change in quantum efficiency per degree Celsius for wavelengths ranging from 600 to 1050 nm.}
\label{fig:SC_temp}
\end{figure}

The effect of the angle of incidence on the solar cell QE was also measured. The solar cell was rotated up to 35 degrees off-axis. The results are shown in Fig.~\ref{fig:SC_angle}. There is a stronger dependence on the incidence angle at wavelengths less than 600 nm, but the change in QE for wavelengths greater than 400 nm is less than \num{5e-4}	 per degree.
\begin{figure}[!h]
\centering
\includegraphics[width=\columnwidth]{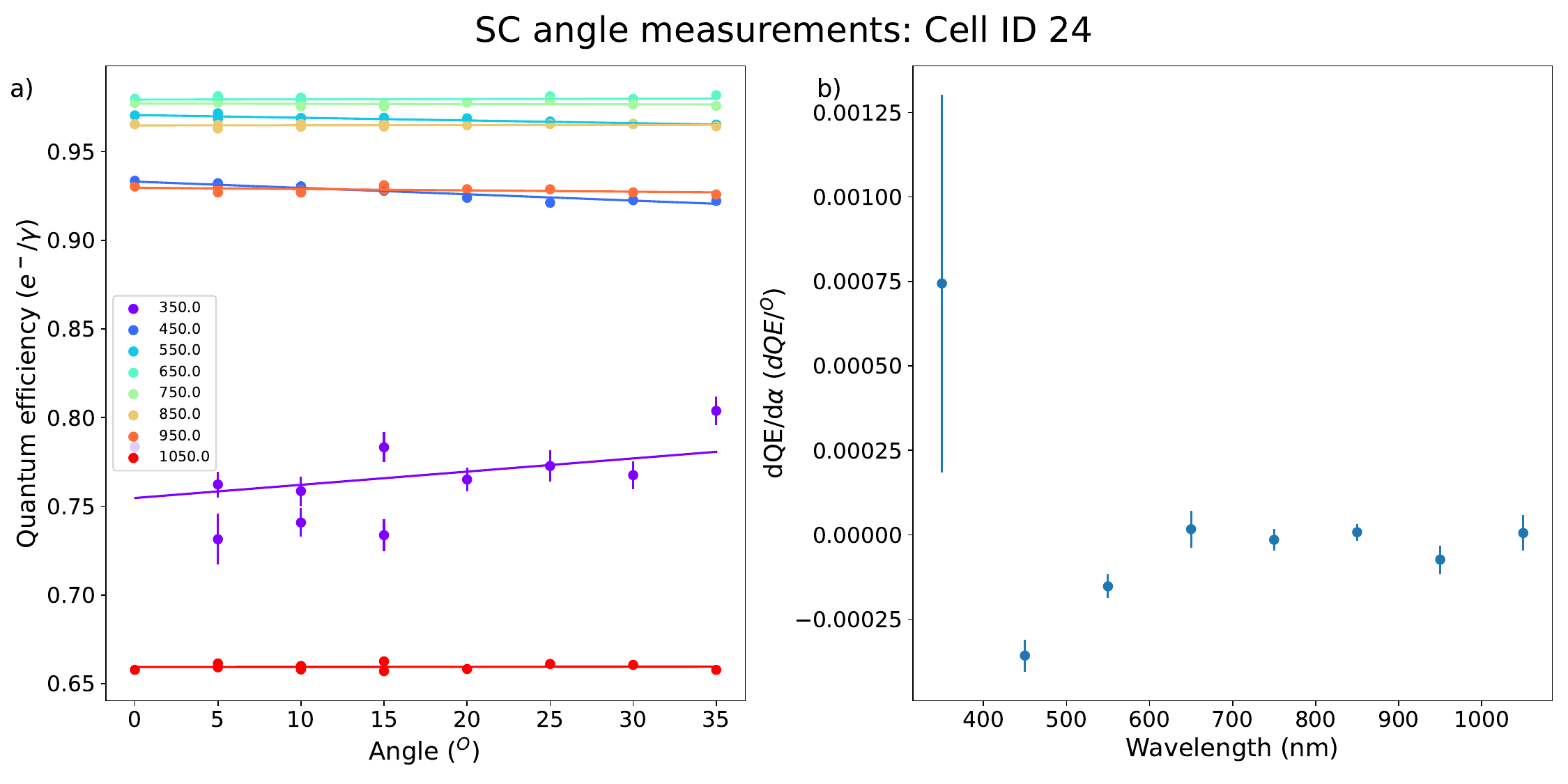}
\caption{(a) Solar cell quantum efficiency vs solar cell tilt relative to normal incidence for different wavelengths. (b) Fitted change in quantum efficiency per degree of solar cell tilt relative to normal incidence vs wavelength.}
\label{fig:SC_angle}
\end{figure}

The solar cell was aligned perpendicularly to the CBP output beam with a precision better than \SI{1}{\degree} and has not been moved throughout the entire measurement campaign. Meanwhile, room temperature was monitored and has not varied more than \SI{2}{\degreeCelsius}.

\subsubsection{Dark current characterization}

Due to the low resistance of the solar cell, we observed a relatively high current of approximately \SI{20}{\nano\ampere} even in dark conditions. Therefore, we built a device that uses a precision voltage source with a tunable voltage divider to inject a counter-current that cancels this contribution. The current value was tuned to observe approximately no drift when using the Keysight in charge mode inside the usual acquisition time window ($< \SI{1}{\minute}$). In doing so, we avoided saturating the electrometer when using the solar cell in dark and laser-on conditions. 

After canceling the dark current drift, the power spectrum of the solar cell dark current revealed a $1/f$ noise, with power line harmonics contributions at \SI{50}{\hertz}, \SI{100}{\hertz}, and \SI{150}{\hertz} (see Figure~\ref{fig:darkcurrentspectrum}). To investigate the source of the $1/f$ component in the power spectrum, we compared the output of the Keysight when connected to a resistor with the same value as the shunt resistance of the solar cell. We found that the two power spectra were identical, leading us to conclude that the fluctuations in the trans-impedance amplifier bias voltage are the source of this $1/f$ noise. These fluctuations cause parasitic currents to flow through the load resistance. By increasing the load resistance, we can decrease the amplitude of the resulting noise. Among a set of calibrated solar cells, we therefore chose the one with the highest shunt resistance at our disposal ($R_{\rm shunt} = \SI{1.8}{\kilo\ohm}$) and limited the burst durations to at most 200 pulses (\SI{200}{\ms}). In doing so, in the burst time window, the Keysight noise is dominated by the power line harmonics, whereas the random $1/f$ noise remains subdominant.

\begin{figure}[h]
\begin{center}
\includegraphics[width=1\columnwidth]{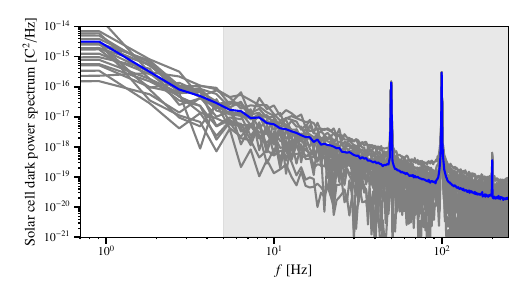}
\end{center}
\caption[]{Solar cell dark current power spectrum for 20 runs without light in the solar cell (gray curves) and their averages (blue). The light gray region encompasses the expected noise spectrum contribution for \SI{200}{\ms} laser bursts.}
\label{fig:darkcurrentspectrum}
\end{figure}

\FloatBarrier   

\subsection{Time synchronisation}
\label{sec:synchro}

The precision of the analysis is significantly improved if the laser pulses can be accurately identified within the photocurrent time series. This is particularly crucial for solar cell measurements, where signal-to-noise ratios are typically lower.

To accomplish such synchronization, we recorded the timing of all three trigger lines of the laser and both electrometers on a single microcontroller. Upon detecting a TTL pulse on any of these lines, the microcontroller registers a 32-bit timestamp using its internal clock counter, enabling synchronization with a resolution of 500 nanoseconds over extended durations (up to 2100 seconds). This custom synchronization device, Arduino-based, whose code is openly accessible \cite{logic_timer}, is referred to as the digital analyzer throughout the remainder of this text.

\subsection{Measurement principle}

Table~\ref{tab:quantities} summarizes the three quantities measured with the two setups in Figure~\ref{fig:cbp_setup}.

\begin{table}
  \centering 
  \caption{Definition of measured quantities in our \SD+CBP setup.}
    \begin{tabular}{M{1cm} M{7cm}} 
        \hline\hline 
        Name & Description \\
        \hline
        \bf{$\Qphot(\lambda)$} [C] & The charge per burst collected by the integrating sphere monitoring photodiode, measured by the Keithley 6514. \\

        \bf{$\Qsolar(\lambda)$} [C] & The charge per burst collected by the solar cell, measured by the Keysight 2987A. \\
        \bf{$\Qccd(\lambda)$} [ADU] & The charge collected by the Andor CCD camera of the \SD telescope. \\
        \hline 
    \end{tabular}
    \label{tab:quantities} 
\end{table}

First, we need to measure the response of the CBP optics $\Rcbp(\lambda)$ by shooting into the calibrated solar cell as shown in Figure~\ref{fig:cbp_setup} left. It is computed with the Equation~\ref{eq:rcbp}. In this equation, we know the quantum efficiency of the solar cell $\Esolar$ from Figure~\ref{fig:sc_qe}, and $e$ is the elemental charge of the electron.

\begin{equation}
    \Rcbp(\lambda) = \frac{\Qsolar(\lambda)}{\Qphot(\lambda) \times \Esolar \times e}.
    \label{eq:rcbp}
\end{equation} 

Once we have this response, we can shoot inside the \SD telescope as shown in Figure~\ref{fig:cbp_setup} right, and obtain $\Rtel(\lambda)$ with the Equation~\ref{eq:rsd}.

\begin{equation}
    \Rtel(\lambda) = \frac{\Qccd(\lambda)}{\Qphot(\lambda) \times \Rcbp(\lambda)}.
    \label{eq:rsd}
\end{equation}

We will focus on $\Rcbp(\lambda)$ and $\Rtel(\lambda)$ respectively in sections \ref{sec:rcbp} and \ref{sec:rsd}.

\subsection{Measurement overview}
\label{sec:strategy}

The schedule for all measurements conducted to determine $\Rtel(\lambda)$ and estimate associated systematic uncertainties is outlined in Table~\ref{tab:schedule}. Each row in the table corresponds to a specific hardware configuration and measurement run. We have assigned labels to each run and specified the following details: (1) the target of the CBP, (2) the pinhole used in the CBP slide, (3) the QSW setting for the laser, (4) the filters used in the \SD camera filter wheel (when applicable), (5) any specificity of the measurement, and (6) the number of runs.

The measurement campaign has started and ended with a calibration of the spectrograph. A calibration Hg-Ar lamp was plugged into the integrating sphere, and its light was measured using a spectrograph. This corresponds to lines No.~1 and No.~13 of the Table~\ref{tab:schedule}.

Two different pinholes in the CBP slides were used for these measurements. When shooting into the \SD telescope, the \spinhole pinhole forms a point-like image of about 10 pixels in diameter well suited for photometry while avoiding issues related to ghosting. When shooting into the solar cell, the \bpinhole pinhole is necessary to achieve a good signal-to-noise ratio. Since $\Rcbp(\lambda)$ depends slightly on the pinhole diameter, we also need to intercalibrate the two responses $R_\mathrm{CBP}^{\mathrm{\SI{5}{\milli\meter}}} (\lambda)$ and $R_\mathrm{CBP}^{\mathrm{\SI{75}{\micro\meter}}} (\lambda)$. This inter-calibration can be performed thanks to the measurements in line No.~8 of Table~\ref{tab:schedule}.

A significant issue with the CBP is that its output light does not illuminate the entirety of the \SD primary mirror as an astrophysical source would do, but only a portion of it. It is necessary to perform a \textit{pupil stitching} to reconstruct the transmission of the mirror by combining the measurement of $\Rtel(\lambda)$ when shooting at different positions on the mirror. Only the point of impact on the mirror is modified when doing so, but the point of incidence on the focal plane is the same. The different positions are shown in Figure~\ref{fig:8_mirror_positions}. The pupil stitching corresponds to lines No.~2 and No.~3 of Table~\ref{tab:schedule}. The method used to perform the pupil stitching is detailed in Section~\ref{sec:pupil_stitching}.

To check the uniformity of the \SD focal plane, a measurement of $\Rtel(\lambda)$ at 16 different positions on the focal plane has been performed. Only the position on the focal plane is modified, while the point of impact on the mirror stays the same. This dataset corresponds to the line No.~12 of Table~\ref{tab:schedule}. 

Finally, some systematic measurements have been carried out. A cap has been placed on the CBP output to measure the room's background for both setup configurations, corresponding to lines No.~9 and No.~10 of Table~\ref{tab:schedule}. A measurement of the scattered light is possible with dataset No.~11 by shifting the position of the solar cell from the CBP output of approximately \SI{16}{\centi\meter}.

\begin{table*}[t]{}
  \centering
  \caption{Detailed schedule of the measurements.}
    \begin{tabular}{M{.25cm} M{3cm} M{1.6cm} M{1.1cm} M{1.4cm} M{3cm} M{4cm} M{1.25cm}}
        \hline\hline
         \bf{N$^{\circ}$} & \bf{Label} & \bf{Target} & \bf{Pinhole} & \bf{QSW} & \bf{\SD bands} & \bf{Specificity} & \bf{Number of runs} \\ 
         \hline
         1 & Wavelength calibration & Spectrograph & - & - & - & Hg-Ar lamp light source & 1 \\ 
         
         2 & Radial pupil stitching & \SD & \SI{75}{\micro\meter} & MAX & \shortstack{u, g, r, i, z, y, \\ EMPTY, GRATING} & 4 mirror radial positions & 1 per position \\
         
         3 & Quadrant pupil stitching & \SD & \SI{75}{\micro\meter} & MAX & EMPTY & 4 mirror  quadrant positions  & 1 per position \\
         
         4 & Repeatability measurement & \SD & \SI{75}{\micro\meter} & MAX & \shortstack{u, g, r, i, z, y, \\ EMPTY, GRATING} & Fixed mirror and focal plane position & 3 \\
         
         5 & CBP response calibration before & Solar cell & \SI{5}{\milli\meter} & 298, MAX & - & CBP monitoring before \SD measurements & 5 \\
         
         6 & \SD main calibration & \SD & \SI{5}{\milli\meter} & MAX & \shortstack{u, g, r, i, z, y, \\ EMPTY, GRATING} & Fixed mirror and focal plane position & 5 \\
                  
         7 & CBP response calibration after & Solar cell & \SI{5}{\milli\meter} & 298, MAX & - & CBP monitoring after \SD measurements & 5 \\
         
         8 & Pinholes inter-calibration & \SD & \SI{75}{\micro\meter}, \SI{2}{\milli\meter}, \SI{5}{\milli\meter} & MAX & EMPTY & 3 different pinhole sizes & 1 per pinhole \\
            
         9 & \SD background measurements & \SD & \SI{5}{\milli\meter} & MAX & EMPTY & Cap on CBP output & 1 \\
            
         10 & Solar cell background measurements & Solar cell & \SI{5}{\milli\meter} & MAX & - & Cap on CBP output & 2 \\
            
         11 & Solar cell distance calibration & Solar cell & \SI{5}{\milli\meter} & 298, MAX & - & 2 solar cell positions at \SI{16}{\centi\meter} relative distance & 1 per position \\
            
         12 & Focal plane measurement & \SD & \SI{75}{\micro\meter} & MAX & EMPTY & 4x4 grid positions on the \SD focal plane & 1 per position \\
           
         13 & Wavelength calibration & Spectrograph & - & - & - & Hg-Ar lamp light source & 1 \\ 
         \hline
    \end{tabular}
    \label{tab:schedule}
\end{table*}

\begin{figure}[!h]
\centering
\includegraphics[width=\columnwidth]{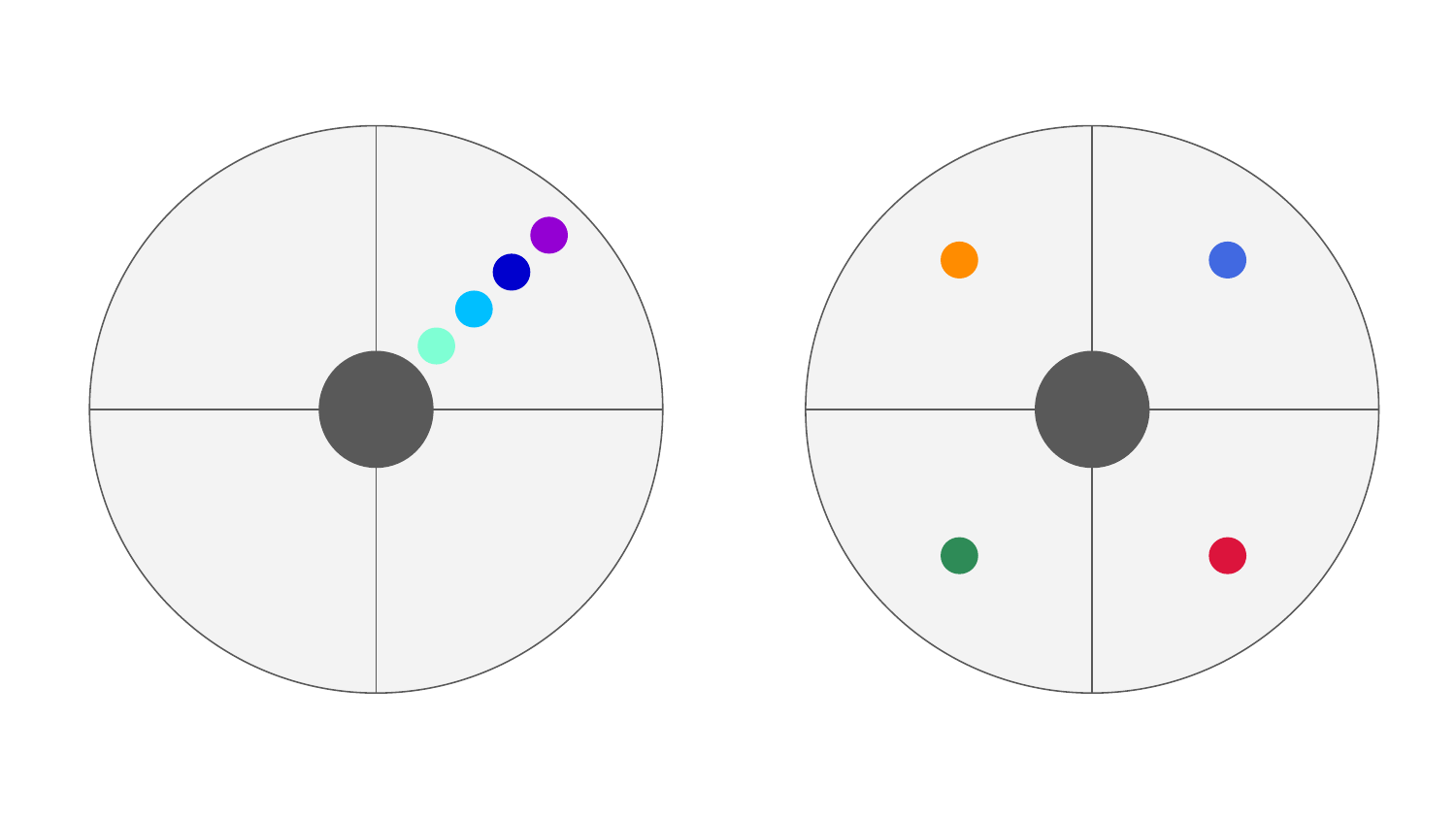}
\caption{Left: Schematic of the 4 different radial relative positions on the primary mirror of the \SD telescope. Right: Schematic of the 4 different quadrant relative positions on the primary mirror of the \SD telescope.}
\label{fig:8_mirror_positions}
\end{figure}

\section{CBP response calibration with a solar cell}
\label{sec:rcbp}

\subsection{Optics setup}

The CBP optics have two degrees of freedom that must be adjusted
before determining its transmission: (1) the focuser distance, which
needs to be tuned so that the pinhole is at the focal plane of the
telescope and produces a collimated beam; and (2) the size of the iris
diaphragm, which needs to be adjusted so that no light misses the CBP
secondary mirror.

To adjust the focuser distance, we re-imaged the CBP pinhole in the
StarDICE camera using the StarDICE telescope set up for infinity
focus. We then adjusted the CBP focus to minimize the size of the
image in the camera and tightened the locking screws. We checked
that the focus was reasonably stable regarding manual changes of the pinhole
slots.\footnote{While this procedure ensures that the two instruments
  are perfectly conjugated, an error in the focus of one is
  compensated by an error in the focus of the other. As this procedure
  was conducted before the first stellar light of the StarDICE
  telescope, we had to use a theoretical value for its focal plane
  position, which proved to be about 2 mm behind its actual focal
  point. As a result, we estimate that the collimated beam was
  diverging by about 2~mrad. While inconsequential on the transmission
  measurement itself, the uncertainty on the exact optical arrangement
  has a practical side effect on the data taking because it affects the
  ghosting pattern which we intended to use to determine the CBP spot
  position on the StarDICE mirror see the discussion in Sect.~\ref{sec:model}}

To adjust the size of the iris diaphragm, we incrementally closed it
while monitoring the flux output of the CBP. A decrease in flux
signaled that the diaphragm was beginning to block collimated
light. Continuing to close the diaphragm would only reduce the
transmission of the CBP without additional advantages in minimizing
scattered light. Therefore, we set the diaphragm at this optimal point
and secured it to maintain stable CBP transmission.

\subsubsection{Alignment of the CBP}

To measure the CBP response, we must ensure that we aim toward the solar cell. For that, we studied the signal collected in the solar cell $\Qsolarmes$ with respect to the CBP mount coordinates in azimuth and altitude. When no measurements are taken, the parking coordinates of the CBP are set as $(\mathrm{alt} = \SI{0}{\degree}, \mathrm{az} = \SI{0}{\degree})$, and the coordinates in the Figure~ \ref{fig:cross_sc} are relative to this origin. With this figure, we can estimate the coordinates at which $\Qsolarmes$ is maximum, corresponding to the solar cell coordinates in the CBP mount frame of reference. Thus we set the solar cell coordinates at $(\mathrm{alt} = \SI{6}{\degree}, \mathrm{az} = \SI{10}{\degree})$, corresponding to the point aimed by the CBP optics for any further solar cell analysis in this paper.

\begin{figure}[h]
    \centering
    \includegraphics[width=\columnwidth]{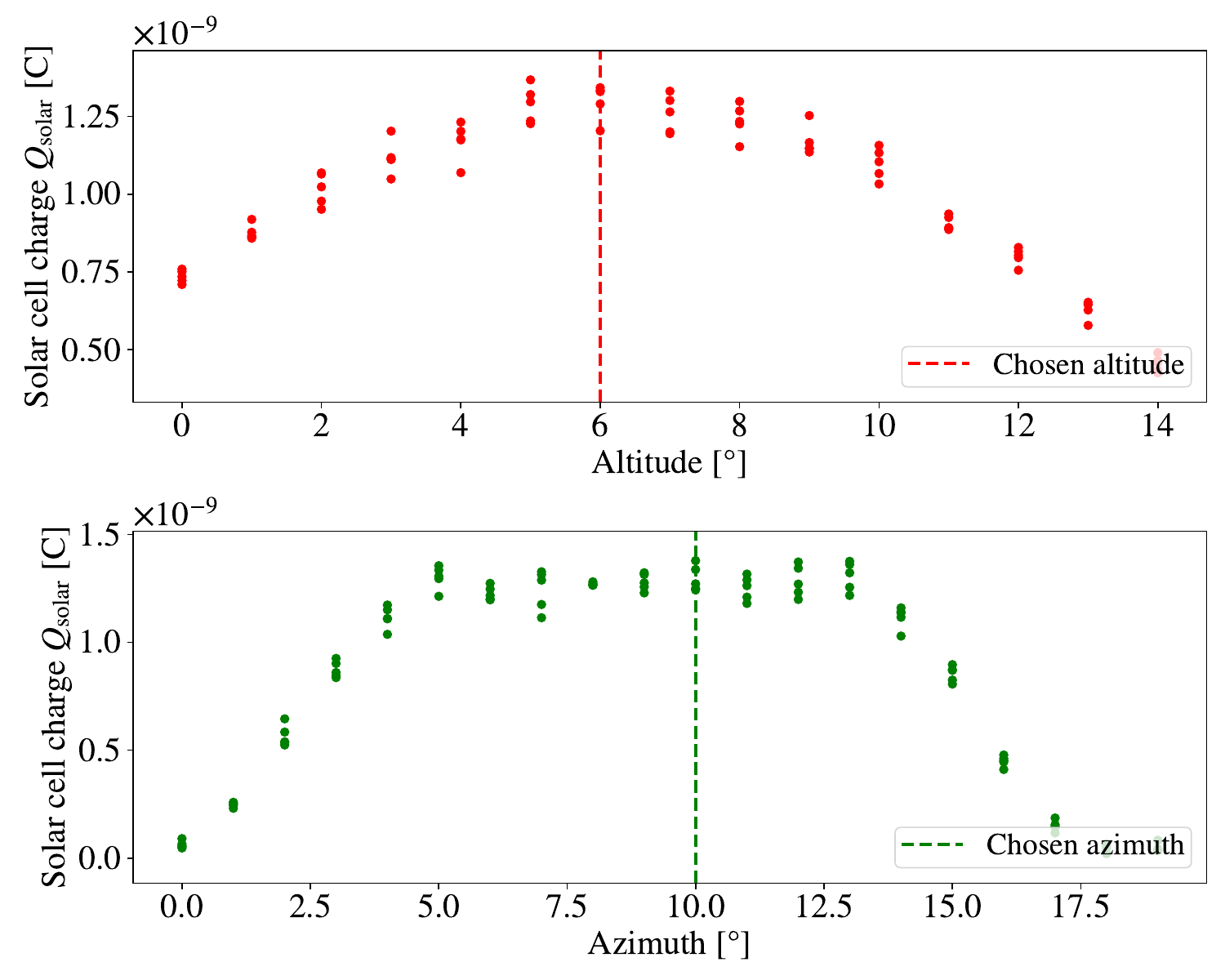}
    \caption{Alignment scan showing the evolution of the beam
      intensity in the solar cell ($\Qsolarmes$) as a function of the
      two coordinates of the CBP mount: altitude (top panel) and
      azimuth (bottom panel). A flux plateau indicates the region
      where the entire beam is contained within the cell
      footprint. The wider plateau in azimuth is due to the smaller
      extent of the beam in the horizontal direction due to the shape
      and orientation of the masked output pupil.  }
    \label{fig:cross_sc}
\end{figure}

\subsection{Description of the CBP data set}
\label{sec:cbp_datadesc}

\begin{figure*}[!h]
\centering
\includegraphics[width=\columnwidth]{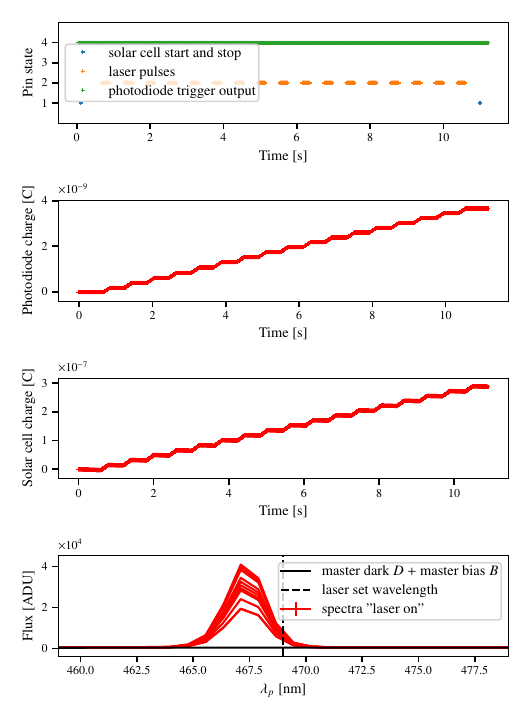}
\includegraphics[width=\columnwidth]{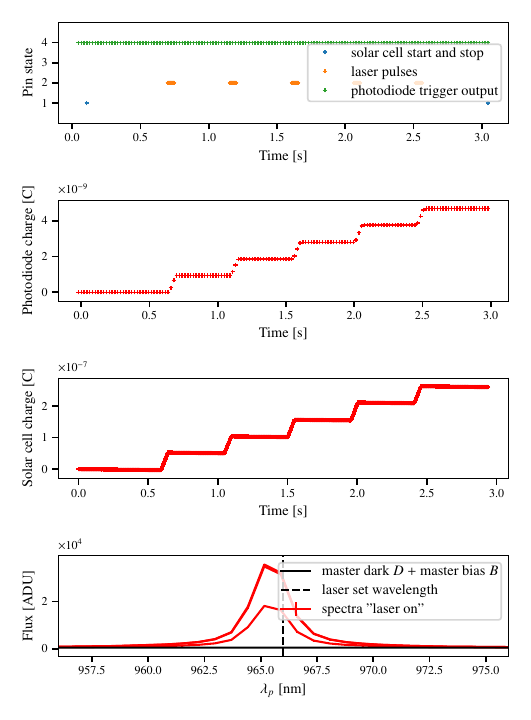}
\caption{Data set examples when targeting the solar cell. From top to bottom: typical data sets for digital analyzer (pin state 4 is the Keithley output, pin state 2 the laser trigger output, pin state 1 the Keysight start and end time acquisition time stamps), charges in the photodiode, charges in the solar cell, flux in the spectrograph. Left: typical data set at \SI{469}{\nm}. Right: typical data set at \SI{966}{\nm}.}\label{fig:sc_dataset_examples}
\end{figure*}

A typical dataset to measure the CBP response is the emission of laser bursts in the solar cell at a given wavelength. Charges in the photodiode and solar cell are recorded jointly, along with the flux in the spectrograph and the time stamps in the digital analyzer (see examples in Figure~\ref{fig:sc_dataset_examples}).

The laser emits pulses at a fixed rate of \SI{1}{\kilo\hertz}, with a power that highly depends on the wavelength. To ensure that all instruments work in their linear regime without saturation and to limit the impact of dark current fluctuations, we decided to shoot light in the solar cell in bursts of pulses separated by dark times at least as long as the burst length. The photodiode being the instrument in common for all our measurements, the number of pulses per burst at each wavelength was adjusted to minimize the change in total flux per burst collected in the photodiode over the entire wavelength range (Figure~\ref{fig:npulses}). With the \SI{5}{\mm} pinhole and the largest laser power mode, the solar cell accumulates around \SI{4}{\nano\coulomb} in a burst. To keep the $1/f$ noise of the solar cell instrumental chain under reasonable bounds, we limited the maximum length of a burst to \SI{200}{\ms}. The CBP system's linearity is checked by varying the laser power as discussed in Section~\ref{sec:sc_linearity}.  The required total flux in the photodiode was then adjusted by accumulating several bursts measured independently by the solar cell.

During a solar cell measurement run, the laser wavelengths range between \SI{350}{\nano\meter} and \SI{1100}{\nano\meter} included with steps of \SI{1}{\nm}, but are randomly chosen to avoid that long-range $ 1/f$ mode in the solar cell dark current correlates neighbored data points of the CBP response. Several runs were accumulated to enhance the signal-to-noise ratio. In particular, five runs (dataset No.~4 from Table~\ref{tab:schedule}) were recorded just before the \SD telescope measurement (dataset No.~5), and five new runs (dataset No.~6) were launched just after.

Runs with different settings have been conducted to estimate systematic uncertainties. We varied the laser global power (QSW) to assess the linearity of the instrumental light (dataset No.~4 and 6). Additionally, we checked the ambient light additive contamination (dataset No.~10) and varied the solar cell distance to estimate the output CBP scattered light (dataset No.~11).

\begin{figure}[!h]
\centering
\includegraphics[width=\columnwidth]{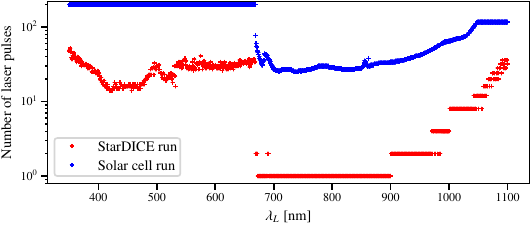}
\caption{Number of laser pulses per burst used for solar cell and StarDICE runs.}\label{fig:npulses}
\end{figure}

\subsection{Spectrograph data analysis}

In our analysis, the spectrograph is used to monitor the laser wavelength and the contamination of extra emission lines in addition to the main laser line. The main components of the contamination are a \SI{532}{\nm} half harmonic and a \SI{1064}{\nm} line from the Yagg laser pump. In the following subsection, we detail the wavelength and flux calibration procedures for this instrument and leave the discussion of the light contamination to section \ref{lightcontamination}

\subsubsection{Spectrograph data reduction}\label{sec:spectro_reduction}

The spectrograph was characterized by taking a series of dark exposures with four different exposure times (the same used in the CBP response measurement) to evaluate its gain and readout noise. The goal is to build an error model for the spectrograph.

For each exposure time, a master dark is constructed, averaging all the spectra. Then, for each spectrograph pixel, we fitted a line through the master dark values as a function of the exposure times. The intercept gave the sensor bias value for each pixel, and a master bias $B(\lambda_p)$ is assembled from the intercept values (corresponding to null exposure time), with $\lambda_p$ the raw spectrograph wavelength value before any calibration associated with each sensor pixel $p$.

The spectrograph could not be synchronized with the laser bursts. Therefore, for a given laser burst, we get spectra with different amplitudes depending on the number of pulses read by the spectrograph (see bottom row Figure~\ref{fig:sc_dataset_examples}.
All spectra are stacked into a curve $S(\lambda_p)$ to get the maximum signal to noise.

We locally fit Gaussian profiles on top of a linear background to detect emission lines in $S(\lambda_p)$. Let's call $\lambda_g$ the line centroid fitted by this Gaussian profile. The fit is unweighted to avoid any dependence of $\lambda_g$ with the line flux, and we verified it was the case. 

The Gaussian profile fit is unweighted, but a statistical uncertainty $\sigma_\lambda^{\rm noise}$ is still estimated for $\lambda_g$ as 
\begin{equation}\label{eq:sigma_lambda_stat}
    \sigma_\lambda^{\rm noise}= \dfrac{1}{\sum_p F_p} \sqrt{\sum_p \left(\lambda_p - \lambda_g\right)^2 F_p}
\end{equation}
with $F_p$ the flux in pixel $p$ after master bias subtraction. The sums are performed over a window of size $\pm 3 \sigma_g$ around $\lambda_g$ where $\sigma_g$ is the fitted line Gaussian profile RMS. Equation~\ref{eq:sigma_lambda_stat} corresponds to the propagation uncertainty formula for the average wavelength weighted by flux $F_p$ in a $\pm 3 \sigma_g$  window around $\lambda_g$. This allows us to consider the shot noise without biasing the $\lambda_g$ fit.

An example of a stacked spectrum with the detection of the laser line at \SI{643}{\nm} and the pump line at \SI{532}{\nm} is shown in Figure~\ref{fig:spectro_reduc_643}, before spectrograph wavelength calibration. 

\begin{figure}[!h]
\centering
\includegraphics[width=\columnwidth]{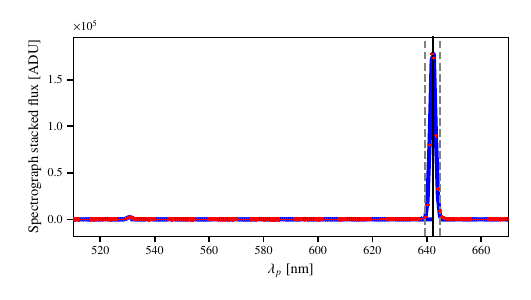}
\caption{Fit of Gaussian profiles (blue lines) in the stacked spectra (red dots) for a laser line set at $\lambda_L=\SI{643}{\nm}$, with raw spectrograph wavelengths on the abscissa axis. The contamination line at \SI{532}{nm} is also visible. The black vertical line gives the fitted laser line centroid. The grey vertical lines delimit a region of $\pm 3\sigma_g$ around the centroid, where $\sigma_g$ is the fitted RMS of the Gaussian profile. The total laser flux is the sum of the pixel values in this region.}\label{fig:spectro_reduc_643}
\end{figure}

\subsubsection{Spectrograph wavelength calibration}

We calibrated the spectrograph according to the manufacturer's specifications before and after the main data acquisition run to measure the CBP and \SD responses. Light from a Hg-Ar lamp was injected into the integrating sphere to illuminate the spectrograph sensor. The spectral lamp data were reduced as described in Section~\ref{sec:spectro_reduction}.

To transform raw sensor wavelengths $\lambda_p$ into calibrated wavelengths $\lambda_c$, we fitted a third-order polynomial function as suggested by the manufacturer to minimize the distance to Hg-Ar tabulated values $\lambda_t$. This was done by minimizing the following function: 
\begin{equation}
    \chi_\lambda^2(a_3, a_2, a_1, a_0) = \sum_{\text{lines}} \frac{\left(\lambda_t-a_3 \lambda_g^3 - a_2 \lambda_g^2-a_1 \lambda_g -a_0\right)^2}{(\sigma_\lambda^{\rm noise})^2}
\end{equation}
over the four polynomial coefficients $a_3, a_2, a_1$ and $a_0$. The sum is performed over lines with high significance (signal-to-noise ratio above 20), and known doublet lines were excluded. Lines are weighted according to their signal to noise. The minimization yields the four best-fit parameters $\hat a_i$ associated with their covariance matrix $\mathbf{C}_a$. 
As the initial Gaussian fit was unweighted, the covariance matrix $\mathbf{C}_a$ is then re-scaled with a global factor $r$ such as we get a final reduced $\chi_\lambda^2$ of one. 
Finally, detected line centroids $\lambda_g$ are transformed into calibrated wavelengths $\lambda_c$ using the third order polynomial function $c(\lambda_g)$ with the four best fit parameters:  
\begin{equation}
    \lambda_c \equiv c(\lambda_g) \equiv \hat a_3 \lambda_g^3 + \hat a_2 \lambda_g^2+\hat a_1 \lambda_g +\hat a_0
\end{equation}
and the $\sigma_\lambda^{\rm cal}$ calibration uncertainties are
\begin{equation}
    \sigma_\lambda^{\rm cal} = \left(r\vec J_a^T \mathbf{C}_a \vec J_a\right)^{1/2},\quad \vec J_a = \left(\lambda_g^3, \lambda_g^2, \lambda_g, 1\right).
\end{equation}

In Figure~\ref{fig:spectro_calib_syst}, we plot the residuals of the fit $c(\lambda_g)-\lambda_t$ in the upper panel, showing the agreement between the re-scaled data uncertainties and the uncertainties propagated to the third order polynomial function $c(\lambda_g)$ using $\mathbf{C}_a$. In the lower panel, the spectrograph calibration systematic uncertainties $\sigma_\lambda^{\rm cal}$ are emphasized: they are lower than $\SI{0.1}{\nm}$ in the entire wavelength range, even lower than \SI{0.025}{\nm} in the visible spectrum.


\begin{figure}[!h]
\centering
\includegraphics[width=\columnwidth]{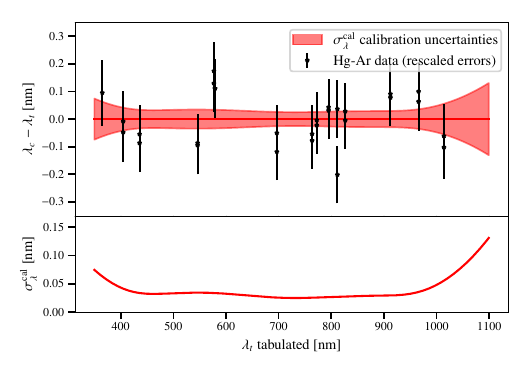}
\caption{Spectrograph calibration plot. Top: difference between the tabulated Hg-Ar emission line wavelengths $\lambda_t$ and the wavelengths computed using the spectrograph calibration function $c(\lambda_g)$. Data taken before and after the data acquisition campaign are superimposed. Their uncertainties were re-scaled with a common factor to get a final reduced $\chi^2$ of one. The red band represents the systematic uncertainty band from the $c(\lambda_g)$ fit on data with the uncertainty rescaling. Bottom: emphasis on the systematic uncertainties of the spectrograph calibration procedure.}\label{fig:spectro_calib_syst}
\end{figure}

\subsubsection{Laser wavelength calibration}

We reduce all spectrograph data sets to calibrate laser wavelengths following the recipe in Section~\ref{sec:spectro_reduction}.

We applied the third-order polynomial $c(\lambda_g)$ to transform $\lambda_g$ sensor wavelengths into calibrated wavelengths $\lambda_c$. We used the \SI{532}{\nm} line to check the quoted statistical errors $\sigma_\lambda^{\rm noise}$ on wavelength. We plotted the fitted \SI{532}{\nm} line position versus time for all data sets. The RMS is compatible with the quoted error bars on wavelength for lower signal-to-noise ratio data sets. However, for the high signal-to-noise ratio data sets, some dispersion was unaccounted for, which is explained by the fact that the Gaussian profile is an incomplete model for high signal-to-noise lines. Therefore, we added a $\sigma_\lambda^{\rm PSF}=\SI{0.012}{\nm}$ statistical uncertainty attributed to PSF modeling on wavelength uncertainty $\sigma_{\lambda}^{\rm noise}$ to get a normal distribution for the residuals normalized by the full statistical uncertainties (see Figure~\ref{fig:wavelength_error_model_consistency}), the latter being
\begin{equation}
    \left(\sigma_{\lambda}^{\rm stat}\right)^2 =  \left(\sigma_{\lambda}^{\rm noise}\right)^2 +  \left(\sigma_{\lambda}^{\rm PSF}\right)^2.
\end{equation}

\begin{figure}[!h]
\centering
\includegraphics[width=\columnwidth]{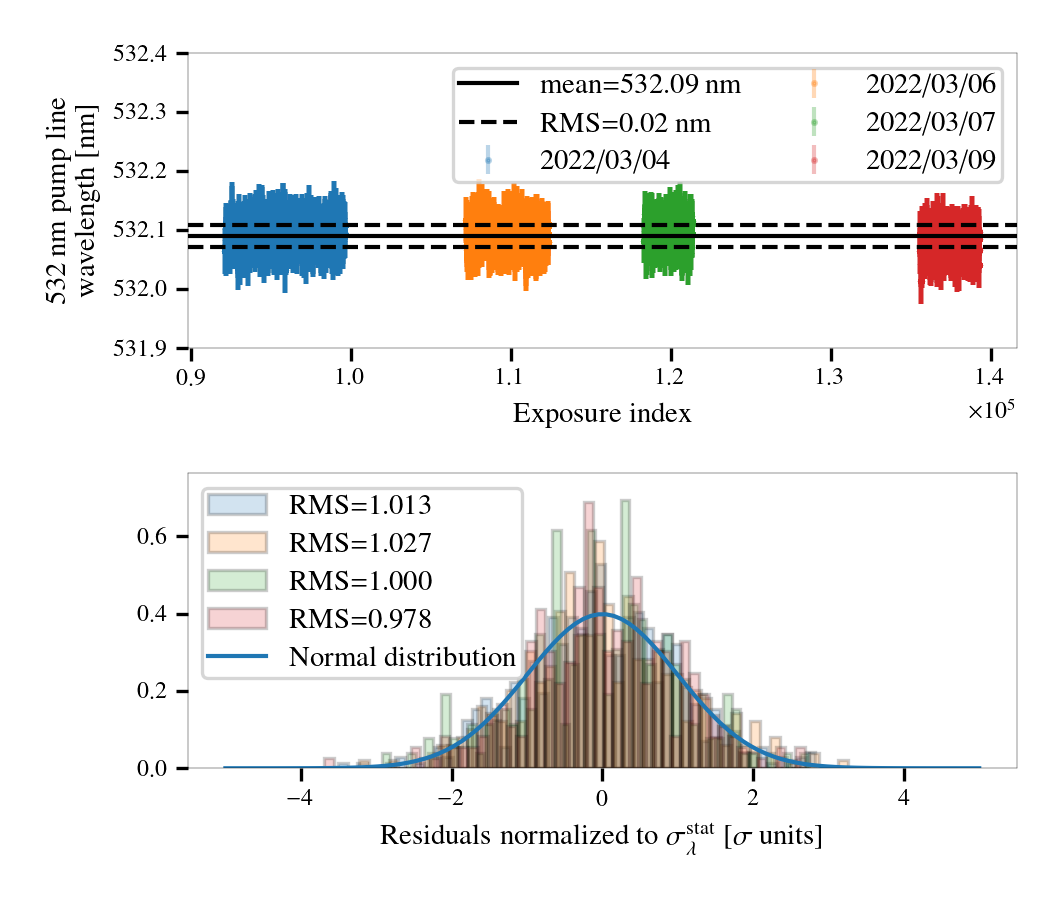}
\caption{Top: all measured \SI{532}{\nm} pump line calibrated wavelengths with respect to exposure index ($\approx\num{250000}$ data points) for the four solar cell runs. Bottom: distributions of residuals to the mean wavelength normalized by $\sigma_{\lambda}^{\rm stat}$ (colored bars) with RMS quoted in legend. A normal distribution of RMS 1 is over-plotted for comparison.}\label{fig:wavelength_error_model_consistency}
\end{figure}

Finally, the wavelength uncertainty on $\lambda_c$ is
\begin{equation}
  \left(\sigma_{\lambda}\right)^2 =  \left(\sigma_{\lambda}^{\rm stat}\right)^2 +  \left(\sigma_{\lambda}^{\rm cal}\right)^2.   
\end{equation}
The composition of the final wavelength error budget is detailed in Section~\ref{sec:wavelength_syst}.

\begin{figure}[!h]
\centering
\includegraphics[width=\columnwidth]{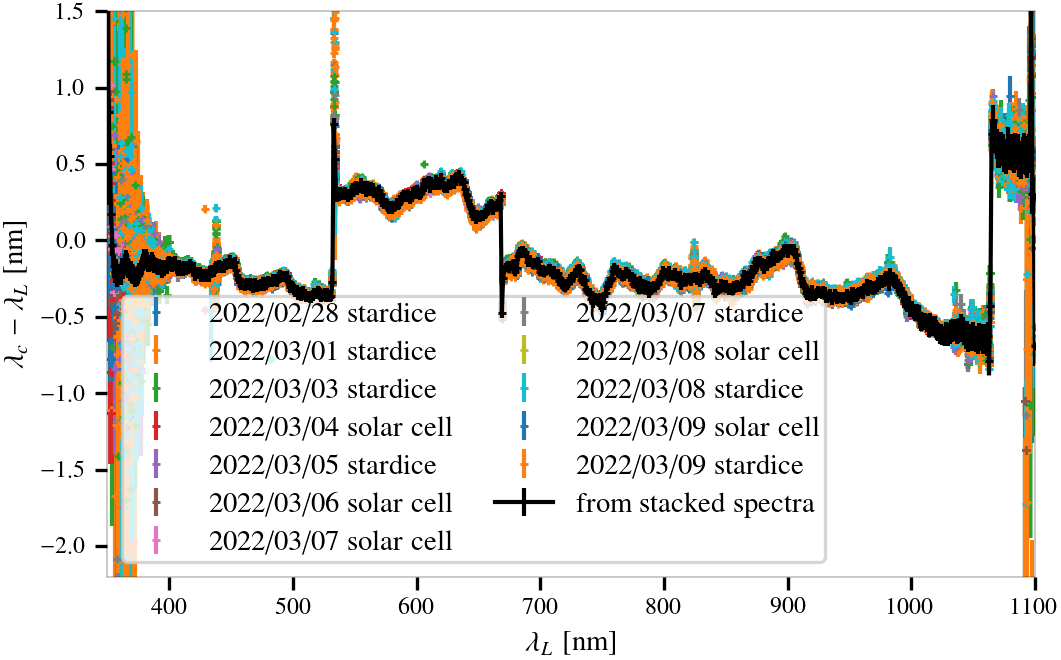}
\caption{Difference between the calibrated $\lambda_c$ and requested laser wavelengths $\lambda_L$, for all the $\approx \num{350000}$ laser line wavelengths acquired during our measurement campaign.}\label{fig:wavelength_stability}
\end{figure}

The realized laser wavelength is never the one a priori asked for, as shown in Figures~\ref{fig:sc_dataset_examples} and~\ref{fig:wavelength_stability}. However, we observed remarkable repeatability of the correspondence between the set wavelength $\lambda_L$ and the realized wavelength $\lambda_c$ in Figure~\ref{fig:wavelength_stability}. This figure represents every $\approx\num{350000}$ laser bursts shot and analyzed during the CBP measurement campaign. The superimposition of all measurements at each laser wavelength $\lambda_L$ exhibits good laser stability with time and of the $\lambda_c$ wavelength fit with line intensity. The RMS is better than \SI{0.02}{\nm} at most wavelengths, except where line determination is ambiguous: close to the contamination lines at $\SI{532}{nm}$ and $\SI{1064}{nm}$, and close to the sensor edges. Given this remarkable stability, to improve the line detection in weak laser regimes (like below \SI{400}{\nm}), we averaged the obtained $\lambda_c$ for each $\lambda_L$. We decided to use this average as our final calibrated wavelength (black curve in Figure~\ref{fig:wavelength_stability}). As the bijection between $\lambda_L$ laser wavelengths and realized wavelengths $\lambda_c$ is unambiguous, in some figures of this paper, we use the set laser wavelength $\lambda_L$ for clarity.

\subsubsection{Light contamination measurements}
\label{lightcontamination}
In addition to using the spectrograph to monitor the wavelength of the laser lines, we also used it in a spectro-photometric mode to estimate light contamination.

The secondary laser lines at \SI{532}{\nm} and from two-photon conversions were our principal suspects of internal laser light contamination. Indeed, we noticed the presence of a \SI{532}{\nm} line in the regime 532 to \SI{669}{\nm} and of a weak line when the laser is set to wavelength above \SI{1064}{\nm}. For the latter, if we call $\lambda_L$ the wavelength at which the laser was set, we observed the production of photons at wavelength $\lambda_{\text{comp}}$, which seems given by
\begin{equation}
 \frac{2}{\SI{1064}{\nm}} \approx \frac{1}{\lambda_L} + \frac{1}{\lambda_{\text{comp}}}
 \end{equation} 
 due to the conversion of two \SI{1064}{\nm} photons into a laser photon at $\lambda_L$ and a complementary photon at $\lambda_{\text{comp}}$ that ended in the laser beam. In practice, we observed a small emission line in the spectrograph when $\lambda_L > \SI{1064}{\nm}$, nearly symmetrical of the laser line with respect to \SI{1064}{\nm}.

The laser line flux $\Qspectromain$, the \SI{532}{\nm} line flux $\Qspectro^{532}$ and the $\lambda_{\rm comp}$ line flux $\Qspectro^{\rm comp}$ flux are computed summing the pixels in a window of $\pm 3 \sigma_g$, after dark and bias subtraction. Associated statistical uncertainties are computed from the standard error propagation of the $\sigma_{i,p}$ values. These fluxes are used to correct solar cell and photodiode charges as well as the \SD photometry from the $\SI{532}{\nm}$ line contamination (see Section~\ref{sec:532_cont}).  

Finally, we needed to build an error model of the measured spectrograph fluxes to measure the level of contamination from photons not belonging to the main laser line. We used the dark exposures acquired with different exposure times to estimate it.

The master bias was subtracted from all dark exposures. We measured the readout noise and sensor gain from those data by doing the following procedure. Let's call $D(\lambda_p)$ the dark exposure minus the master bias value for pixel $p$. For each exposure time, the variance $\sigma_p^2$ and average $\bar{D}(\lambda_p)$ of each $D(\lambda_p)$ pixel were computed. The variance evolution with the average is well described by a second-order polynomial function, parameterized as follows:
\begin{equation}\label{eq:spectro_error_model}
\sigma^2_p =\sigma_{ro}^2 +  \bar{D}(\lambda_p)/G + \sigma_G^2 \bar{D}^2(\lambda_p)/G^2
\end{equation}
with $\sigma_{ro}$ the readout noise, $G$ the sensor gain and $\sigma_G$ a statistical noise on the gain itself. The fit of this model to data led to the following values (Figure~\ref{fig:spectro_ptc}):
\begin{align}
    & \sigma_{ro} = 1.26\ \mathrm{ADU}, \\
    & G = 25.8\ e^-/\mathrm{ADU} ,\\
    & \sigma_G/G = 0.7\%.
\end{align}
The first two values are compatible with the sensor specifications given by the spectrograph vendor\footnote{For the QE65000 fiber spectrograph, the vendor specifies a read-out noise of 1.5 ADU (or 40 electrons) and thus a gain of $26\ e^-/\mathrm{ADU}$.}. 
With this error model, error bars are compatible with the observed dispersion in the dark parts of the spectra. 

\begin{figure}[!h]
\centering
\includegraphics[width=\columnwidth]{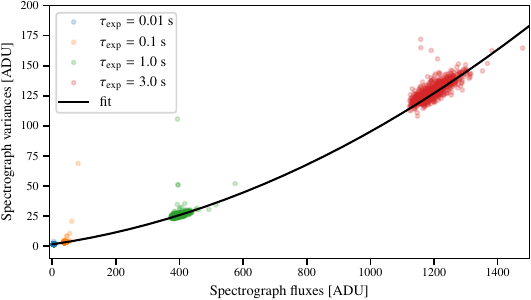}
\caption{Spectrograph photon transfer curve to estimate gain and read-out noise. Pixel variance from dark spectra is represented versus their averages for four different exposure times $\tau_{\mathrm{exp}}$.}\label{fig:spectro_ptc}
\end{figure}

\subsection{Electrometer data reduction}

In one data set, we accumulated several laser bursts, and a per-burst analysis was conducted to be able to remove outliers more easily. All bursts are combined only at the end of the analysis to get the CBP and StarDICE responses.

\subsubsection{Photodiode data reduction}
\label{sec:pd_reduction}

Photodiode data are charge time series with stair-case shape. Each step is a laser burst, and the step height roughly gives the charge $\Qphotmes$ accumulated during a laser burst (see Figures~\ref{fig:sc_dataset_examples} and~\ref{fig:pd_reduc}). The length of the charge rise is the laser burst duration $\tau_b$ while flat sequences are dark times. For the photodiode, the time stamps come directly from the digital analyzer clock, with a sampling at \SI{50}{\hertz}. The digital analyzer clock also provides the time stamps of the laser pulses. 

For each burst, we fit straight (almost horizontal) lines in the charge sequence during the dark times before and after a laser burst, removing the closest points to the burst. During those intervals, the charge evolution measures the charge leak corresponding to the dark current we need to account for.

We call $t_1$ and $t_2$ the time stamps of the beginning and end of the laser burst, respectively, given by the laser trigger output itself. 
The accumulated charge $\Qphotmes$ during a burst is then
\begin{equation}
q_{\rm phot}^{\rm dark}(t) = a_{\rm phot} t + b_{\rm phot}
\end{equation}
\begin{align}
\Qphotmes = \, &q_{\rm phot,2}^{\rm dark}(t_2) - q_{\rm phot,1}^{\rm dark}(t_1)\\ & - \frac{1}{2} \left[ q_{\rm phot,1}^{\rm dark}(t_2) -  q_{\rm phot,1}^{\rm dark}(t_1) + q_{\rm phot,2}^{\rm dark}(t_2) - q_{\rm phot,2}^{\rm dark}(t_1)  \right]   \notag 
\end{align}
where $q_{\rm phot,j}^{\rm dark}(t_i)$ is the line fit of the dark part $j$ ($j=1$ before the burst, $j=2$ after) evaluated at time $t_i$. The term in brackets subtracts an estimation of the dark current contribution during the burst itself. After this procedure, the subtraction $q_{\rm phot,2}^{\rm dark}(t_2) - q_{\rm phot,1}^{\rm dark}(t_1)$ gives the raw height of the burst step in the charge sequence\footnote{Moreover, doing the subtraction removes systematic inaccuracy coming from the Keithley electrometer.} while the terms in brackets remove the averaged contribution of the dark current using both dark times before and after the burst. Note that we do not model anything during the burst time, as the laser power stability does not guarantee that this can be modeled with a simple mathematical function. The only model assumption is that dark sequences are modeled with straight lines.

The fit of the two parameters $a_{\rm phot}$ and $b_{\rm phot}$ of each $q_{\rm phot,j}^{\rm dark}(t)$ line models is performed via a standard $\chi^2$ minimization, using the \texttt{curve\_fit} method from python library \texttt{scipy}. Uncertainties on the data points, equally weighted, are tuned so that the final reduced $\chi^2$ is one. As a result, we assume that the fit residuals are only due to Gaussian instrumental noise, as justified by Figure~\ref{fig:pd_reduc}. The analysis of the residuals also confirms our choice to use order 1 polynomial functions to model the dark sequences.

\begin{figure}[!h]
\centering
\includegraphics[width=\columnwidth]{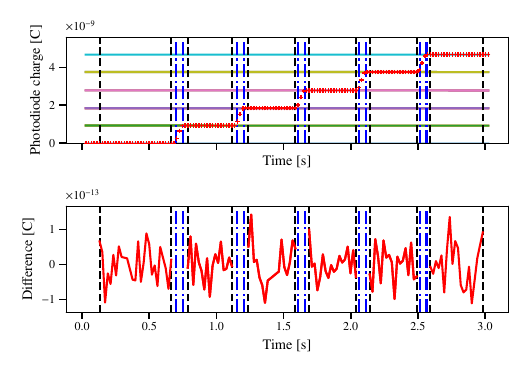}
\caption{Photodiode charge sequence reduction process at wavelength $\lambda_L=\SI{469}{\nm}$. Vertical blue lines indicate the laser starts and stops, while black lines flank the dark sequences. Colored horizontal lines are fitted during dark times. The top panel shows the raw charges acquired with the photodiode, while the bottom panel shows the residuals of the linear fits during dark times.}\label{fig:pd_reduc}
\end{figure}

Covariance matrix uncertainties from all linear model parameters are then propagated to compute the statistical uncertainty $\Sphotstat$ of $\Qphotmes$ per burst. They are typically of the order of the residual RMS, around \SI{5e-5}{\nano\coulomb} (Figure~\ref{fig:pd_reduc}), more than three orders of magnitude below the typical $\Qphotmes$ values. We tested the fitting procedure on pure dark sequences and found an unbiased null measurement $\Qphot^{\rm dark}$ with a pull distribution of RMS $\;\approx 1$ whatever the burst duration $\tau_b$: computed statistical uncertainties $\Sphotstat$ nicely covers the data Gaussian noise (Figure~\ref{fig:charge_pull}).

\begin{figure}[!h]
\centering
\includegraphics[width=\columnwidth]{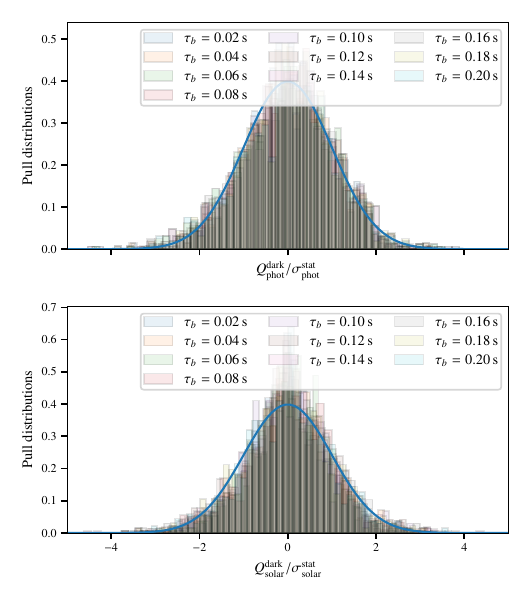}
\caption{Top: pull distributions for photodiode charge measurements during pure dark time series $\Qphot^{\rm dark} / \Sphotstat$, with different burst durations $\tau_b$. The cyan curve represents a Gaussian distribution of mean 0 and RMS 1. Bottom: same but for the solar cell case $\Qsolar^{\rm dark} / \Ssolarstat$.}\label{fig:charge_pull}
\end{figure}

\subsubsection{Solar cell data reduction}
\label{sec:solar_reduction}

Solar cell charge time series are very similar to photodiode time series (see Figures~\ref{fig:sc_dataset_examples} and~\ref{fig:sc_reduc}). However, they are affected by two supplementary contributions as seen in the noise power spectrum: a random $1/f$ noise and power line harmonics mainly at \SI{50}{\hertz} and \SI{100}{\hertz} (Figure~\ref{fig:darkcurrentspectrum}). Time-stamps come directly from the Keysight electrometer. However, as this device sends triggers at the start and end of the acquisition, the electrometer clock is re-scaled using the digital analyzer. In doing so, all electrometers are synchronized via the digital analyzer's internal clock.

\begin{figure}[!h]
\centering
\includegraphics[width=\columnwidth]{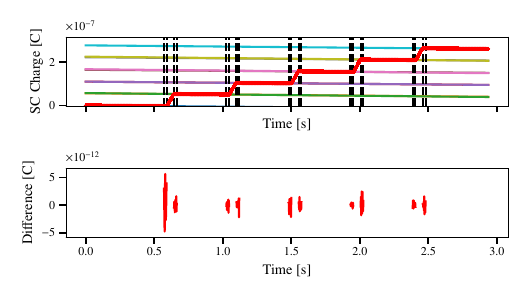}
\caption{Solar cell charge sequence reduction process at wavelength $\lambda_L=\SI{469}{\nm}$. Pairs of vertical black at the left and right of a burst encompass the fitted dark sequences. Colored curves are the dark model fitted within these dark times. The top panel shows the raw charges acquired with the solar cell, while the bottom panel shows the residuals to the dark model fit during dark times only.}\label{fig:sc_reduc}
\end{figure}

As for the photodiode, we modeled the dark current by estimating the charge decrease over the regions before and after each laser burst by fitting the slope of their departure from horizontal. The solar cell dark model is the sum of a linear function plus two sinusoidal functions at fixed frequency \SI{50}{\hertz} and \SI{100}{\hertz}:
\begin{align}
    q_{\rm solar}^{\rm dark}(t) = a_{\rm solar}t + b_{\rm solar} & + A_{50} \sin \left( 100 \pi t + \phi_{50}\right)  \notag \\  & +  A_{100} \sin \left( 200 \pi t + \phi_{100}\right)
\end{align}
where $a_{\rm solar}, b_{\rm solar}, A_{50}, A_{100}, \phi_{50}$ and $\phi_{100}$ are free parameters fitted on data. 
Again, we call $t_1$ and $t_2$ the time stamps of the beginning and end of the laser burst, respectively, given by the laser trigger output itself.
The accumulated charge $\Qsolarmes$ during a burst is then
\begin{align}\label{eq:qsolar}
\Qsolarmes  = & q_{\rm solar,2}^{\rm dark}(t_2) - q_{\rm solar, 1}^{\rm dark}(t_1) \\  &  - \frac{1}{2} \left[q_{\rm solar,1}^{\rm dark}(t_2) - q_{\rm solar,1}^{\rm dark}(t_1) + q_{\rm solar,2}^{\rm dark}(t_2) - q_{\rm solar,2}^{\rm dark}(t_1)  \right]    \notag
\end{align}
where the indices $1$ and $2$ refer again to data before and after the burst, respectively.  Free $q_{\rm solar}^{\rm dark}(t)$ parameters were fitted, minimizing a $\chi^2$ with a Newton-Raphson gradient descent. Power lines are well fitted by the model (no more periodic oscillations in the residuals) as shown in Figure~\ref{fig:sc_reduc_zoom} where $A_{50}$ was about $\SI{50}{\pico\coulomb}$.
The $1/f$ noise is not modeled in $q_{\rm solar}^{\rm dark}(t)$ as it is sub-dominant, but the lowest frequency modes are captured by the values of $a_{\rm solar}$ and $b_{\rm solar}$. To get a close estimate of their contributions during the burst, we fit the dark sequences only during $\tau_b/2$ around the burst to rely on their extrapolation inside the burst window (with a minimum of 12 data points to encompass at least one \SI{50}{\hertz} period). 
Indeed, as exhibited in Figure~\ref{fig:sc_reduc_zoom}, fits of $q_{\rm solar,2}^{\rm dark}(t)$ after the first burst (orange) and of $q_{\rm solar,1}^{\rm dark}(t)$ before the second burst (green) do not superimpose because of the departure from linearity due to the long-range $1/f$ noise. So restricting the fits in two windows of size $\max(\SI{22}{\ms},\tau_b/2)$ permits to capture contaminating $1/f$ modes no longer than $\tau_b$. 

The charge value $\Qsolarmes$ is computed for each laser burst following Equation~\ref{eq:qsolar}. Concerning estimating its statistical uncertainties $\Ssolarstat$, we add in quadrature the contributions from the parameter covariance matrix and the RMS of the residuals (to account for uncaptured $1/f$ modes). We trained the fitting procedure on pure dark data and noticed that $\Ssolarstat$ was too small to account for $\Qsolar^{\rm dark}$ null measurement dispersion. The RMS of the pull distributions was linearly dependent on the burst duration $\tau_b$, showing $\Ssolarstat$ did not capture all the long-range $1/f$ noise. Therefore, we corrected all $\Ssolarstat$ values with a multiplicative factor dependent on $\tau_b$. Doing so, we found an unbiased null measurement $\Qsolar^{\rm dark}$ with a pull distribution of RMS $\;\approx 1$ whatever the burst duration $\tau_b$ (Figure~\ref{fig:charge_pull}). 

\begin{figure}[!h]
\centering
\includegraphics[width=\columnwidth]{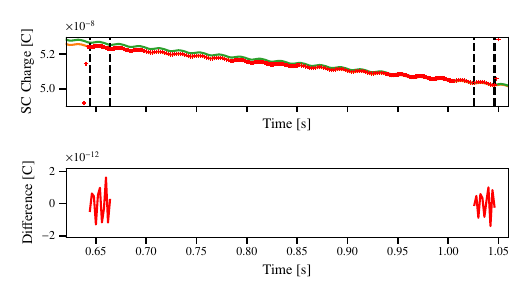}
\caption{Same as Figure~\ref{fig:sc_reduc} but zoomed on the second dark sequence. The orange model is fitted on dark data on the left of the plot, while the green model is fitted on dark data on the right.}\label{fig:sc_reduc_zoom}
\end{figure}

\subsection{CBP ratio of charges}

The ratio of charges $r_{\rm CBP}^{\rm mes} = \Qsolarmes/\Qphotmes$ is computed to check the statistical uncertainties and then analyze systematic uncertainties. It is presented in Figure~\ref{fig:cbp_charge_ratio}. Each black point is a ratio of charge measurements from one burst at one wavelength $\lambda_L$. They follow a smooth curve in $\lambda_L$. The $r_{\rm CBP}^{\rm mes}$ statistical uncertainties 
\begin{equation}
    \sigma_{\mathrm CBP}^{\rm stat} = r_{\rm CBP}^{\rm mes} \sqrt{\left(\frac{\Ssolarstat}{\Qsolarmes}\right)^2 +  \left(\frac{\Sphotstat}{\Qphotmes}\right)^2 }
\end{equation}
are higher in the blue as the laser is weaker in this regime (around $0.1\%$) than in the red (around $0.01\%$). They were estimated as the difference between all data points and a spline interpolation, normalized by the statistical uncertainties. As expected, they follow a Gaussian distribution of mean 0 and RMS$\;\approx 1$. 

\begin{figure}[!h]
\centering
\includegraphics[width=\columnwidth]{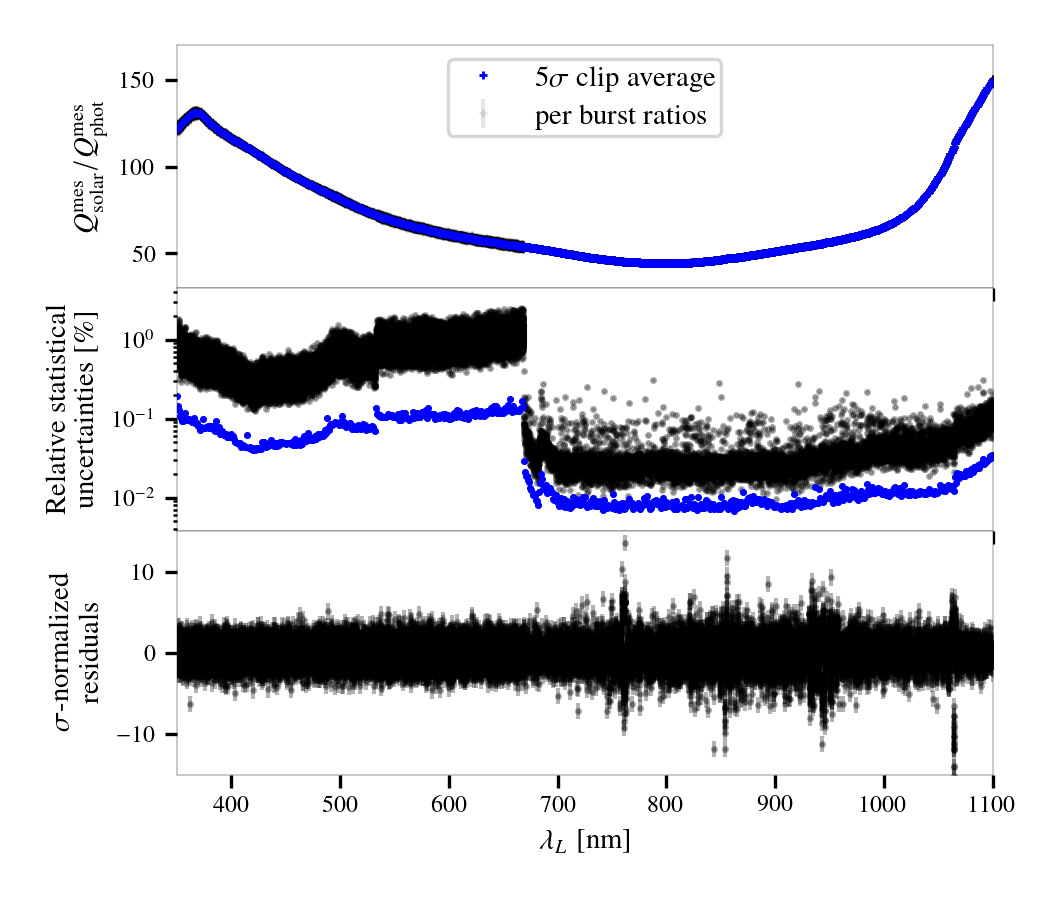}
\caption{CBP charge ratio $\Qsolarmes/\Qphotmes$ as a function of $\lambda_L$. Top: every black point is a charge measurement ratio from one burst, and the blue curve is the $5\sigma$-clipped average. Middle: relative uncertainties on $\Qsolarmes/\Qphotmes$. Bottom: pull distribution after spline subtraction.}\label{fig:cbp_charge_ratio}
\end{figure}

\subsection{Systematics}

\subsubsection{Wavelength calibration}\label{sec:wavelength_syst}

Figure~\ref{fig:wavelength_error_budget} details the total error budget from wavelength calibration for three different runs: two runs shooting at StarDICE with two pinholes and one run shooting at the solar cell. Uncertainty on wavelength $\sigma_\lambda$ is primarily dominated by wavelength calibration uncertainties in the range 400 to \SI{1080}{\nm}. Statistical uncertainty dominates around $\SI{532}{\nm}$ and close to the spectrograph sensor edges. Except in these cases, $\sigma_\lambda$ is well below $\SI{0.1}{\nano\meter}$. There are no noticeable differences despite the different experimental conditions for the three runs presented in Figure~\ref{fig:wavelength_error_budget}.

\begin{figure}[!h]
\centering
\includegraphics[width=\columnwidth]{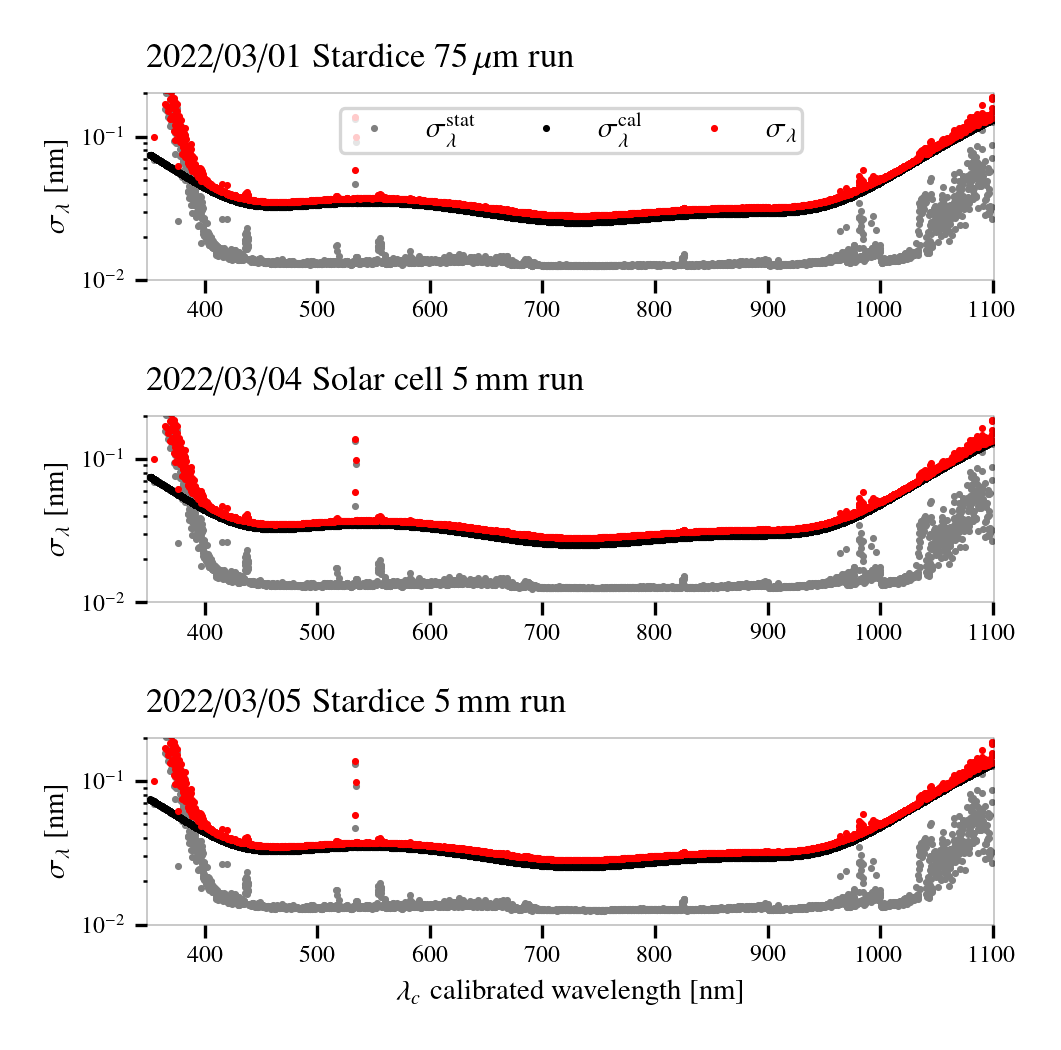}
\caption{Total error budget (red) of calibrated wavelengths $\lambda_c$ for three different runs (from top to bottom: StarDICE run with \SI{75}{\um} pinhole, solar cell run with \SI{5}{mm} pinhole, StarDICE run with \SI{5}{mm} pinhole) as a function of the laser set wavelength $\lambda_L$. Detection uncertainties (grey) represent PSF modeling uncertainties and Gaussian fit uncertainties, while calibration uncertainties (black) come from the Hg-Ar lamp calibration procedure. }\label{fig:wavelength_error_budget}
\end{figure}

To evaluate the systematic uncertainty on the CBP response due to the wavelength calibration, we computed $r_{\rm CBP}$ at $\lambda_c+\sigma_{\lambda}^{\rm cal}$ and $\lambda_c-\sigma_{\lambda^{\rm cal}}$. The difference between both CBP responses is well below the statistical uncertainty since the CBP response varies slowly in $\lambda$ (Figure~\ref{fig:cbp_charge_ratio}) and the whole $\sigma_\lambda$ < \SI{0.1}{\nano\meter} (Figure~\ref{fig:wavelength_error_budget}). The wavelength measurement uncertainty has a systematic effect mainly in determining the sharp edges of the filter pass-bands.


\subsubsection{Ambient light}\label{sec:sc_linearity}

\begin{figure}[h]
    \centering
    \includegraphics[width=\columnwidth]{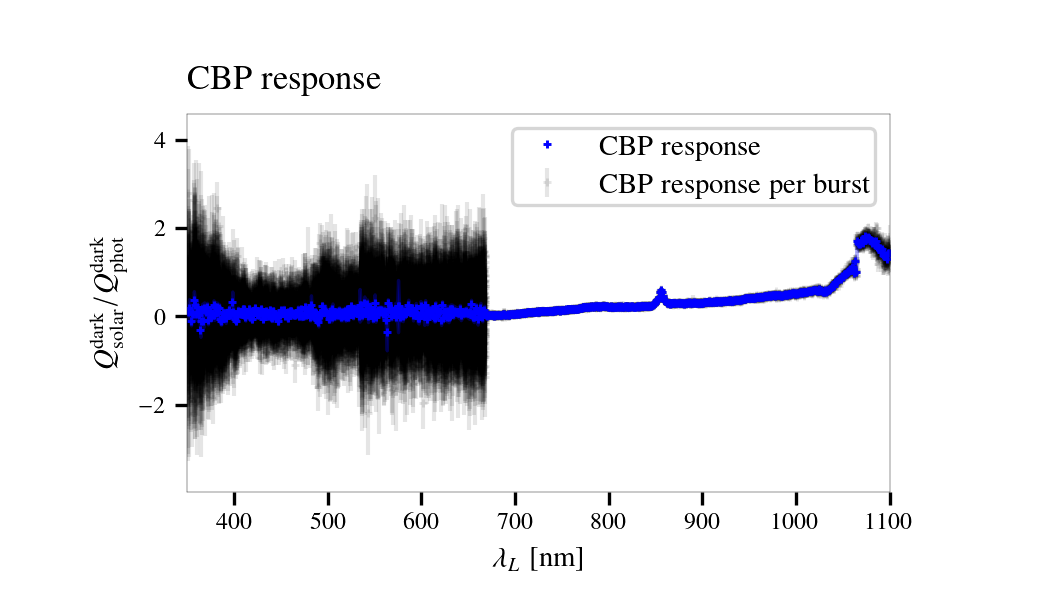}
    \caption{CBP charge ratio $\Qsolar^{\rm dark}/\Qphot^{\rm dark}$ as a function of $\lambda_L$: every black point is a charge measurement ratio from one burst, the blue curve is the $5\sigma$-clipped average.}
    \label{fig:sc_dark}
\end{figure}

We ran a solar cell acquisition with a cap on the CBP telescope to measure the ambient light contamination. This setup mimics a regular solar cell acquisition without direct light, therefore specifically measuring the laser light leaks during operation. The measurement of this indirect ambient light $\Qsolar^{\rm dark}$ constitutes a "dark" for our CBP calibration. From this specific run, we built a dark CBP response $r_{\mathrm{CBP}}^{\rm dark} = \Qsolar^{\rm dark} / \Qphot^{\rm dark}$ where $\Qsolar^{\rm dark}$ (resp. $\Qphot^{\rm dark}$) is the burst charge collected in the solar cell (resp. the photodiode). For each $\lambda_L$, burst ratios are averaged with a $5\sigma$ clipping, giving the blue curve $r_{\mathrm{CBP}}^{\rm dark}(\lambda_L)$ in Figure~\ref{fig:sc_dark}. The high dispersion below \SI{670}{\nano\meter} corresponds to the wavelength regime where the laser power is low, inducing a faint ambient light poorly measured. The dark contribution in our solar cell data is then evaluated as follows:
\begin{equation}
    \Qsolar^{\rm dark} = r_{\mathrm{CBP}}^{\rm dark}(\lambda_L) \times \Qphotmes(\lambda_L)
\end{equation}
and subtracted from all our measurements. This correction is the main contribution to instrumental non-linearity we identified.

\subsubsection{Laser light contamination}
\label{sec:532_cont}

As described in Section~\ref{sec:cbp}, the light source used is a tunable laser using a pump laser at \SI{532}{\nano\meter}, which has different regimes. When we operated within the range [532 - 644] nm, we detected in the spectrograph a contamination light at \SI{532}{\nano\meter} for all wavelengths in this range. 

We must account for this light contamination to get the true amount of charges $\Qsolarcal$ coming from the main laser line. We built a model for the \SI{532}{\nano\meter} contribution observed in the range [532 - 644] nm. The total charge measured in the solar cell $\Qsolarmes$ is the sum of the charges from the main wavelength $\lambda_L$ and the charges from contaminations, like the \SI{532}{\nm} contamination. The same applies to the total charges measured in the photodiode $\Qphotmes$ with $\Qphot$ and $\Qphot^{532}$ respectively, the charges from the main laser line and the charges from the \SI{532}{\nm} contamination:
\begin{align}
\Qphotmes(\lambda_L) & = \Qphot(\lambda_L) + \Qphot^{532}(\lambda_L) \label{eq:qphot_mes}
\end{align}

In the spectrograph, we measured two fluxes from two separated peaks: the one from the main wavelength $\Qspectromain$, and the one from the \SI{532}{\nm} contamination $\Qspectro^{532}$. As the light is homogeneous in the integrating sphere, the proportion of contamination light and laser light is the same for both instruments. This translates into:
\begin{equation}
    \frac{\Qspectro^{532}}{\Qspectromain} \times \frac{\Espectro(\lambda_L)}{\Espectro(532)} = \frac{\Qphot^{532}}{\Qphot} \times \frac{\Ephot(\lambda_L)}{\Ephot(532)}
    \label{eq:prev_alpha}
\end{equation}
where $\Ephot(\lambda_L)$ and $\Espectro(\lambda_L)$ are the quantum efficiencies of the photodiode and the spectrograph sensor with its optical fiber, respectively. The ratio:
\begin{equation}\label{eq:eta}
\eta(\lambda) = \frac{\Espectro(\lambda)}{\Ephot(\lambda)} = \frac{\Qspectromain(\lambda)}{\Qphot(\lambda)}
\end{equation}
can be obtained either as a dimensionless quantity from the manufacturer data sheets (Figure~\ref{fig:QEs}) or directly measured in spectrograph ADU per photodiode unit (Figure~\ref{fig:QEs}). In the following, we used the spline interpolation of the measured ratio and quoted as uncertainty the RMS of the resulting residuals.


We denote the level of contamination in the photodiode $\Qphot^{\lambda_L}/\Qphot^{532}$ by the ratio $\alpha(\lambda)$ that reads:
\begin{equation}
    \alpha(\lambda_L) = \frac{\Qphot^{532}(\lambda_L)}{\Qphot(\lambda_L)} = \frac{\Qspectro^{532}(\lambda_L)}{\Qspectromain(\lambda_L)} \times \frac{\Espectro(\lambda_L)\Ephot(532)}{\Ephot(\lambda_L)\Espectro(532)} 
    \label{eq:alpha}
\end{equation}

\begin{figure}[h]
    \centering
    \includegraphics[width=\columnwidth]{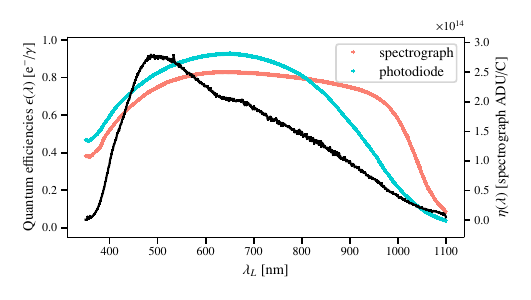}
    \caption{Quantum efficiencies of the spectrograph (red) and the photodiode (cyan), with the $\eta(\lambda)$ ratio (black).}
    \label{fig:QEs}
\end{figure}
    
We measured $\alpha(\lambda_L)$ at every wavelength using the spectrograph fluxes for the main line and the \SI{532}{nm} contaminant (Figure~\ref{fig:alpha_532}). We rejected data where $\sigma_{\lambda} > \SI{0.1}{\nm}$ to remove instances where the main line and the \SI{532}{nm} contaminant are poorly separated. In the 532 - \SI{540}{nm} range, where an unambiguous separation between the main line and the contaminant proved infeasible, we extrapolated the $\Qspectro^{532}$ values by a linear interpolation over the entire range where $\Qspectro^{532}$ was available.
The resulting $\alpha(\lambda_L)$ yield a rather flat curve between 540 and \SI{644}{\nm} with small wiggles around a global slope of $\sim 1\% $.
Uncertainties are propagated in the extrapolation range using the standard error propagation formula, considering the covariances between the 2 model parameters. With this $\alpha$ model, we deduced the true photodiode charges coming from the main laser line at $\lambda_L$
\begin{equation}
        \Qphotcal(\lambda_L) \equiv  \frac{\Qphotmes(\lambda_L)}{1 + \alpha(\lambda_L)} \quad\text{if}\ \lambda_L \in \left[532, 644\right]\mathrm{\,nm}\label{eq:qphot_cal532}
\end{equation}
We introduce here the notation $\Qphotcal$ as the final calibrated amount of charges detected in the photodiode per laser burst. 

Light detected by the solar cell has gone through the CBP optics, with a response $\Rcbp$.
Then, the contribution of the \SI{532}{\nano\meter} photons collected by the solar cell is computed as
\begin{equation}
\begin{aligned}
    \Qsolar^{532}(\lambda_L) & = \Rcbp(532)  \Qphot^{532}(\lambda_L) \\ 
    & = \Rcbp(532) \frac{ \alpha(\lambda_L) }{1+ \alpha(\lambda_L)} \Qphotmes(\lambda_L).
    \label{eq:qsolar_cal532}
\end{aligned}
\end{equation}
Thanks to the spectrograph data, using Equations~\ref{eq:qphot_cal532} and~\ref{eq:qsolar_cal532}, we can correct all our measurements from the \SI{532}{\nano\meter} in both the photodiode and the solar cell. The impact of this correction is illustrated in Section~\ref{sec:cbp_summary}.

\begin{figure}[h]
    \centering
    \includegraphics[width=\columnwidth]{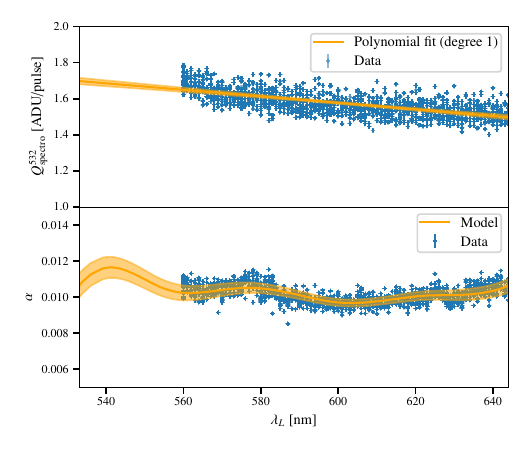}
    \caption{Top: all available data points for $\alpha$ computed from the spectrograph measurements (blue crosses) and linear fit (orange) with its uncertainty (shaded orange band). Bottom: difference between data and model.}
    \label{fig:alpha_532}
\end{figure}

We did the same correction for the complementary wavelength line $\lambda_{\rm comp}$ appearing after $\lambda_L > \SI{1064}{\nm}$. The correction coefficient $\beta$ analogue to $\alpha$ is
\begin{equation}
    \beta(\lambda_L) = \frac{\Qphot^{\lambda_{\rm comp}}(\lambda_L)}{\Qphot(\lambda_L)} = \frac{\Qspectro^{\lambda_{\rm comp}}(\lambda_L)}{\Qspectromain(\lambda_L)} \times \frac{\Espectro(\lambda_L)\Ephot\contcomp}{\Ephot(\lambda_L)\Espectro\contcomp} 
    \label{eq:beta}
\end{equation}
and is represented Figure~\ref{fig:beta}. The $\Qspectro^{\lambda_{\rm comp}}$ data points were modeled by a linear function to allow extrapolating $\beta$ in the range [1064 - 1070]\,nm. Similarly, the calibrated amount of charges in the photodiode corrected from the $\lambda_{\rm comp}$ photons is
\begin{equation}
        \Qphotcal(\lambda_L) \equiv  \frac{\Qphotmes(\lambda_L)}{1 + \beta(\lambda_L)} \quad\text{if}\ \lambda_L > \SI{1064}{\nano\meter}
        \label{eq:qphot_cal1064}
\end{equation}
and their contribution $\Qsolar^{\lambda_{\rm comp}}$ in the solar cell is given by
\begin{equation}
\begin{aligned}
    \Qsolar^{\lambda_{\rm comp}}(\lambda_L) & = \Rcbp(\lambda_{\rm comp})  \Qphot^{\lambda_{\rm comp}}(\lambda_L) \\ 
    & = \Rcbp(\lambda_{\rm comp}) \frac{ \beta(\lambda_L) }{1+ \beta(\lambda_L)} \Qphotmes(\lambda_L).
    \label{eq:qsolar_cal1064}
\end{aligned}
\end{equation}
Both corrections' impact are illustrated in Section~\ref{sec:sc_linearity} and Figure~\ref{fig:SCqswlinearity}, and Section~\ref{sec:sd_contaminations}.

\begin{figure}[h]
    \centering
    \includegraphics[width=\columnwidth]{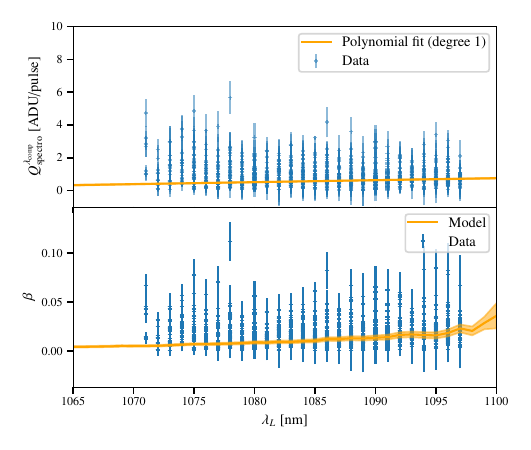}
    \caption{Same as Figure~\ref{fig:alpha_532} but for $\beta$ correction coefficient.}
    \label{fig:beta}
\end{figure}

\subsubsection{Integrating sphere fluorescence}\label{sec:fluorescence}

\begin{figure}[h]
    \centering
    \includegraphics[width=\columnwidth]{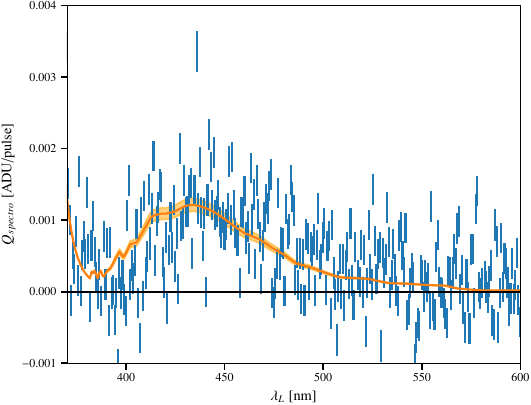}
    \includegraphics[width=\columnwidth]{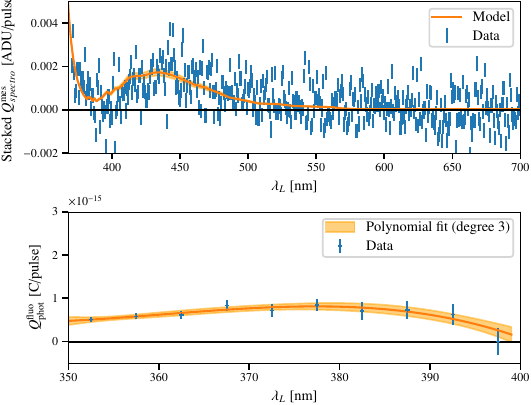}
    \caption{Top: integrating sphere fluorescence spectrum from a stack of all solar cell run data with $\lambda_L < \SI{370}{\nano\meter}$ (blue crosses), with the best-fit model (orange line). Bottom: estimated fluorescence contribution in photodiode measured charges $\Qphot^{\rm fluo}$ as a function of wavelength (blue crosses) with a fitted third-order polynomial function (orange line).}
    \label{fig:fluo}
\end{figure}

Our integrating sphere appeared to be fluorescent at laser wavelengths below \SI{400}{\nano\meter}. The fluorescence of integrating spheres is studied in~\cite{shaw2007ultraviolet}. The fluorescent signal is visible in the \SD camera using the g filter or the grating but is very weak in the spectrograph. To visualize and model it, we stacked all our spectra in bins of \SI{5}{\nano\meter} in $\lambda_L$ (Figure~\ref{fig:fluo} top). The fluorescence signal spans a range of wavelength between $\approx \SI{400}{\nano\meter}$ and $\approx\SI{500}{\nano\meter}$, with an emission peak around \SI{450}{\nano\meter}. 

To estimate the contamination from fluorescence photons, we fitted a fluorescence spectrum model taken from Figure~9 of \cite{shaw2007ultraviolet}, with a constant background and a Moffat profile for the laser line for each stacked spectra. The fluorescence spectrograph flux is converted into photodiode charges $\Qphot^{\mathrm{fluo}}$ using the $\eta(\lambda)$ conversion factor and normalized by the total number of laser pulses (Figure~\ref{fig:fluo} bottom)\footnote{Contrary to the \SI{532}{\nano\meter} line contamination correction, we can not normalize by the flux in the main laser line as it is often noise-dominated in un-stacked spectra when $\lambda_L < \SI{400}{\nano\meter}$.}. We observed that the fluorescence spectrum cancels at $\lambda_L \geq \SI{400}{\nano\meter}$. For every wavelength $\lambda_L < \SI{400}{\nano\meter}$, we evaluate and subtract the contribution from the fluorescence contamination in the photodiode using the $\Qphot^{\mathrm{fluo}}(\lambda_L)$ model from Figure~\ref{fig:fluo}:
\begin{equation}
        \Qphotcal(\lambda_L) \equiv  \Qphotmes(\lambda_L) - \Qphot^{\mathrm{fluo}}(\lambda_L) \quad\text{if}\ \lambda_L < \SI{400}{\nano\meter}
        \label{eq:qphot_calfluo}
\end{equation}
We perform identically for $\Qsolarmes$ multiplying by the CBP response at \SI{450}{\nano\meter}: 
\begin{equation}
\begin{aligned}
    \Qsolar^{\rm fluo}(\lambda_L) & = \Rcbp(450)  \Qphot^{\rm fluo}(\lambda_L)
    \label{eq:qsolar_calfluo}
\end{aligned}
\end{equation}
This correction's impact is illustrated later in Section~\ref{sec:sd_contaminations}.

After the fluorescence, \SI{532}{\nano\meter} and $\lambda_{\mathrm{comp}}$ corrections, we updated our $\eta(\lambda)$ estimate and iterated several times to refine the light contamination subtractions.

\subsubsection{Instrumental chain linearity check}\label{sec:sc_linearity}
For the two solar cell runs we undertook, we varied the laser output power by a factor of around 2, namely QSW at maximum and QSW set at 298. The ratio of the two CBP charge ratios before and after dark subtraction is presented in Figure~\ref{fig:SCqswlinearity}. No corrections lead to a $\approx 5\,$permil deviations of the two CBP responses with respect to wavelength. Both CBP responses agree at $\approx 0.5\,$permil for wavelengths above \SI{669}{\nano\meter} when applying dark subtraction. Then, laser contamination correction makes the two CBP responses agree in the [532, 669]\,nm range better than 0.1 permil.

To assess the value of systematic uncertainties on the CBP response due to non-linearities, we take the absolute distance of the binned ratio to unity in four different ranges of wavelengths after dark subtraction and laser contamination correction (red segments in the bottom plot of Figure~\ref{fig:SCqswlinearity}).

\begin{figure}[h]
    \centering
    \includegraphics[width=\columnwidth]{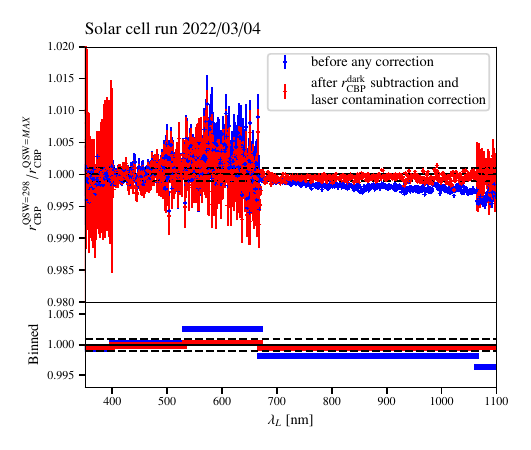}
    \caption{Ratios of the CBP charge ratio for two different QSW values as a function of $\lambda_L$ coming from the 2022/03/04 solar cell run. Blue is for raw data, while red is used for data corrected by dark contribution and laser contaminations. Black dashed lines encompass the permil precision region. Top: ratios for each $\lambda_L$. Bottom: binned ratio for four different wavelength ranges. A similar plot is obtained for the 2022/03/06 solar cell run.}
    \label{fig:SCqswlinearity}    
\end{figure}

\subsubsection{CBP scattered light varying the solar cell distance}

To measure the influence of scattered light in the CBP beam, we measured the CBP throughput by putting the solar cell \SI{16}{\cm} farther and compared it to the initial value. At this new position, we measured again the CBP dark from solar cell $\Qsolar^{\rm dark}$. Dark subtraction and laser contamination correction are applied. The comparison of both transmissions is presented in Figure~\ref{fig:sc_distance}. There is a decrease of the total light of about 3\textperthousand\ with a chromatic effect of about 2.5\textperthousand\ difference between \SI{350}{\nano\meter} and \SI{1100}{\nano\meter}, quantified by the linear fit in Figure~\ref{fig:sc_distance}. This constitutes the dominant systematics in the CBP throughput measurement.

\begin{figure}[h]
    \centering
    \includegraphics[width=\columnwidth]{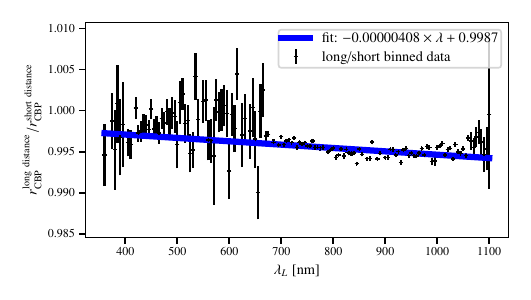}
    \caption{Ratios of the CBP charge ratio for two different distances to the solar cell as a function of $\lambda_L$ coming from the 2022/03/08 solar cell run. The long distance is \SI{16}{\cm} larger than the short distance. Black points are the binned ratio for each $\lambda_L$, and the blue line is a fit whose equation is in legend.}
    \label{fig:sc_distance}
\end{figure}

\subsubsection{Repeatability}

Finally, we measured the value of the CBP response three times during our measurement campaign. For run $i$, we computed the CBP charge ratio $r_{\rm CBP}^{\mathrm{run}\ i}$, applying dark subtraction and laser contamination correction, binned in \SI{1}{\nano\meter} intervals in $\lambda_L$. Then, we computed the mean CBP response $\overline{r_{\rm CBP}}$ as the mean of the three different runs. We observed $\approx 1$\,permil differences between the three CBP charge ratios and $\overline{r_{\rm CBP}}$ (Figure~\ref{fig:SCrepeatability}), depending slightly with wavelength.

To assess the value of systematic uncertainties on the CBP response due to its stability, we take the maximum absolute distance of the binned ratio to unity in four different ranges of wavelengths (segments the farther from 1 in the bottom plot of Figure~\ref{fig:SCrepeatability}). The strongest systematics comes from the scattered light.

\begin{figure}[h]
    \centering
    \includegraphics[width=\columnwidth]{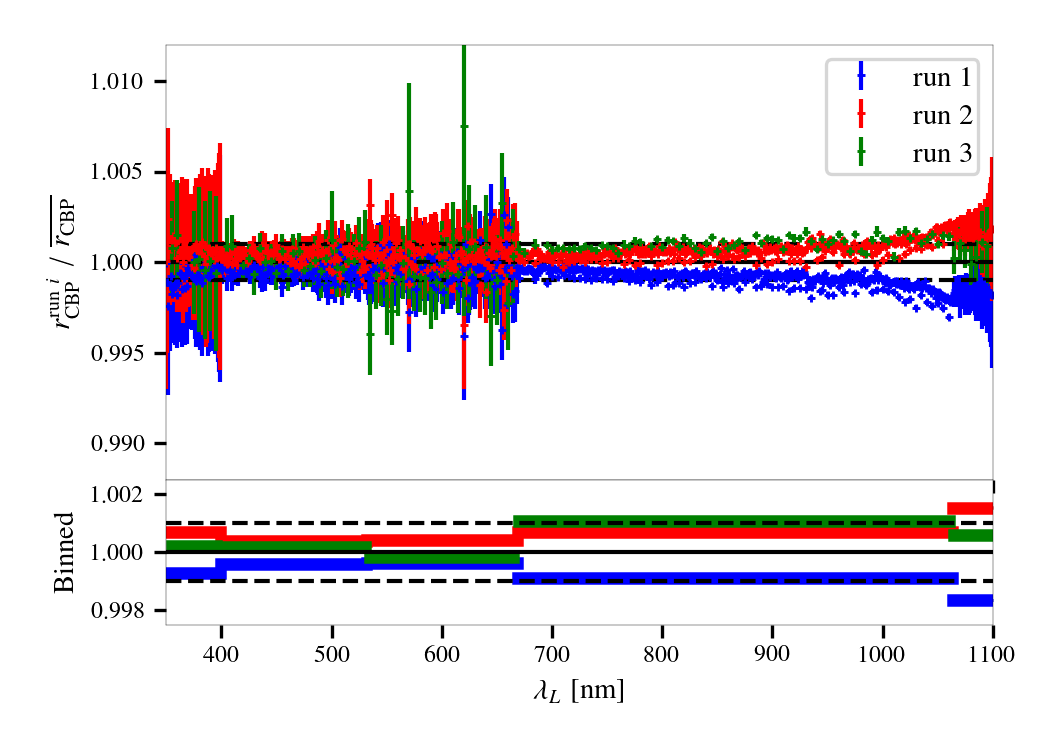}
    \caption{Ratios of the CBP charge ratio $r_{\rm CBP}^{\mathrm{run}\ i}\;/\;\overline{r_{\rm CBP}}$ for three different runs as a function of $\lambda_L$. Black dashed lines encompass the per-mil precision region. Top: ratios for each $\lambda_L$. Bottom: binned ratio for four different wavelength ranges.}
    \label{fig:SCrepeatability}
\end{figure}

\subsubsection{Summary}\label{sec:cbp_summary}

In summary, we defined the calibrated amount of charges in the solar cell as:
\begin{equation}
\Qsolarcal \equiv \Qsolarmes - \Qsolar^{\rm dark} - \Qsolar^{532} - \Qsolar^{\lambda_{\rm comp}} - \Qsolar^{\rm fluo}
\end{equation}
and in the photodiode as:
\begin{equation}
\Qphotcal(\lambda_L) = \left\lbrace
\begin{array}{ll}
          \Qphotmes(\lambda_L) - \Qphot^{\mathrm{fluo}}(\lambda_L) &\ \text{if}\ \lambda_L < \SI{400}{\nano\meter} \\
         \Qphotmes(\lambda_L)(1 + \alpha(\lambda_L)) &\ \text{if}\ \lambda_L \in \left[532, 644\right]\mathrm{\,nm} \\
        \Qphotmes(\lambda_L)(1 + \beta(\lambda_L)) &\ \text{if}\ \lambda_L > \SI{1064}{\nano\meter} \\
        \Qphotmes(\lambda_L)&\ \text{elsewhere}
\end{array}\right. 
\end{equation}
All uncertainties from the evaluation of all these terms were propagated. The summary of the error budget on the CBP response is decomposed in Figure~\ref{fig:cbp_budget} as a function of laser wavelength $\lambda_L$. Systematics coming from the wavelength calibration are not represented. Indeed, as the CBP response varies slowly with wavelength, it is negligible compared to others.  In the visible range, scattered light systematics dominates. In the near-infrared, the subtraction of the $\lambda_{\mathrm{comp}}$ photons is the main systematic uncertainty, while in the UV range, fluorescence correction systematic dominates.

\begin{figure}[h]
    \centering
    \includegraphics[width=\columnwidth]{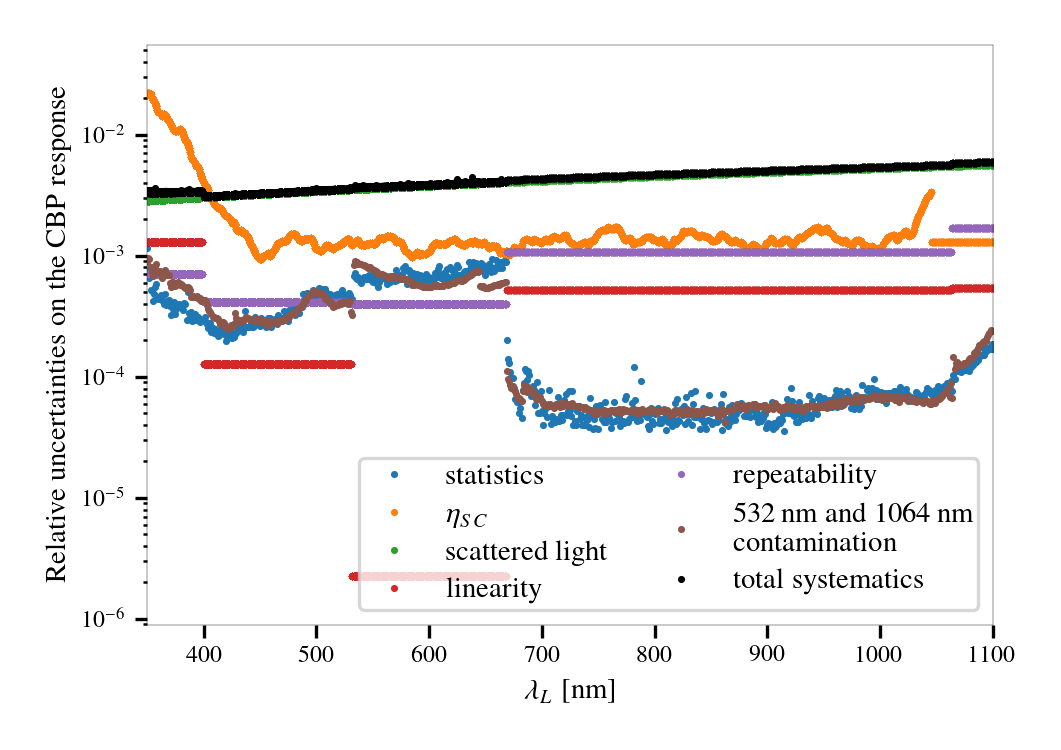}
    \caption{Total error budget for CBP response. Total systematic budget (black dots) results from the quadratic sum of all the CBP systematic uncertainties (the statistical uncertainties and the solar cell QE uncertainties are excluded here).}
    \label{fig:cbp_budget}
\end{figure}

\subsection{CBP response}

The final CBP response in output photons per Coulomb unit in the photodiode is
\begin{equation}
    \Rcbp(\lambda_c) = \frac{\Qsolarcal(\lambda_c)}{\Qphotcal(\lambda_c) \times \epsilon_{\mathrm{solar}}(\lambda_c) \times e}.
    \label{eq:rcbp2}
\end{equation} 
It can be computed for each laser burst. We averaged the values to increase the signal-to-noise ratio and get the red smooth curve presented in Figure~\ref{fig:cbp_response}:
\begin{equation}
    \overline{\Rcbp(\hat{\lambda}_c)} = \left\langle\frac{\Qsolarcal(\lambda_c)}{\Qphotcal(\lambda_c) \times \epsilon_{\mathrm{solar}}(\lambda_c) \times e}\right\rangle_{\lambda_c\in [\lambda,\lambda+\delta \lambda]}.
    \label{eq:rcbp3}
\end{equation} 
The average is performed on every burst of the three runs, with a $5\sigma$ clipping, and $\hat{\lambda}_c$ the mean calibrated wavelength in a $\delta \lambda = \SI{1}{\nano\meter}$ bin. All uncertainties are combined in quadrature (Figure~\ref{fig:cbp_response} bottom).

\begin{figure}[h]
    \centering
    \includegraphics[width=\columnwidth]{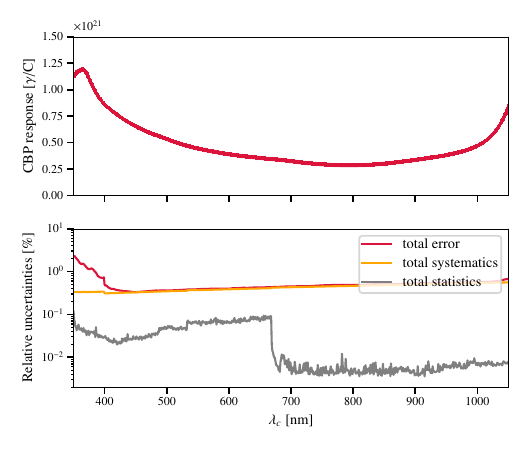}
    \caption{Top: CBP response $\overline{\Rcbp(\hat{\lambda}_c)}$ obtained with the \bpinhole{} pinhole and $\delta \lambda = \SI{1}{\nano\meter}$. Bottom: relative uncertainties.}
    \label{fig:cbp_response}
\end{figure}


\section{Measurement of the \SD telescope response}
\label{sec:rsd}
\label{sec:sd_datadesc}

This section presents the measurement of the \SD{} telescope response $\Rtel(\lambda)$ from CBP shoots using the \spinhole pinhole. One difficulty is the inter-calibration between the \spinhole pinhole and the \bpinhole pinhole with which the CBP transmission has been determined in the previous section. Both produce images that are contained in the \SD sensor as shown in Figure~\ref{fig:ccd_examples}, but the latter form an image of about 250 pixels in radius while the former form an image of about 4 pixels in radius. Their photometry is thus very differently affected by issues such as ghost reflection and light scattered in the tails of the PSF. In this section, we start by building a model of the PSF of \SD+ CBP, including light's reflection in the optics. We use this model to correct the inter-calibration ratio between the two pinhole sizes, measured with run No.~8 data. We then describe the generic photometry method applied to all other runs using the \spinhole pinhole data and present the resulting response measurements.   

\begin{figure}[h]
    \centering
    \includegraphics[width=\columnwidth]{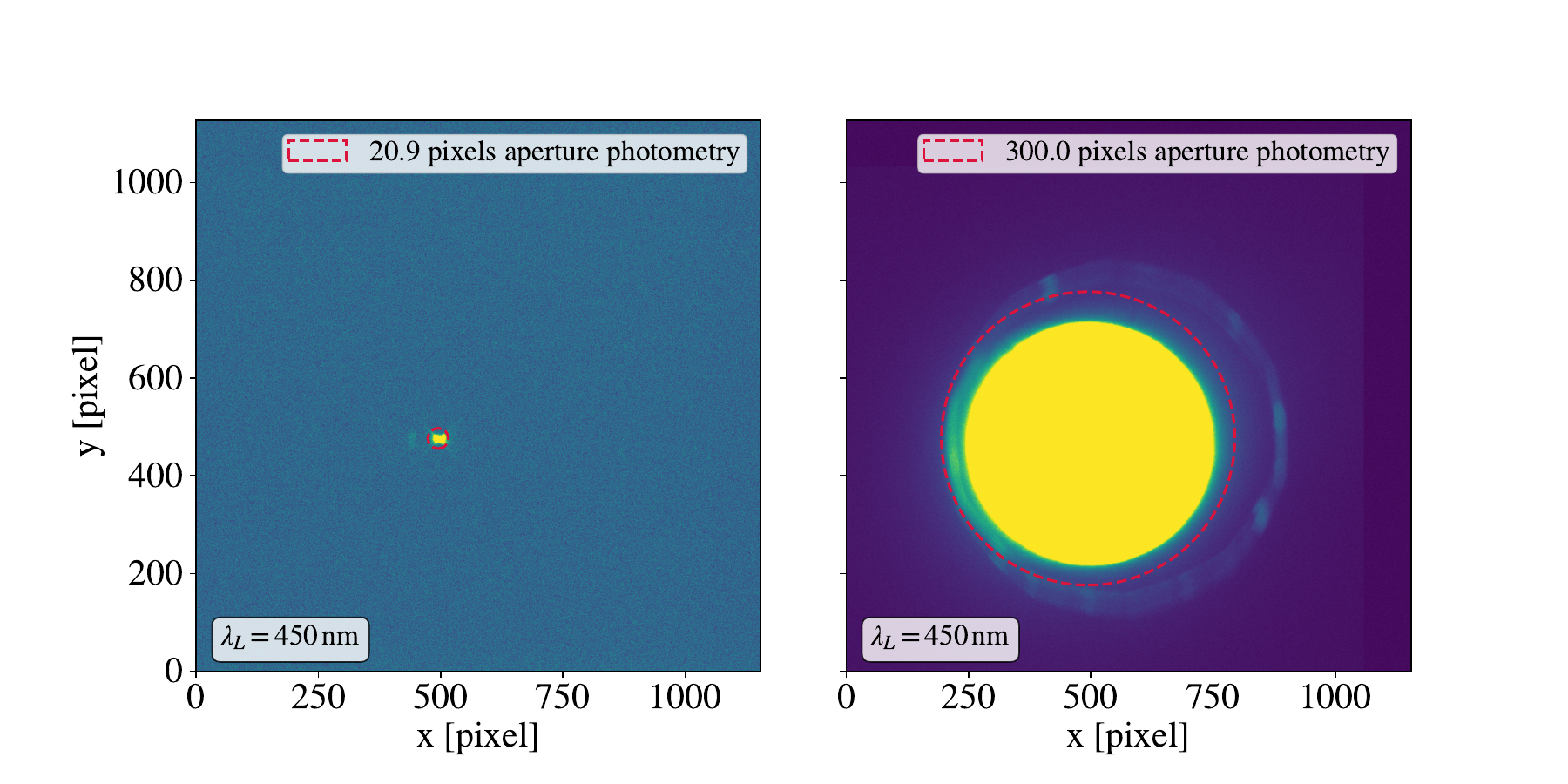}
    \caption{Examples of images obtained when the CBP shoots in the \SD telescope at $\lambda_L=\SI{450}{\nm}$ with the \bpinhole pinhole on the right and \spinhole pinhole on the left. The ghost reflection is visible for both images at the left of the main spot. In the \bpinhole pinhole image, a large annulus around the main spot is visible and corresponds to light diffusion around the mechanical iris at the input of the CBP.}
    \label{fig:ccd_examples}
\end{figure}

\subsection{Modelisation of the \SD PSF on \spinhole pinhole data}
\label{sec:modelisation-sd-psf}

Figure~\ref{fig:ghost_contrast} shows stacks of images obtained with the \spinhole pinhole without any filter. The illumination of a section of the primary StarDICE mirror results in a superimposition of the \SD telescope PSF and a set of additional fainter images that we call ghosts. The ghosts result from undesired but unavoidable reflections on optical surfaces. The most visible ones come from the beam reflection on the CCD surface and back reflection on its covering window, as described in Figure~\ref{fig:schema_ghost}.\\

\begin{figure}[h]
    \centering
    \includegraphics[width=\columnwidth]{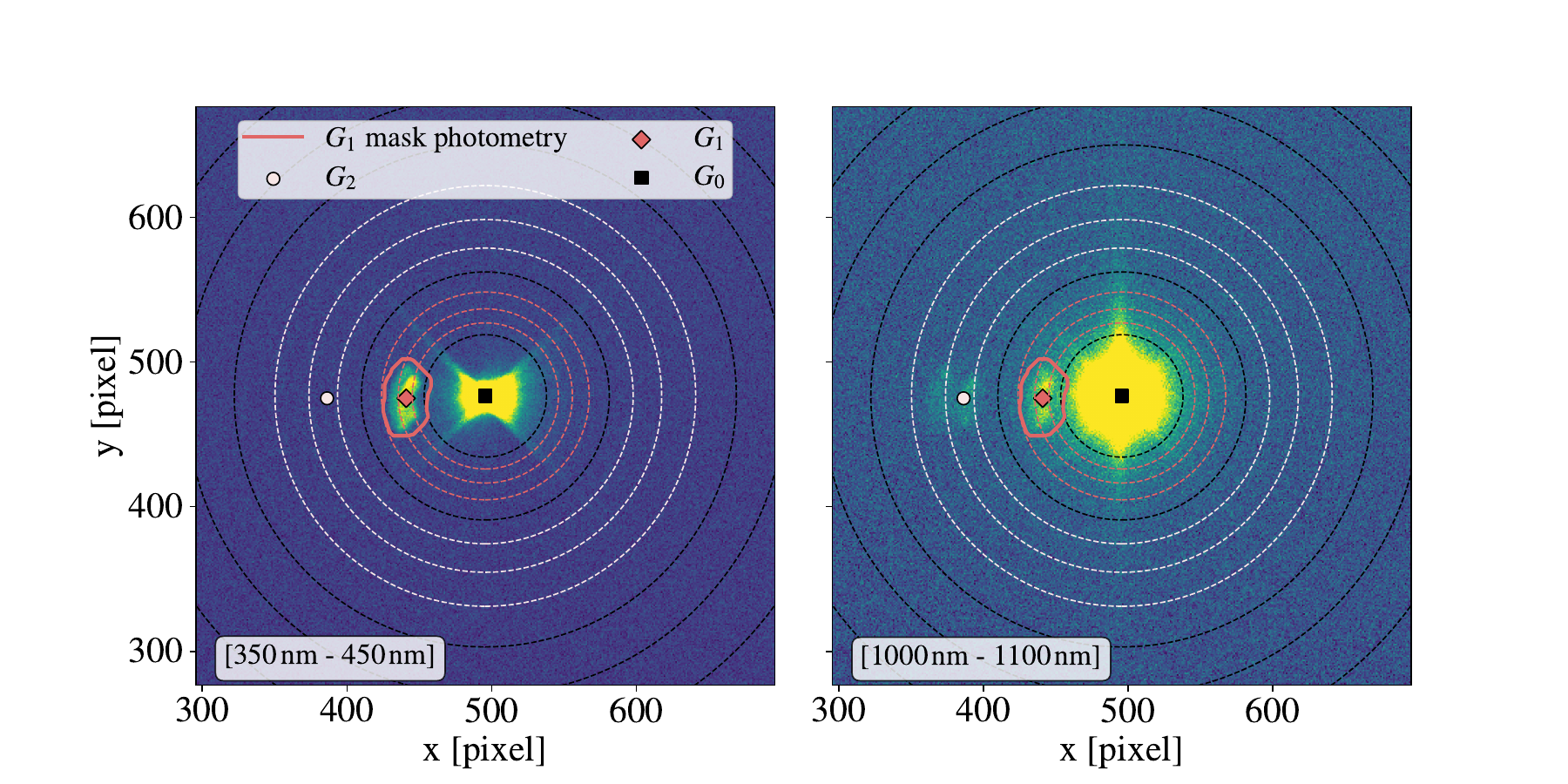}
    \caption{Stack of images at different wavelengths, with one image per nanometer. \textit{Left:} Stack of images between \SI{350}{\nano\meter} and \SI{450}{\nano\meter}. The scale is set on the ZScale from IRAF to make faint features visible. The spot of interest $G_0$ is at the center, and the ghost reflections are at its left. The circles correspond to the aperture photometry at different radii. When an annulus contains a ghost contribution, it is colored according to the ghost's order. The red area around the 1$\up{st}$ order ghost represents the area where the photometry of the ghost is measured. \textit{Right:} Same image but for a stack of images between \SI{1000}{\nano\meter} and \SI{1100}{\nano\meter}. We note that the 2$\up{nd}$ order ghost is not visible for the stack in the UV, where the 1$\up{st}$ order is maximal, while it is visible in the IR, where the 1$\up{st}$ order is lower.}
    \label{fig:ghost_contrast}
\end{figure}

For each image in this dataset, we subtract a bias pattern, estimated on a column and row basis, by computing the mean of the horizontal and vertical overscans. We compute the centroid of the signal and then perform aperture photometry by summing pixels within a radius $r=\SI{20.9}{pixels}$, and then for successive annulus of external radius from \SI{24.9}{pixels} to \SI{419.1}{pixels}. These radii are regularly spaced on a logarithm scale shown in Figure~\ref{fig:ghost_contrast}.

We model the \SD telescope PSF as the sum of a Moffat function, a constant background level, and contributions from ghosts at specific distances from the main spot. The Moffat function \citep{moffat}, integrated in an aperture of radius $r,$ reads:
\begin{equation}
M(r, \lambda)= 1 - \left( 1+\frac{r^2}{\alpha(\lambda)^2} \right)^{1-\beta(\lambda)},
\end{equation}
with $\alpha(\lambda)$ and $\beta(\lambda)$ the scale and exponent parameters of the Moffat distribution, which depend on the wavelength. The flux $F(r, \lambda)$ measured in the CCD with aperture photometry can then be modeled as: 
\begin{equation}
F(r, \lambda) = A(\lambda) \times \frac{M(r, \lambda) + \Kghostfit(r, \lambda)}{1 + \Kghostfit(r \rightarrow +\infty, \lambda)} + \pi r^2 bkg(\lambda),
\label{eq:moffat_model}
\end{equation}
with $A(\lambda)$ the total amplitude, $\Kghostfit(r, \lambda) = \frac{\sum_{n=1}^{+\infty} G_n(\lambda)}{A(\lambda)}$ the relative contribution of the sum of all the ghosts $G_n(\lambda)$, and $bkg(\lambda)$ the background level in ADU/pixel. 

The evolution of the Moffat distribution is smooth with respect to wavelength. We therefore develop the parameters $\alpha(\lambda)$ and $\beta(\lambda)$  on a B-spline basis with 15 wavelength nodes regularly spaced between \SI{350}{\nano\meter} and \SI{1100}{\nano\meter}. The same applies to the function $\Kghostfit(r, \lambda)$ as it corresponds to light reflections on surfaces in the optical path. This reduces the number of free parameters and allows us to perform outlier rejection of a few annulus whose photometry was affected by a cosmic ray (9 photometric points affected out of 13122 measurements). The fit results are shown in Figure~\ref{fig:result_params}. 

\begin{figure}[h]
    \centering
    \includegraphics[width=\columnwidth]{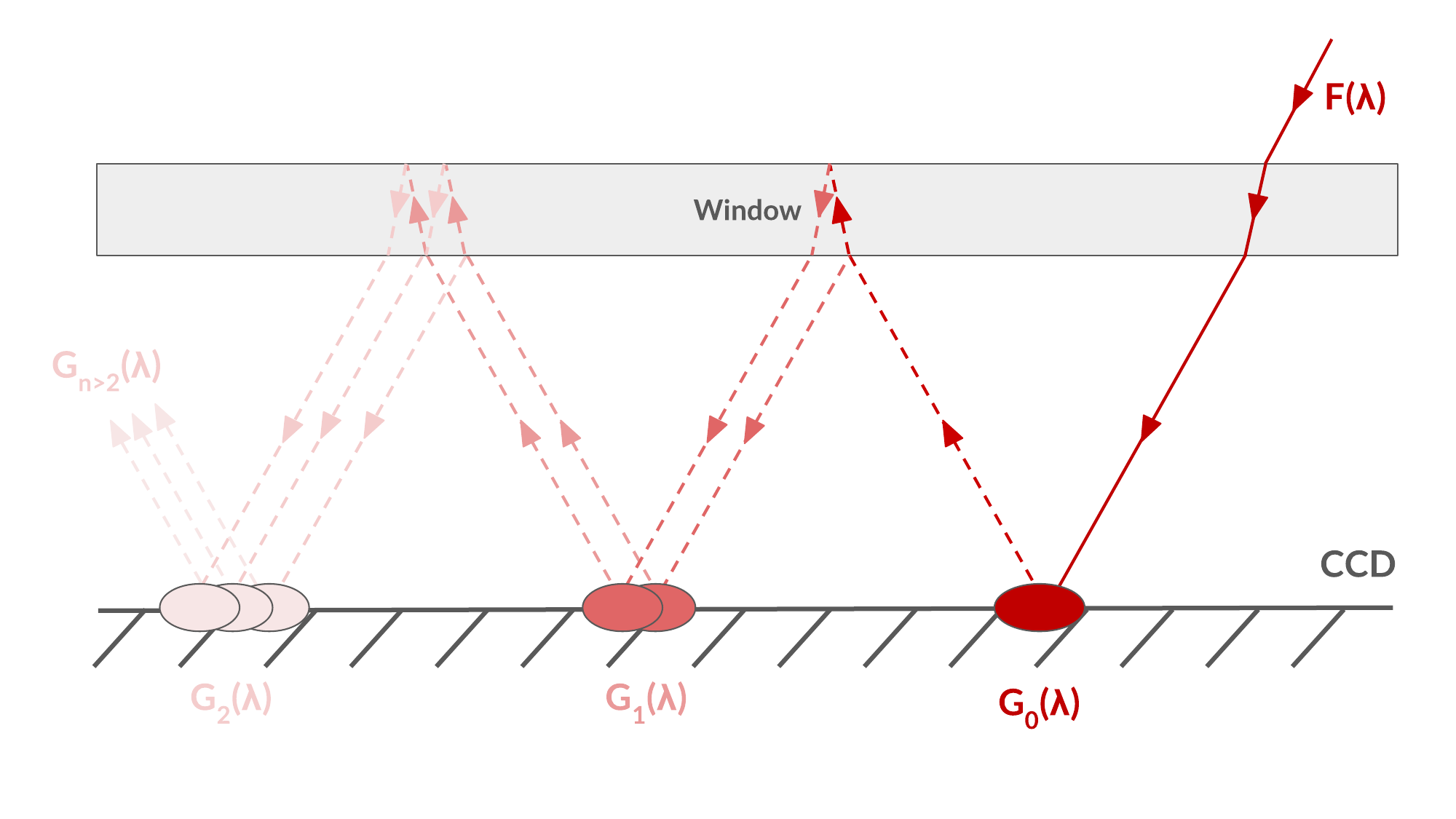}
    \caption{Schematic of the light reflections that generate the ghosts, which are defocused and less intense images at other positions in the focal plane. A fraction of the light is reflected at the different interfaces, and $G_0(\lambda)$, $G_1(\lambda)$ and $G_2(\lambda)$ are respectively the main spot, the 1\up{st} order ghost, and the 2\up{nd} order ghost.}
    \label{fig:schema_ghost}
\end{figure}

The most striking result is a significant degradation of the PSF
after \SI{950}{nm}, apparent in the steep decrease of the value of
the $\beta$ Moffat parameter and the increase of the $\alpha$
parameter presented in the first and second panels. This effect is also
visible when comparing the two stacks in
Fig.~\ref{fig:ghost_contrast}. We were not able to pin down the exact
origin of this degradation. A potential explanation could be that one
of the reflective surfaces becomes partially transparent at these
wavelengths and generates diffused light. The consequence of this
effect on photometry will be further discussed in Section \ref{sec:photometry_small}.

The $\alpha(\lambda)$ and $\beta(\lambda)$ best-fit parameters are
otherwise fairly stable between \SI{350}{\nano\meter} and
\SI{900}{\nano\meter}. The third panel shows the background level
reconstructed across all wavelengths. With a mean of
$\mu_\mathrm{bkg, fitted}=\SI{0.267}{ADU/pixel}$ for an exposure time
of \SI{1.1}{\second}, it is very consistent with the values measured
on dark images. Two datasets of dark images have been studied, one
with the laser turned off and a second by masking the CBP output with
a cap (dataset No.~9 in Table~\ref{tab:schedule}). Both datasets show
no trend in time or wavelength and have a mean value of
$\mu_\mathrm{dark, photometry}=\SI{0.252}{ADU/pixel}$ and a standard
deviation $\sigma_\mathrm{dark, photometry}=\SI{0.059}{ADU/pixel}$,
represented by the black dashed line and shaded area in this third
panel.

The fourth panel presents the relative contribution of the first order
ghost $G_1(\lambda)$,
$\Kghostfitfirst(\lambda) = \frac{G_1(\lambda)}{A(\lambda)}$ in
percent. We compared these reconstructed values with direct photometry
of the visible first-order ghost pattern performed in 2D images as
detailed in Appendix~\ref{sec:ghost_photometry}. This more direct
determination of the ratio $\Kghost$ is shown as the dashed black
line. The two different methods match perfectly up to \SI{950}{nm}
showing that the identified ghost constitutes the only significant
deviation from the Moffat profile. Beyond \SI{950}{nm}, the second method
does not give accurate photometry for the ghost because it starts
to be mixed with the core of the PSF.


The fit reduced chi-squared of 1.86 is explained by the fact that the Moffat
shape does not describe the complex shape of the PSF core. To mitigate this
issue, and as we are mainly interested in modeling the tails of the PSF, we
start our profile modeling at a rather large inner aperture of 20.9 pixels, where
residuals no longer display visible structure, and the chi-square displays no
trend in wavelength. Going to a larger inner aperture increases the quality of
the fit but decrease its stability as the scale parameter $\alpha$ is no longer
properly constrained.

\begin{figure}[h]
     \centering
     \resizebox{\hsize}{!}{\includegraphics{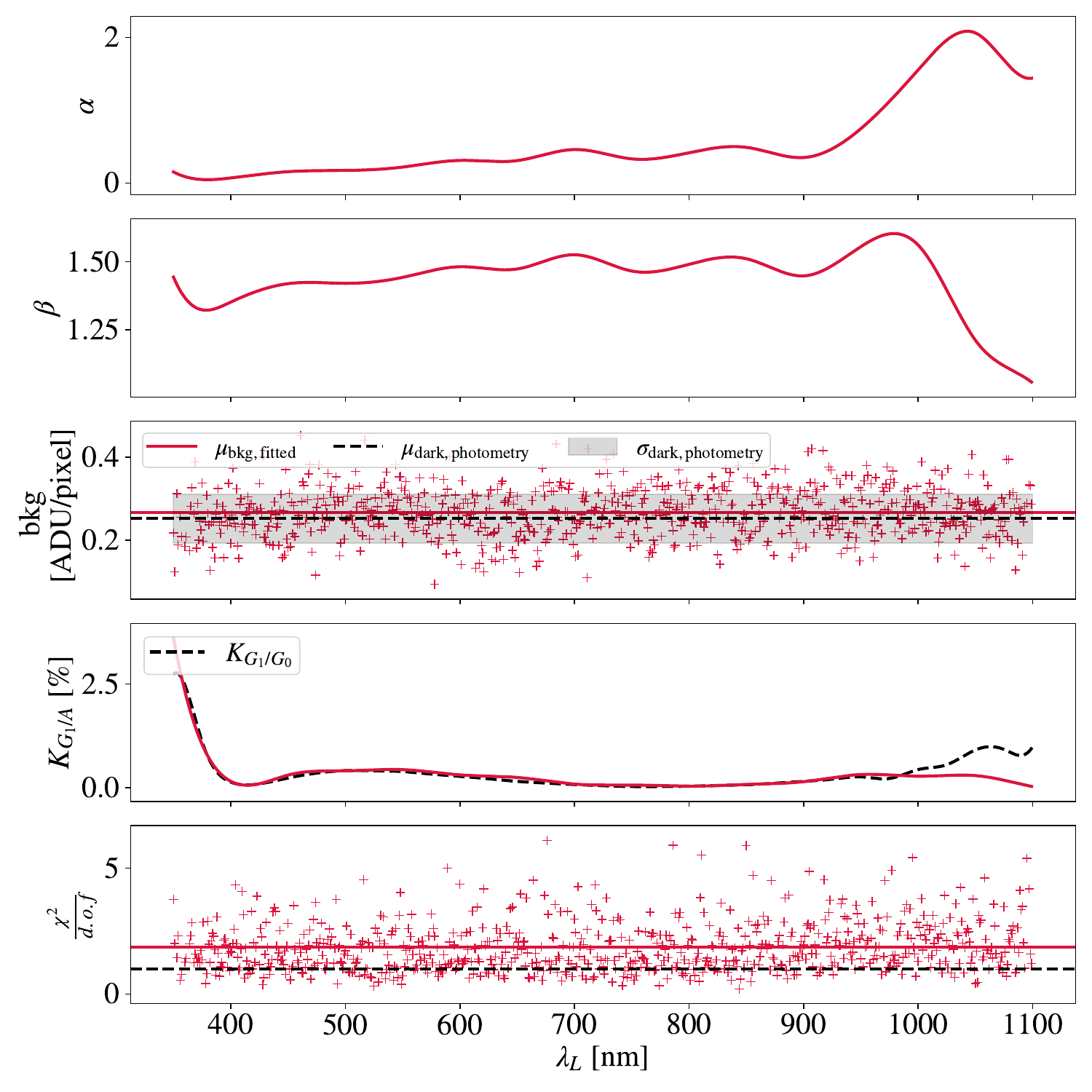}}
     \caption{Best fitting parameters for the model of Equation~\ref{eq:moffat_model}. Red plain lines correspond to the result of the fit, and black dashed lines, when applicable, correspond to the values expected. From top to bottom: the first and second panels represent respectively the $\alpha(\lambda)$ and $\beta(\lambda)$ parameters from the Moffat distribution; the third panel shows the background contribution per pixel and its mean value in red, while the dark dashed line is the mean value of the dark; the fourth panel represents $\Kghostfitfirst(\lambda)$, and $\Kghost$; the fifth panel represents the reduced chi-squared of the fit.}
     \label{fig:result_params}
\end{figure}

%
%
%
%

\subsection{Accounting for CBP light contamination}\label{sec:sd_contaminations}

As for the photodiode and the solar cell (see Sections~\ref{sec:532_cont} and~\ref{sec:fluorescence}), the total signal measured in the \SD camera $\Qccdmes$ is the sum of the flux from the main laser line $\Qccdcal$ and the flux from the \SI{532}{\nm} contamination $\Qccd^{532}$, the $\lambda_{\rm comp}$ contamination $\Qccd^{\lambda_{\rm comp}}$, and the integrating sphere fluorescence $\Qccd^{\mathrm{fluo}}$, as follows:
\begin{equation}
    \Qccdmes = \Qccdcal + \Qccd^{532} + \Qccd^{\lambda_{\rm comp}} + \Qccd^{\mathrm{fluo}}
    \label{eq:qccd_mes}
\end{equation}
The contamination light in the \SD camera can be estimated by multiplying $\Qphot^{532}$, $\Qphot^{\lambda_{\rm comp}}$ and $\Qphot^{\mathrm{fluo}}$ measured with the CBP photodiode by a first estimation of $\Rcbp(\lambda)$ and $\Rtel(\lambda)$. We then refine the measurement of $\Rtel(\lambda)$ and correct iteratively the effect of those contaminations.

The effects of this procedure are most clearly demonstrated by its impact on the \SD $g$ filter transmission measurement, as presented in Figure~\ref{fig:g_filter_532}. This measurement is particularly susceptible to contamination from integrating sphere fluorescence and the \SI{532}{\nano\meter} line. Our results indicate that contamination is responsible for the apparent out-of-band transmission of the filter, which completely disappears once contributions at contaminating wavelengths are corrected.

\begin{figure}[h]
    \centering
    \includegraphics[width=\columnwidth]{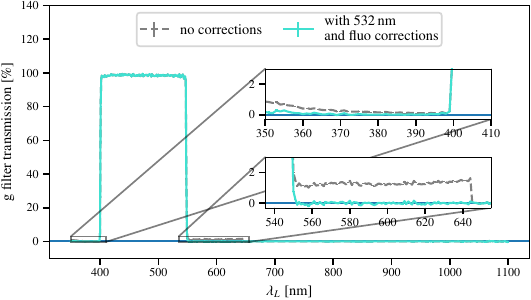}
    \caption{\SD g filter transmission as a function of the set laser wavelength $\lambda_L$, with and without the \SI{532}{\nm} and fluorescence correction. For clarity, we added zooms on the out-of-band transmissions where contamination light is transmitted while the main wavelength is blocked.}
    \label{fig:g_filter_532}
\end{figure}

\subsection{Pinhole intercalibration}

Given that we utilize different pinholes for measuring the CBP and \SD responses, it is necessary to intercalibrate the CBP response with both pinholes. The CBP response $\Rcbp$, as defined in Equation~\ref{eq:rsd}, can be expressed as follows:
\begin{equation}
  \label{eq:K}
	\Rcbp(\lambda) \equiv \Rcbp^{\spinhole}(\lambda) = \Rcbp^{\bpinhole}(\lambda) \times \Kpinholes,
\end{equation}
with $\Kpinholes$ the inter-calibration term and $\Rcbp^{\spinhole}$ (resp. $\Rcbp^{\bpinhole}$) the CBP response measured with the \spinhole pinhole (resp. \bpinhole pinhole). 

We estimate the correction term $\Kpinholes$ as the ratio of the \SD responses obtained with both pinholes in runs No.~8. The \bpinhole pinhole photometry is performed as follows. The background is estimated with dark exposures from the dark datasets, and the spatial mean of all these dark images is computed to obtain a master dark. When performing aperture photometry on the \bpinhole image, we subtract a background equivalent to the aperture photometry of the master dark at a similar position and radius, obtaining $\Qccd^{\bpinhole}$. The optimal radius is evaluated at 300 pixels for the \bpinhole pinhole to contain the main spot and the 1\up{st} order ghost. Figure~\ref{fig:stardice_5mm_response} shows the \SD response obtained with the \bpinhole pinhole and no filter. 


\begin{figure}[h]
    \centering
    \includegraphics[width=\columnwidth]{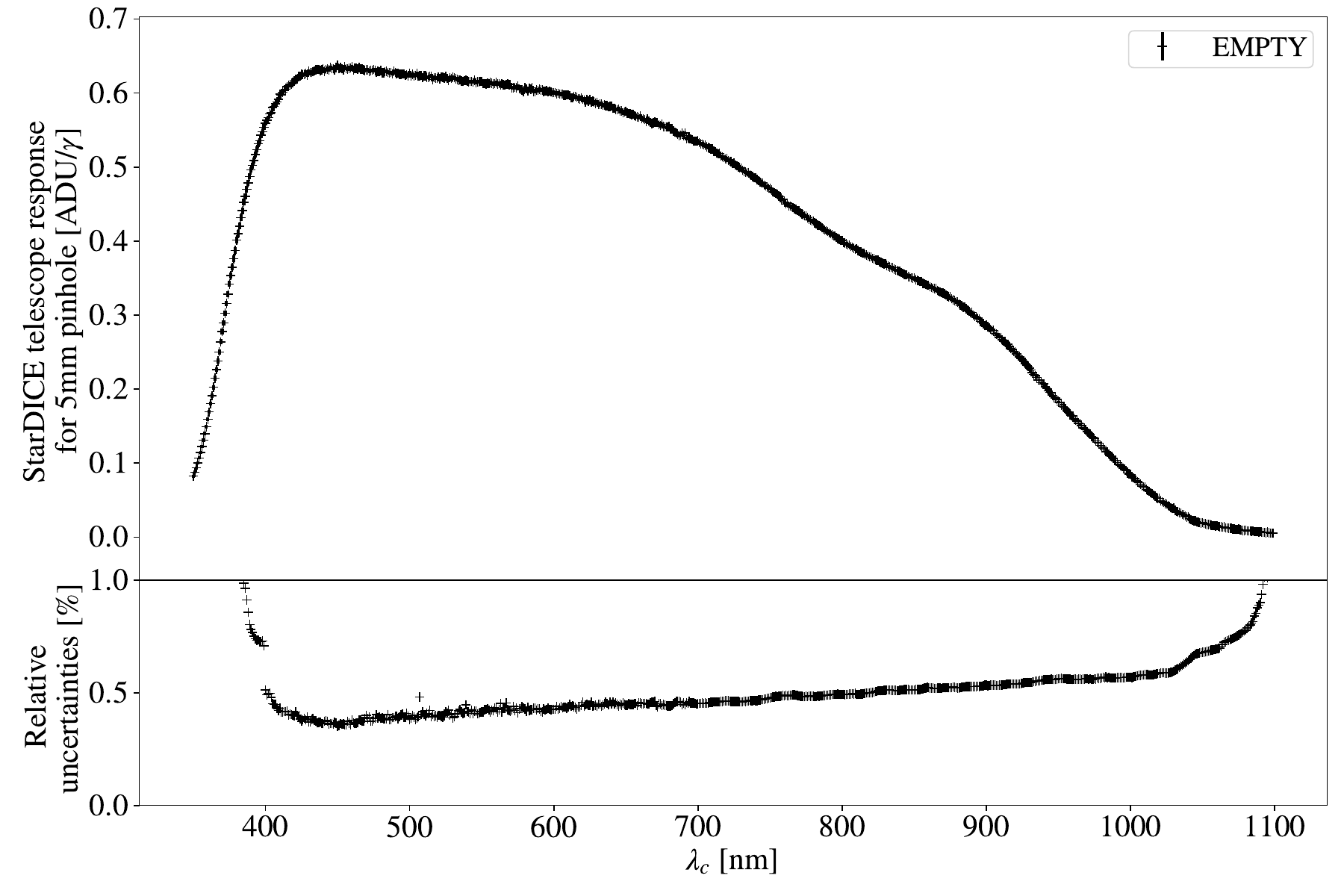}
    \caption{\textit{Top:} \SD response with no filter and \bpinhole pinhole with respect to wavelength in nanometer. \textit{Bottom:} Uncertainties over the \SD response measurement with respect to wavelength in nanometer.}
    \label{fig:stardice_5mm_response}
\end{figure}



We estimate the pinhole flux, $F(300, \lambda)$, as the flux obtained by integrating the best-fit model (Equation~\ref{eq:moffat_model}) within a 300-pixel aperture. This approach ensures that we compare the \SD responses for both pinholes at an equivalent radius, thereby incorporating identical features of the point spread function (PSF) while maintaining a favorable signal-to-noise ratio for the fainter pinhole. The ratio $\Kpinholes=\frac{F(300)}{\Qccd^{\bpinhole}}$ is presented in Figure~\ref{fig:ratio_pinholes}.

The noticeable oscillations in wavelength observed in this ratio starting at \SI{900}{nm} are attributable to fringes present in the CCD, which significantly impact the \spinhole pinhole measurement but have a lesser effect on the large pinhole as most average out over the large surface of the pinhole image. Apart from this phenomenon, which is unrelated to a change in the CBP transmission, a slight chromatic change in the measured ratio cannot be attributed to differences in photometry. We attribute this change to an effective modification in $\Rcbp(\lambda)$ upon switching pinholes and model it as a linear function of wavelength. We fit this function in the range \SIrange{400}{900}{\nano\meter} where the ratio measurement is clean. This function is utilized for recalibrating the CBP transmission with the \spinhole pinhole, as described in Equation~(\ref{eq:K}), throughout the remainder of this analysis.

\begin{figure}[h]
    \centering
    \includegraphics[width=\columnwidth]{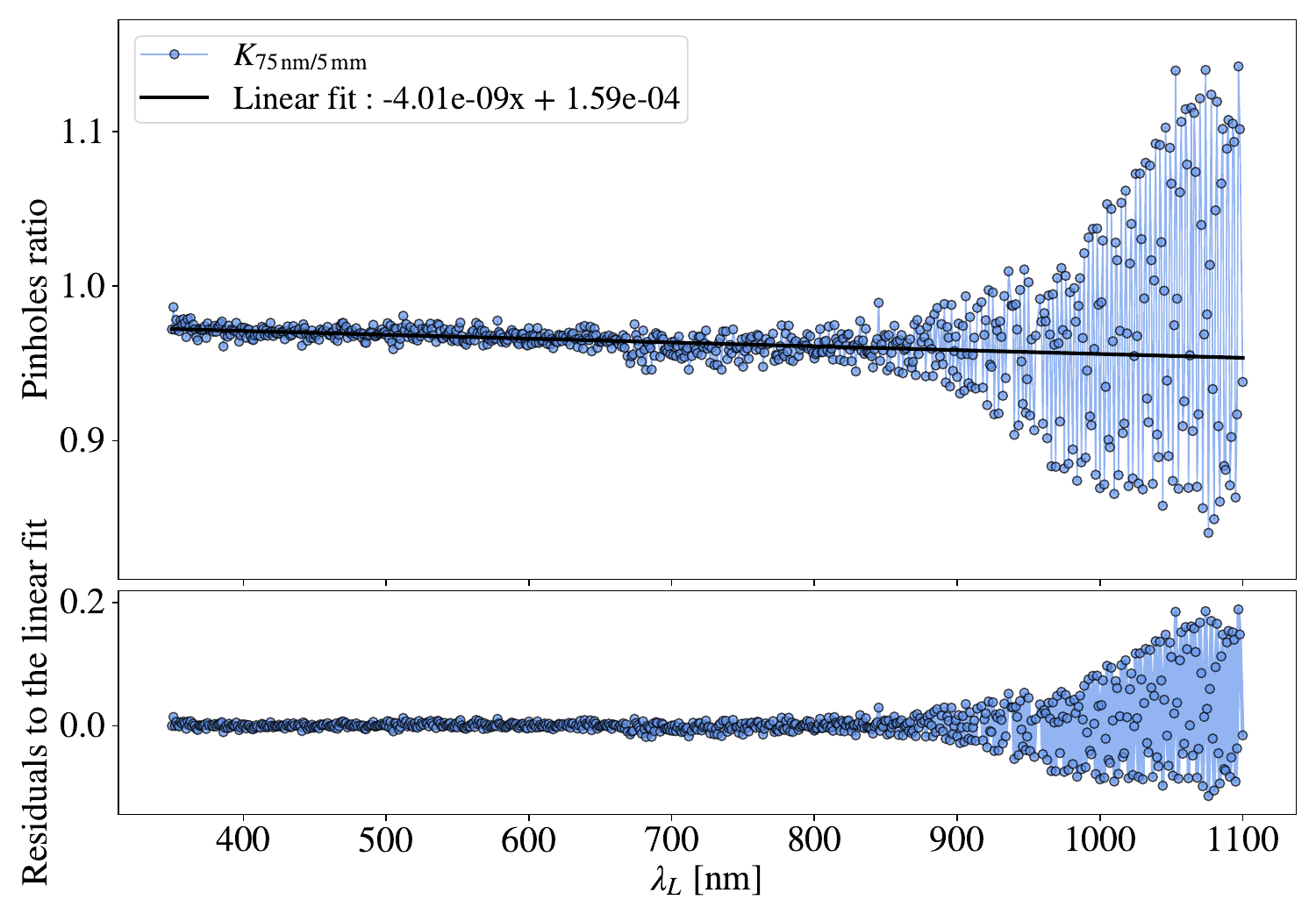}
    \caption{Ratio $\Kpinholes$ as a function of wavelength, relative to the prediction given by the ratio of the nominal area of the pinholes. The black line corresponds to a linear fit between \SI{400}{\nm} and \SI{900}{\nm}.}
    \label{fig:ratio_pinholes}
\end{figure}

\subsection{Photometry applicable to on-sky data}
\label{sec:photometry_small}

The measurement of the PSF at a long distance from its core is possible in CBP images because of the shallow level of background light in these images. It is impractical to port this method to on-sky data. We choose to carry out the throughput measurement for an aperture photometry and a background estimation algorithm applicable to on-sky data. 

The photometry is performed as follows. To estimate the background contribution, the sources in the image need to be detected and masked. We compute the standard deviation of an image $\sigma$, and every pixel with a signal higher than 5$\sigma$ is masked, as well as all the surrounding pixels with a signal higher than 2$\sigma$. Then we proceed to a segmentation of the masked image into boxes of $129\times132$ pixels (corresponding to \textasciitilde $\SI{3.6}{arcmin} \times \SI{3.6}{arcmin}$). We compute the mean and the standard deviation of the background in each of these boxes, and we interpolate their values in a 2D map to get our estimation of the background. This background estimation is subtracted from the image. Then, the centroid of the spot of interest is computed as the Gaussian-weighted first moment of the flux to pursue aperture photometry at this position. $\Qccd$ is measured with aperture photometry at a radius of \SI{20.9}{pixels} for every image of the dataset with the \spinhole pinhole.

Based on the model built in Sect.~\ref{sec:modelisation-sd-psf}, we can quantify the fraction of the total flux measured in \spinhole pinhole images with this aperture photometry method. We define the fraction of flux missed as the ratio between the value measured with aperture photometry for a given radius and wavelength and the total amplitude $A$ fitted with the method detailed in the previous section for the same wavelength. The result is the blue curve in Figure~\ref{fig:bias_aperture}. About 98\% of the flux is collected between \SI{400}{\nano\meter} and \SI{900}{\nano\meter}. The ghost contribution, which is missed when using an aperture of \SI{20.9}{pixels}, contributes about another extra percent to the missed flux below \SI{400}{\nano\meter}. Above \SI{900}{\nano\meter} where the PSF is degrading fast, the fraction of flux missed increases by one order of magnitude and reaches more than 75\% at \SI{1050}{nm}.

While most of this effect is directly explained by the aperture correction, a small contribution also comes from a bias in the background estimate caused by the pollution of the background map by the source PSF tails. Given that the model of the PSF predicts the value of the aperture correction (red curve in Fig.~\ref{fig:bias_aperture}), we can estimate the contribution of the background bias as the difference between the two curves. This curve, presented in the bottom panel of Fig.~\ref{fig:bias_aperture}, shows that the background bias contributes about 0.2\% of the total flux.

\begin{figure}[h]
     \centering
     \resizebox{\hsize}{!}{\includegraphics{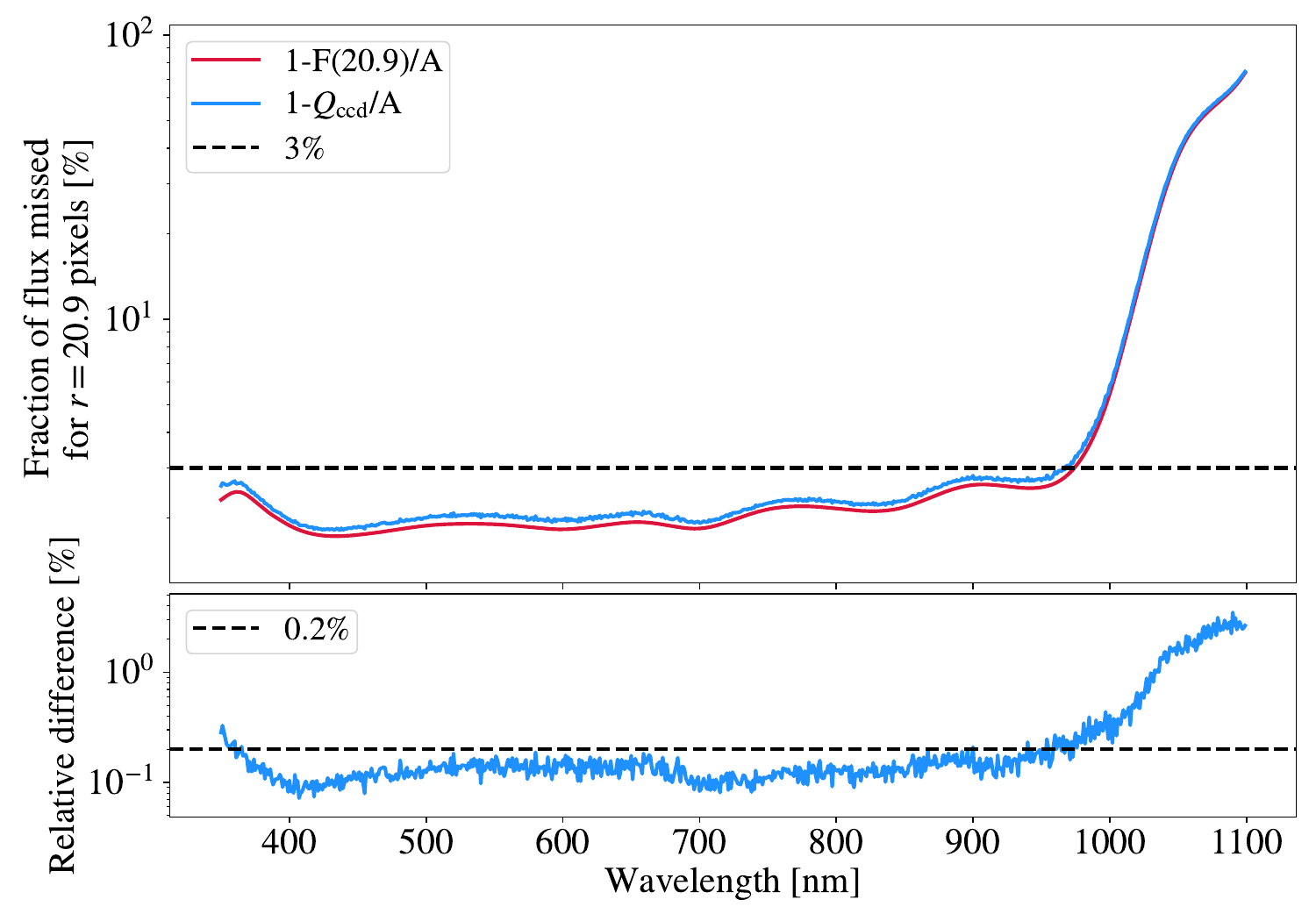}}
     \caption{\textit{Top:} Fraction of missing flux against $\lambda_L$ when measuring \spinhole pinhole dataset with aperture photometry at \SI{20.9}{pixels} rather than taking the total amplitude $A$ fitted. The red curve corresponds to the estimation of the fit and the blue one to aperture photometry when the background is estimated with the method described in Section~\ref{sec:photometry_small}. \textit{Bottom:} Relative difference in percent between the flux measured for aperture photometry with the two background estimations and the model estimation at \SI{20.9}{pixels}.}
     \label{fig:bias_aperture}
\end{figure}

%
%
We conclude that the instrument response determined for our large aperture photometry of the \spinhole pinhole is likely to apply to similar star photometry with minimal aperture corrections in the range between \SI{400}{\nano\meter} and \SI{900}{\nano\meter}. Below \SI{400}{\nano\meter}, there will be a need to consider the ghosting pattern for the full aperture, which will mix partially with the photometry aperture. Above \SI{900}{\nano\meter}, our method is unlikely to give a representative determination of the instrument's response to star flux. At this stage, we think that the \SD telescope is responsible for the PSF degradation, with one possible explanation being that the primary mirror becomes partially transparent and generates diffused light. Further discussion of the applicability of the measured transmission to on-sky data will require a study of the PSF on stellar images, which is left for future work.

%
%
%
%

\subsection{Results}
\subsubsection{\SD response and filter transmissions}

We measured the \SD response with the empty slot of the filter wheel, the transmission of the $ugrizy$ filters, and the grating transmission with dataset No.~6, using the \spinhole pinhole, shooting at a fixed mirror and focal plane position. Figure~\ref{fig:stardice_75um_response} shows the results obtained, following the procedure based on aperture photometry. We note that the wavelength coverage does not scan the entire passband of the $u$ filter. Filling the gap between our measurements and the atmospheric UV cut-off would need additional measurements that are out of the scope of this paper.

After accounting for the relative uncertainties, we demonstrate a precision of \textasciitilde\SI{0.5}{\%} per bin of \SI{1}{nm} for all the filter transmissions, the empty filter wheel slot, and the 1$^\mathrm{st}$ order of the diffraction grating from \SIrange{400}{950}{\nano\meter}. The \SI{0.2}{nm} wavelength accuracy of our measurements is illustrated by the fine resolution of the sharp filter edges.

\begin{figure}[h]
    \centering
    \includegraphics[width=\columnwidth]{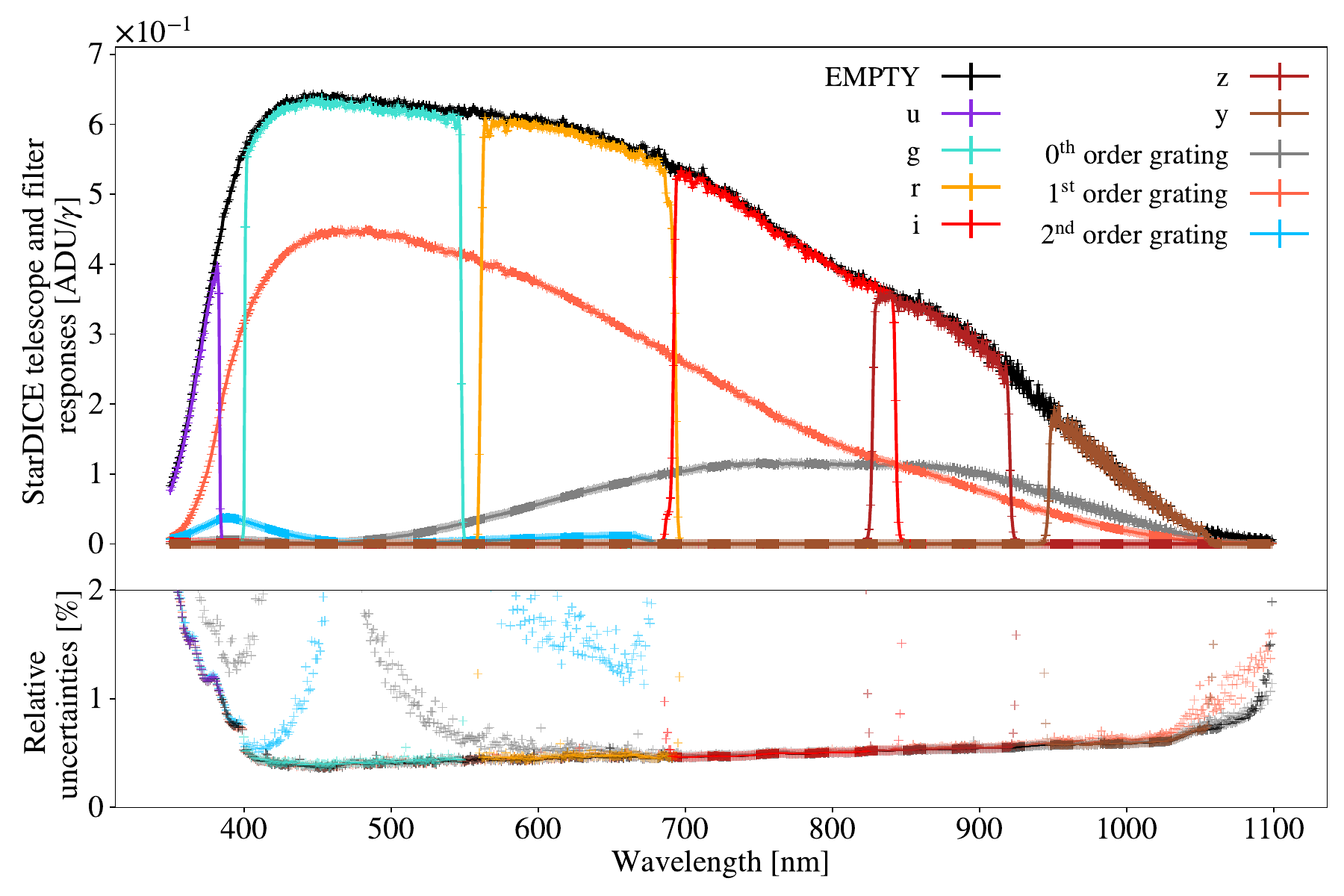}
    \caption{\textit{Top}: \SD response against wavelength in nanometer, with the empty filterwheel slot; all $ugrizy$ filters; and the 0$^\mathrm{th}$, 1$^\mathrm{st}$ and 2$^\mathrm{nd}$ order diffraction of the grating. All of these responses have been measured with the \spinhole pinhole. \textit{Bottom:} Relative uncertainties over the \SD response measurements against wavelength in nanometer.}
    \label{fig:stardice_75um_response}
\end{figure}


\subsubsection{\SD Responses variations}

Since the CBP only illuminates a portion of the primary mirror, we need several measurements at different positions to span the full transmission of the \SD telescope. Dataset No.~2, where we point at 4 different positions along the primary mirror radius, and No.~3, where we point at every one of the four quadrants of the primary mirror, have been acquired specifically to investigate these variations. We also explored the small-scale variations over the focal plane with dataset No.~12, where we made slight pointing changes around a fixed position.

Figure \ref{fig:radial_positions} is divided into two panels. The left panel exhibits the results of transmission measurements taken at various radii (corresponding to dataset No.~2). In contrast, the right panel displays the spatial variations in these measurements across the focal plane (representing dataset No.~12).

\begin{figure*}[h!]
    \centering
    \includegraphics[width=\columnwidth]{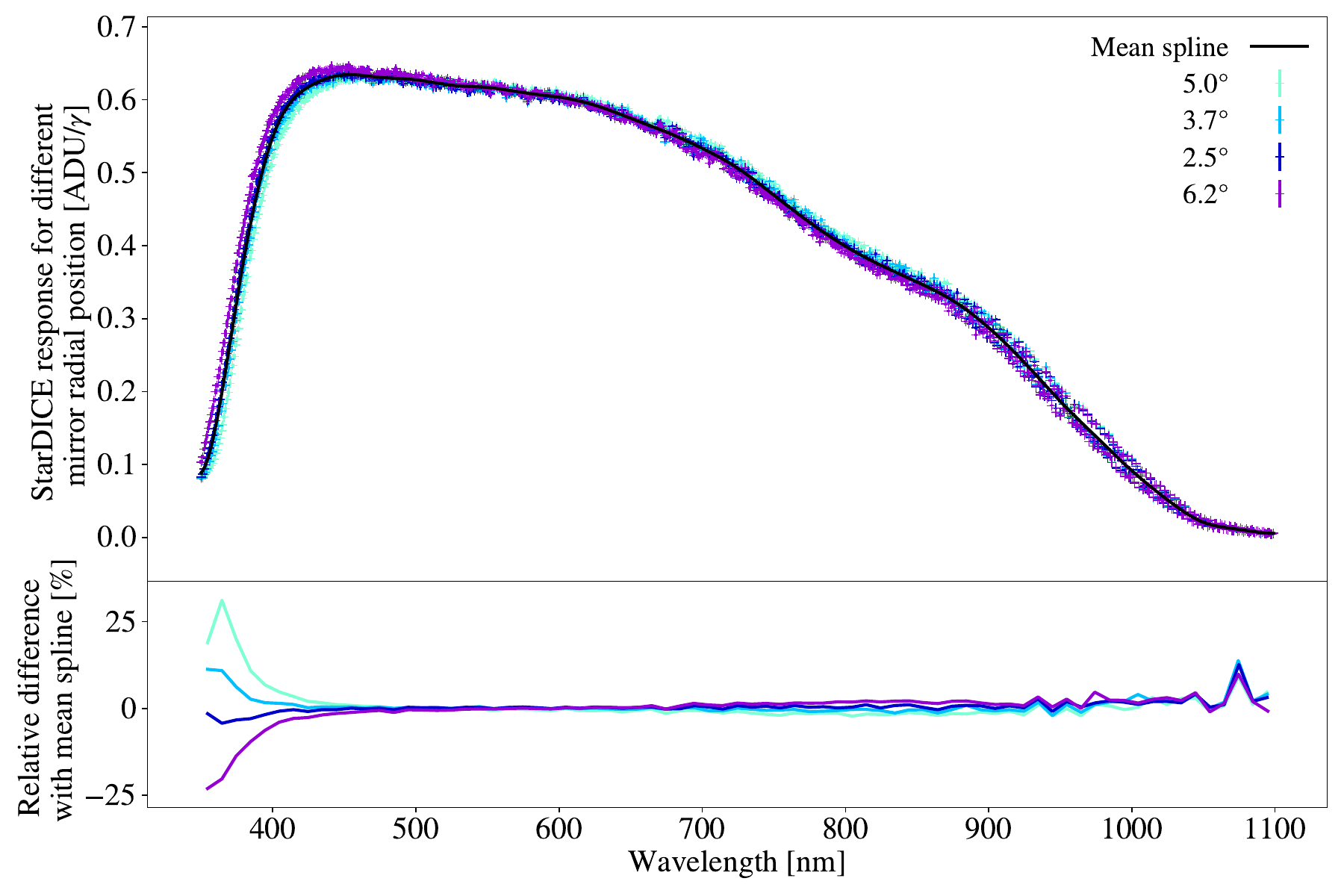}
    \includegraphics[width=\columnwidth]{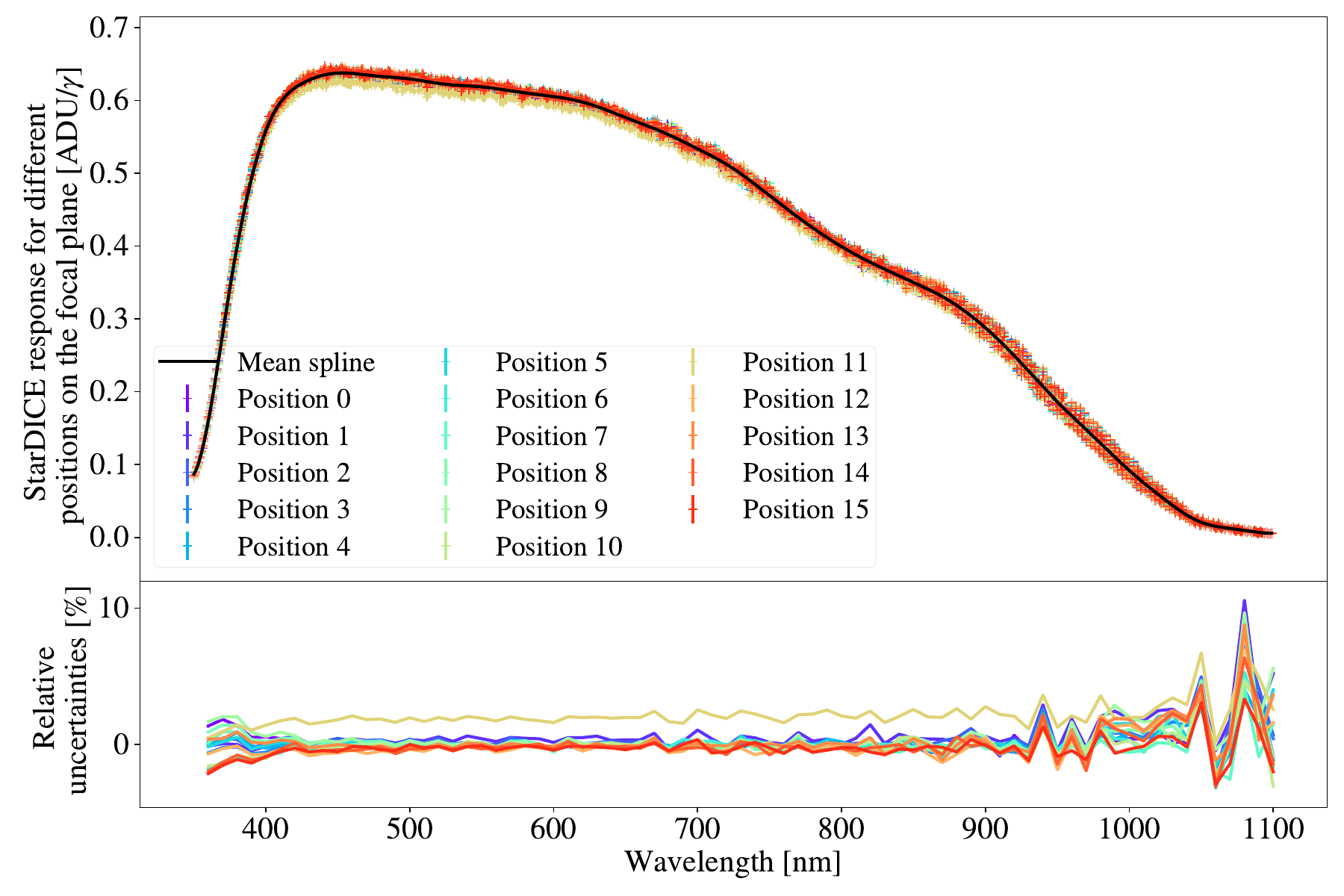}
    \caption{\textit{Top:} \SD response for the different radial positions on the mirror with a fixed focal plane position (left) and different focal plane positions with a fixed mirror position (right). The colors in the left plot match with positions illustrated in Figure~\ref{fig:8_mirror_positions}, while each color in the right plot represents a position on the focal plane. In both plots, the black dashed line corresponds to the mean spline. \textit{Bottom:} Relative difference between the data and the mean spline, binned to a \SI{10}{\nano\meter} resolution to smooth the effect of fringing.}
    \label{fig:radial_positions}
\end{figure*}

We display the average behavior for both datasets by fitting a smooth spline through the different measurements. The variations above \SI{900}{\nano\meter} are due to small-scale variations caused by interference fringing in the detector. As expected, the impact is larger when the optical path differences between measurements are smaller, i.e., when we explore the variations over the focal plane (right panel). We also note the yellow curve being off the mean spline by \textasciitilde 1\%, corresponding to a local non-uniformity in the focal plane, but we haven't investigated this aspect yet. 

The most striking feature of Figure~\ref{fig:radial_positions} needing discussion is the significant (\SI{20}{\%}) variation of transmission measured in the different quadrants of the primary mirror below \SI{400}{\nano\meter}. We didn't reach a firm conclusion concerning the explanation of this behavior. Still, the more solid assumption is that it might be explained by the Hilux coating\footnote{\url{https://www.orionoptics.co.uk/optical-coatings/}} on the \SD mirrors, whose reflectivity varies with the light incidence angle. This significant effect prevented us from making good use of the measurement in the 4 telescope quadrants (dataset No.~3) as the radius corresponding to those measurements could not be determined with sufficient precision, which impact will be quantified in Section~\ref{sec:final_uncertainty_budget}.

In the 400-\SI{900}{\nano\meter} region, the variations over the focal plane and
between primary mirror positions are below the percent level. This relative
homogeneity of the telescope response allows us to pursue the modeling of the full
pupil response of the instrument expecting a reasonable accuracy without the
need of additional data, as discussed further in section \ref{sec:pupil_stitching}.

Also, regarding the modeling of the full pupil transmission,
figure~\ref{fig:blueshift} shows the measurement of the expected interference
filter edges shift caused by the incident light angle variation when we
illuminate the primary mirror at different radii. The figure shows the edges of
the $g$ and $r$ filter in the top panel, and of the $z$ and $y$ filters in the
bottom panel as measured when illuminating four different radial positions in
dataset No~2. The edges are noticeably bluer at higher radial positions, which
correspond to higher incidence angles, illustrating how the accuracy of those
measurements allow us to consider their interpolation to obtain the full
pupil transmission of the telescope.

\begin{figure}[h]
    \centering
    \includegraphics[width=1\columnwidth]{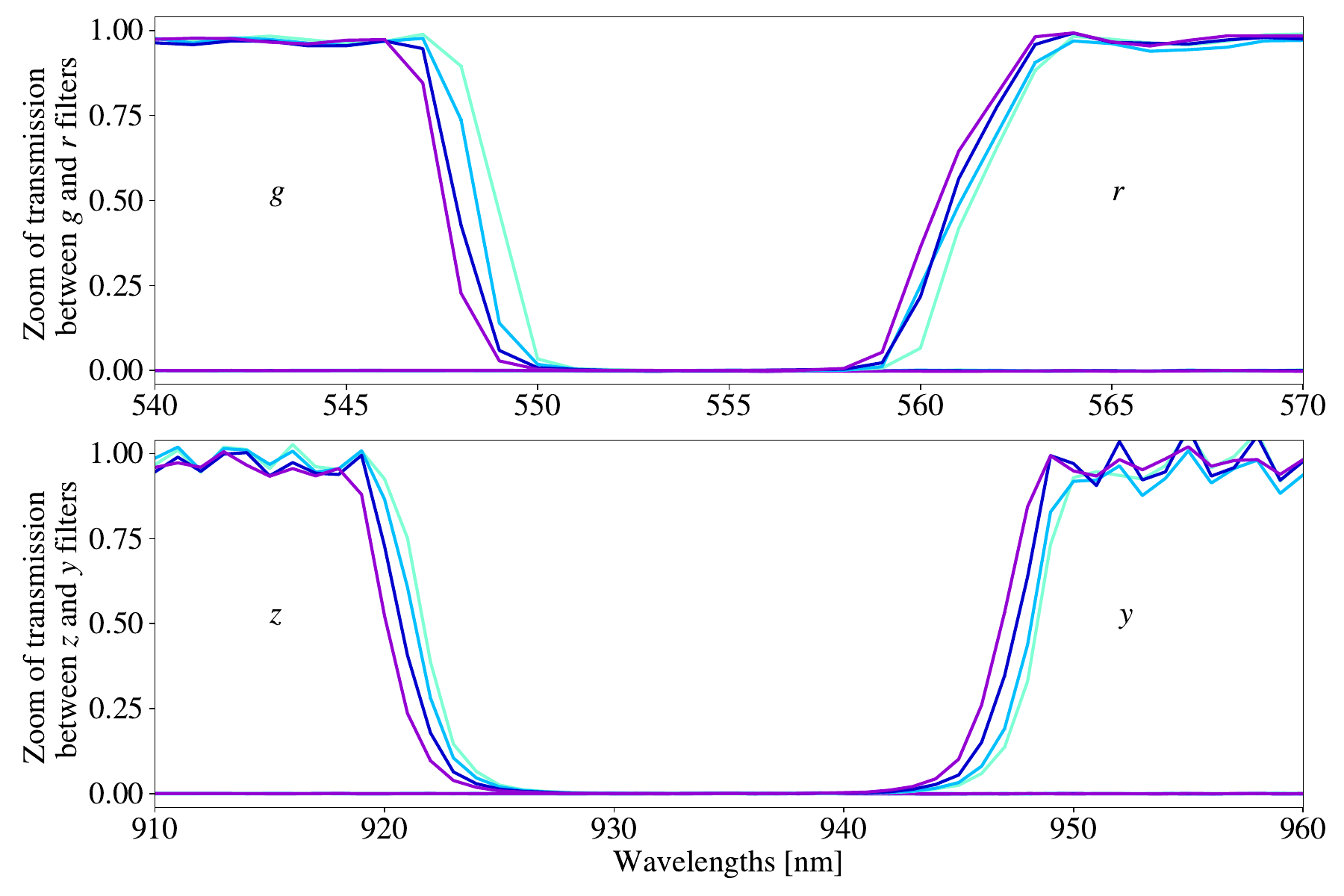}
    \caption{Zoom on the \SD filter transmissions measured at different radial positions on the mirror, between the $g$ and $r$ filters (top panel) and the $z$ and $y$ filters (bottom panel). The colors correspond with the radius labeled in Figure~\ref{fig:8_mirror_positions}.}
    \label{fig:blueshift}
\end{figure}

%
%

\section{Synthesis of the equivalent transmission for full pupil illumination}
\label{sec:pupil_stitching}

The telescope's reflectivity and the filters' transmission exhibit a distinct radial dependency. Although the filter
transmission dependence is expected due to its interferometric nature,
the origin of the telescope reflectivity dependence remains
unclear. Despite this, we can construct an empirical model that
assumes smooth transitions between measurements and calculate the
theoretical "full pupil" transmission by averaging the model over the
illuminated portion of the primary mirror. These two steps are
necessary to achieve sub-percent color accuracy and sub-nanometer
precision in central wavelengths determination.

We also estimate statistical and systematic errors in the final
transmission curves and propagate them through the analysis process. A
better understanding of these errors in interpreting broadband
photometry is obtained using the spectrophotometric standard star
G191B2B as a sample star.

\subsection{Radial model of the instrument transmission}
\label{sec:model}

The open transmission of the telescope is modeled as a smooth 2D
function of wavelength and incidence angle. The function is defined on
a basis of cubic B-splines, with \num{35} regularly spaced wavelength
nodes covering the range from \SIrange{350}{1100}{nm}, and two nodes
at angles corresponding to the inner and outer edges of the
occultation-free primary mirror, ranging from \SIrange{1.97}{7.24}
{\degree}. These angles represent radii of \SIrange{55}{203}{mm} for a
primary mirror with a focal length of \SI{1600}{mm}.

For data acquired using a filter, we multiply the open transmission
model by a model of the interference filter transmission, defined as
follows:
\begin{equation}
  \label{eq:filtertransmission}
T(\lambda, \theta) = \mathcal T\left(\frac{\lambda}{\sqrt{1 -
    (\sin(\theta) / n_\text{eff})^2}}\right)\,.
\end{equation}
Here, $n_\text{eff}$ is an effective index for the filter,
and $\mathcal{T}$ is a piece-wise linear function of wavelength. The piece-wise
linear function is initially created with \num{150} regularly spaced
nodes between \SIrange{350}{1100}{nm}, providing a general resolution
of \SI{5}{nm}. In cases where more precision is needed, we further
refine the grid by equally splitting intervals where the local mean
chi-square exceeds the global mean chi-square by more than three
standard deviations. This process is repeated four times to ensure
that filter fronts are typically modeled with up to approximately
\SI{0.3}{nm} resolution.

In the specific case of the  grating zeroth and first orders, the photometry is
adjusted using a cubic B-spline model with \num{105} nodes in
wavelength.

The composite model is fit to dataset No.~2, which includes data
from four different radii and successive observations without filters
or with all seven filters and the grating. The baseline model has
\num{1979} free parameters: \num{148} for the open transmission,
approximately $6\times165$ for each of the $ugrizy$ filters, and
\num{420} for each of the grating orders. Unfortunately, the blue edge
of the $u$-band filter cannot be accurately measured with this
data. Fit results are displayed in
Fig.~\ref{fig:lambdathetafitresults}.

\begin{figure*}
  \centering
  \includegraphics[width=1\linewidth]{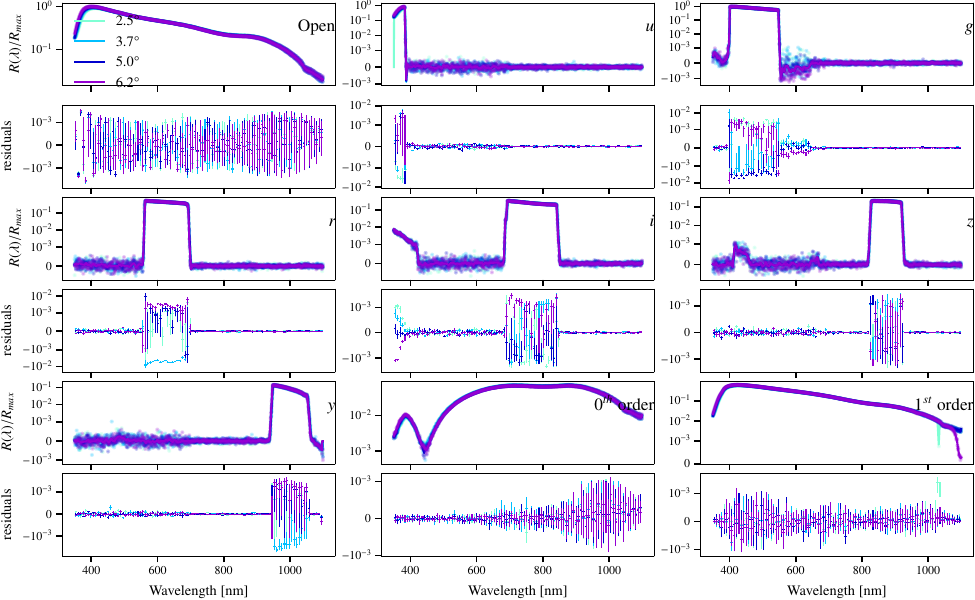}
  \caption{Model of the wavelength and radial dependency of the
    StarDICE response to CBP illumination $R(\lambda)$. Each panel
    display the raw measurements at the 4 sampled positions for each of
    the filter configurations: no filters (Open), with one of the 6
    photometric filters ($ugrizy$) or with the grating looking either
    at the zeroth order or the first order spots. The panel
    immediately below each panel display the residuals to the
    model as a fraction of the data. For easy comparison, all panels are normalized to the peak
    of the response, which occurs in the open configuration at
    $\lambda = \SI{398}{nm}$ and $\theta = \SI{6.2}{\degree}$.
 }
  \label{fig:lambdathetafitresults}
\end{figure*}

\begin{figure}
  \centering
  \includegraphics[width=1\linewidth]{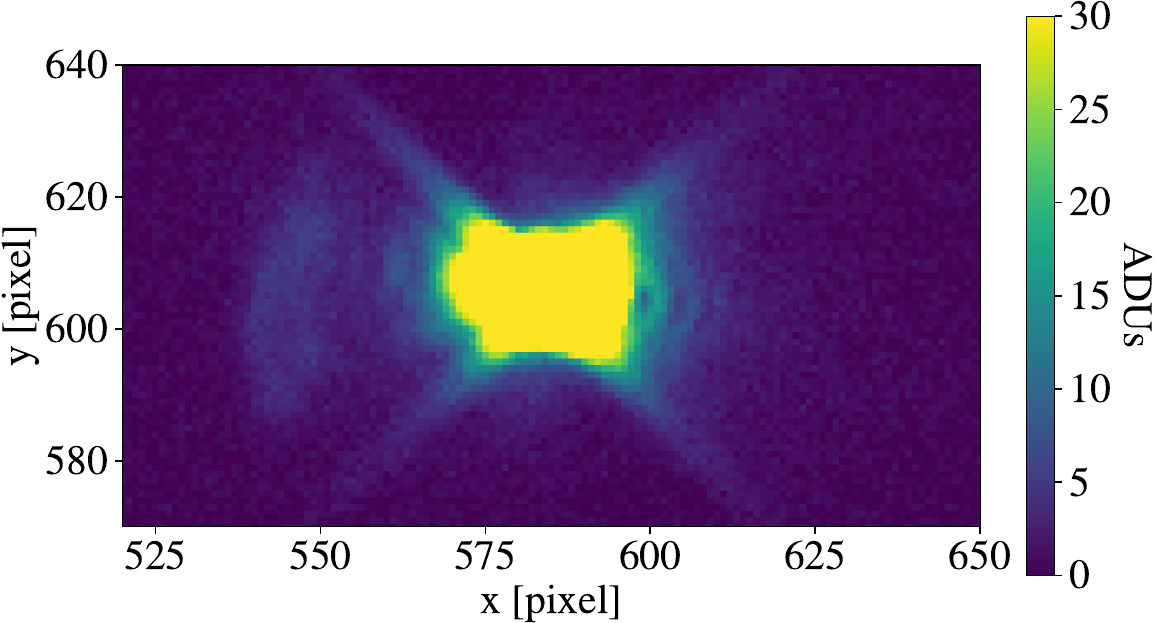}
  \caption{Diffraction rings due to the presence of a dust spot on the
    r filter, intercepting the CBP beam.}
  \label{fig:dust}
\end{figure}

Overall, the model provides a satisfactory description of the
dataset. The most significant discrepancy is a noticeable gray
decrease in the transmission of the $r$ filter for the sample
measurement at \SI{3.7}{\degree} with respect to the average of the
other three.  After investigating this issue, we identified that the
corresponding images displayed a diffraction figure consistent with
the presence of a dust particle on the filter surface intercepting the
CBP beam, visible in Figure~\ref{fig:dust}. Based on the approximate area of the beam spot
($\sim$\SI{12}{\milli\metre\squared}), and the estimated particle size range (between
\SIrange{200}{300}{\micro\metre}), we calculated a potential decrease
in transmission for this specific partial illumination of the primary
mirror of up to \SI{0.6}{\percent}. This value is in good agreement
with the observed decrease. Similar discrepancies were found for some
observations in $u$, $g$, and $y$ bands, although determining particle
sizes from diffraction features was not possible in these cases. We
attribute these discrepancies to dust contamination on the filter
surface as well.

The top panel of Fig.~\ref{fig:metrics} shows the difference between
the transmission integral for measurements and the model at each
position, helping readers better visualize the 'gray'
discrepancies. The standard deviation of the model/measurement
discrepancies is \SI{10.2}{mmag}. Considering this value as an
estimate of dust-induced dispersion, we can deduce that the
corresponding uncertainty for per-filter normalization of the model
(averaging 4 independent samples) is approximately $\SI{5}{mmag}$.

This noise could be decreased by averaging more sample measurements of
the mirror. In our case 4 additional positions were sampled as part
of run 3 (see table~\ref{tab:schedule}). However, while measurements of run No.~2 were carefully taken to sample the mirror from the outer
edge to the inner edge along a radius with separations provided by the
mount encoders, the other measurements were positioned semi-randomly
with the idea that the position of the CCD-window ghost in the images
will enable precise determination of the position of the beam in the
mirror. Two issues were overlooked at this stage which complicated the
determination of the geometry from the position of the ghosts: (1) The
wedge of the CCD window and (2) the ghost position dependence on the
conjugation relation between the CBP and the telescope which is only
approximately known. The complexity of including those parameters in a
full model of the telescope+CBP sets the analysis of those data out of
the scope of the current paper.

The bottom panel in Fig.~\ref{fig:metrics} displays the difference
between the measured central wavelength and the model-predicted value
for all 4 radius positions. The model accurately reproduces the
central wavelengths of all filters with a standard deviation of only
$\SI{0.13}{nm}$. This represents an improvement by an order of
magnitude compared to a model that neglects wavelength shift
considerations due to the light incidence angle on the filter, highlighting the importance of including this effect
in the StarDICE response model.

\begin{figure}
  \centering
  \includegraphics[width=1\linewidth]{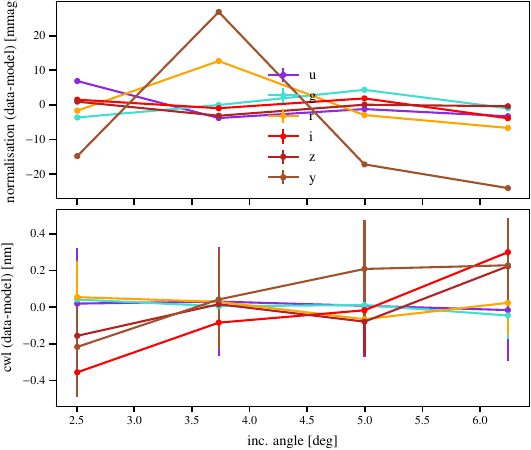}
  \caption{Discrepancies between model and raw measurements at
    different mirror locations summarized according to 2 metrics:
    \emph{Top:} relative difference in the integral of the passband,
    expressed in millimagnitudes; \emph{Bottom:} relative difference
    in central wavelength computed as the barycenter of the passband.}
  \label{fig:metrics}
\end{figure}

\subsection{Full pupil synthetic transmission curves}

The full pupil transmission is synthesized by numerically averaging
the above model assuming that the pupil is a perfect annulus with an
inner radius of \SI{55}{mm} and an outer radius of \SI{203}{mm},
corresponding to an effective mirror area of \SI{1202}{cm^2}. The
rectangle rule with 100 evenly sampled points in radius has been used
for the averaging. The curves have been normalized using the CBP
response from Sect.\ref{sec:cbp}. The resulting transmission curves
are shown in Fig.~\ref{fig:fullpupiltrans}.
\begin{figure}
  \centering
  \includegraphics[width=1\linewidth]{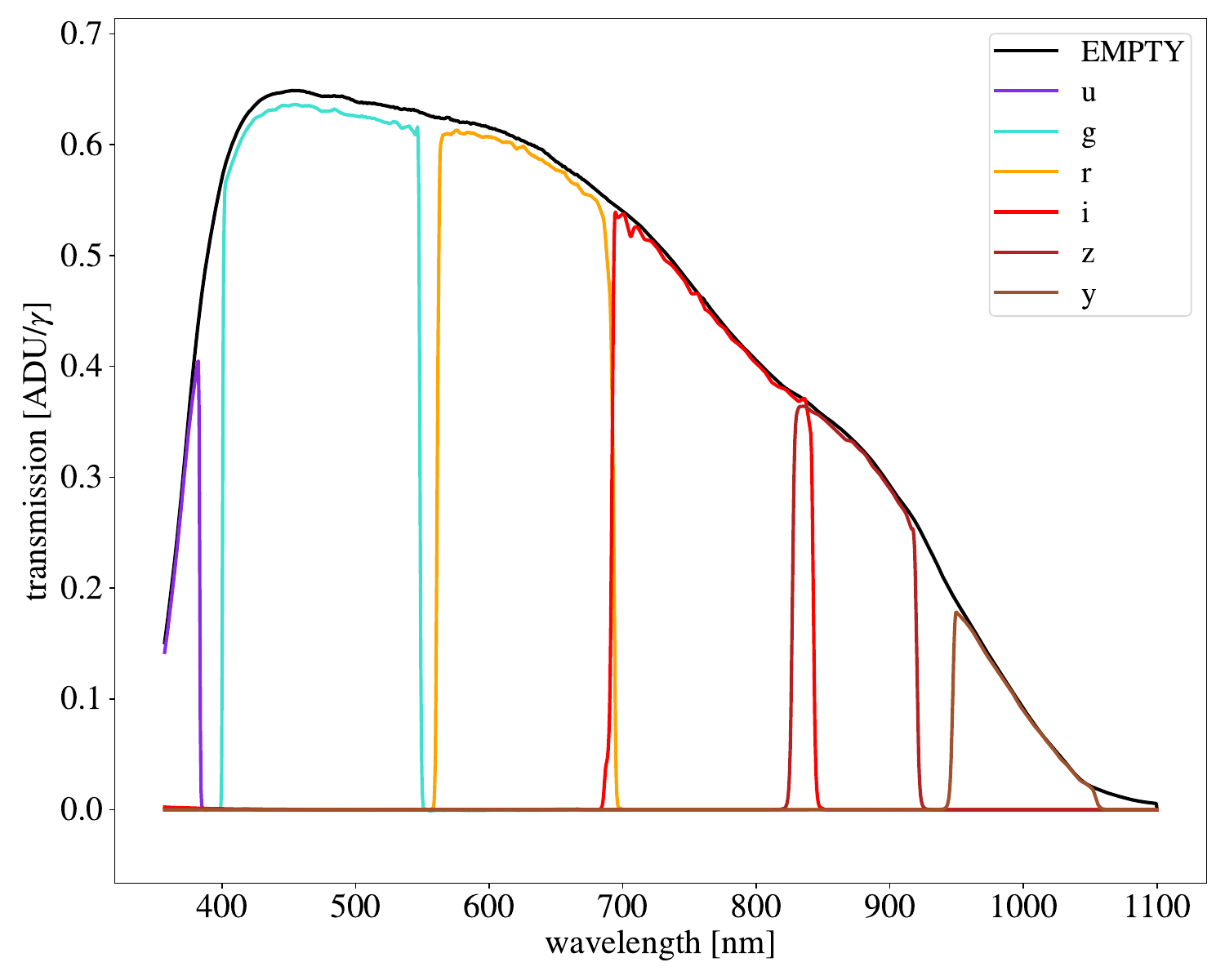}
  \caption{Full-pupil transmission curves for the StarDICE instruments.}
  \label{fig:fullpupiltrans}
\end{figure}

\subsection{Final uncertainty budget}
\label{sec:final_uncertainty_budget}

The impact of the uncertainties in our determination of the full-pupil
transmission curves depend on the application. We first consider the
use of the CBP as the sole source for absolute calibration of the
instrument response and then its use in conjunction with a broadband
star-like calibration light source.

\subsubsection{Uncertainties in absolute fluxes}
\label{sec:absolute}

The first obvious application is to take
each curve as an absolute calibration of the instrument throughput and
use it to interpret broadband fluxes measured by this instrument,
according to the equation:
\begin{equation}
  \label{eq:mb}
  \phi_b = \int_\lambda  R_b(\lambda) A(\lambda) S(\lambda) \lambda d\lambda
\end{equation}
where $R_b$ is the full pupil response curve of the instrument
synthesized previously, $S$ is the top of the atmosphere spectrum of
the target and $A$ is the atmospheric transmission for this
observation. Setting aside the question of the atmospheric
transmission, the error $\delta R_b$ on the passband will translate
into an error on the synthesized flux $\delta \phi$, whose magnitude
depends on the spectrum of the object. As an illustration we
propagated all our uncertainties to the synthetic fluxes of the
primary spectrophotometric standards G191B2B. We report the results in
Table~\ref{tab:budget} as the relative uncertainty on the broadband
flux $\sigma(\delta \phi)/\phi$ for each band, with one line per
contributions. The table does not include the uncertainty on the gray
scale, mainly coming from the uncertainty in the effective area of the
telescope which cannot be determined accurately with such a setup. 

\begin{table}
  \centering
  \caption{Relative uncertainty in the synthetic broadband fluxes of
    G191B2B, split by contributions, in permil.}
  \label{tab:budget}
  \begin{tabular}{@{}l@{}rrrrrr@{}}
    \toprule
    \toprule
    Source & $u^\dag$ & $g$ & $r$ & $i$ & $z$ & $y$ \\
    & $[\text{\textperthousand}]$ & $[\text{\textperthousand}]$ & $[\text{\textperthousand}]$ & $[\text{\textperthousand}]$ & $[\text{\textperthousand}]$ & $[\text{\textperthousand}]$\\
    \midrule
    StarDICE & 0.5 & 0.2 & 0.2 & 0.3 & 1.0 & 3.8 \\
    $\Esolar$ & 1.6 & 0.1 & 0.1 & 0.1 & 0.1 & 0.1 \\
    CBP & 2.2 & 0.5 & 1.2 & 0.0 & 0.0 & 0.1 \\
    \midrule
    Stat (total) & 2.7 & 0.5 & 1.2 & 0.3 & 1.0 & 3.8 \\
    \midrule
    Scattered light & 2.7 & 3.1 & 3.8 & 4.3 & 4.8 & 5.3 \\
    Repeatability & 0.8 & 0.4 & 0.5 & 1.1 & 1.1 & 1.1 \\
    Linearity & 0.2 & 0.2 & 0.3 & 0.5 & 0.5 & 0.5 \\
    Contamination & 4.4 & 0.6 & 0.8 & 0.1 & 0.1 & 0.1 \\
    Mirror sampling noise & 5.4 & 5.4 & 5.4 & 5.4 & 5.4 & 5.4 \\
    NIST photodiode & 0.2 & 0.1 & 0.0 & 0.0 & 0.0 & 0.1 \\
    Solar cell temperature & 0.0 & 0.0 & 0.0 & 0.0 & 0.0 & 3.9 \\
    Wavelength calibration & 0.7 & 0.6 & 0.5 & 0.4 & 0.3 & 0.3 \\
    \midrule
    Syst (total) & 7.6 & 6.3 & 6.7 & 7.1 & 7.4 & 8.7 \\
    \midrule
    Total & 8.0 & 6.3 & 6.8 & 7.1 & 7.5 & 9.4 \\
    \bottomrule
  \end{tabular}
  \tablefoot{$^\dag$Our measurement does not capture the blue side of the $u$ band filter. We propagate uncertainties as if the transmission was dropping to 0 at the edge of the measurement to illustrate the performances that would be obtained from a complete measurement. For all practical purposes, however, the actual $u$ band transmission cannot be determined from the existing measures.}
\end{table}

The dominant contributor to the uncertainty in our flux reconstruction
is the mirror sampling noise. With the CBP beam sampling only a small
fraction of the primary mirror, and therefore a small fraction of the
filters, any inhomogeneity in the transmission of the filter ends up
causing a noise in the band-to-band ratio of transmissions. This noise
decreases with more sampling points on the primary mirrors, at a large
observational cost.

The next significant issue is the difficulty of getting rid of
non-collimated (scattered) light during the calibration of the CBP on
the solar-cell array. A natural way to deal with this issue would be
to significantly increase the solar-cell CBP distance. In our case, however, the beam created by the \bpinhole pinhole already barely fit
inside the footprint of the solar cell. Increasing the solar-cell CBP
distance would have caused issues for the collimated beam calibration in exchange
for getting rid of the non-collimated light.

Some caution needs to be paid to changes in the working temperature
of photodiodes when calibrating the $y$ band. In our current measurement,
the uncertainty in the difference in temperature between the
calibration of the solar cell and its use to calibrate the CBP is
quite large (\SI{1.6}{\celsius}). This has easily been fixed in the latter
iteration of the solar-cell design by including temperature sensors
directly cell enclosure.

Our mitigation procedure for the contamination of the beam by light
at different wavelengths is found satisfactory for all bands but $u$. This
is due to the remaining uncertainties in the determination of the
phosphorescence contribution in the integrating sphere.

\begin{figure}
  \centering
  \includegraphics[width=1\linewidth]{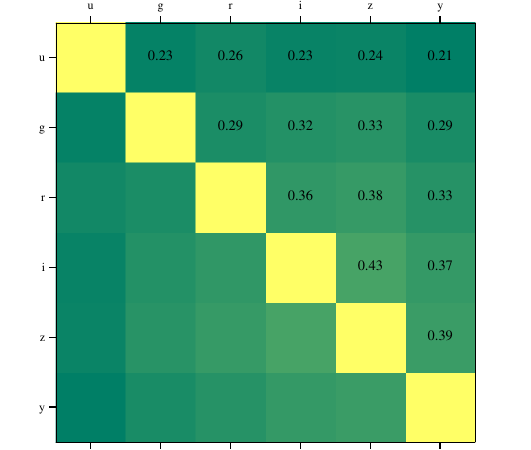}
  \caption{Correlation of the uncertainties on the synthetic fluxes of G191B2B.}
  \label{fig:correlation}
\end{figure}

Lastly, most systematic uncertainties have modes coherent across
wavelengths. As a result, the errors induced on broadband magnitudes
are correlated across bands, with typical correlation levels in the
20-40\% range. The correlation matrix resulting from the full
propagation of uncertainties identified in Table~\ref{tab:budget} is
displayed in Fig.~\ref{fig:correlation}. Again, we do not account for
the uncertainty in the gray scale which is perfectly correlated
between all bands and would dominate the uncertainty budget.

\subsubsection{Uncertainties in relative fluxes}
\label{sec:relative}

The more common application of telescope transmission measurements is
to rely on the transmission curves to predict actual
observations of a spectrophotometric standard of known flux and
determine an independent re-calibration factor for each band. The
errors on the interpretation of observations of another object will
then depend on the difference in color between the object and the
spectrophotometric standard, canceling for objects whose spectrum is
very similar to the spectrum of the standard.

For the StarDICE telescope, the role of photometric standard is played
by a collection of narrow-spectrum LEDs, whose observations set the
absolute normalization of each band. The telescope is then used to
observe CALSPEC standard stars and precisely measure their broadband
fluxes, anchoring them to the LEDs absolute calibration.
In this operation, all uncertainties which mostly impact the
normalization of the passbands cancels out. In order to illustrate the
impact of the uncertainties in our passband determination in this
case, we modeled the spectrum of 6 calibration LEDs as Gaussian shapes
centered on the central wavelength of each filter with a full-width
half-maximum of 7\% in wavelength. 
To simulate the effect of the passband recalibration on LED
observations, we propagate the uncertainties on the \emph{ratios} of
broadband fluxes between the LEDs and G191B2B. The results are
presented in Table \ref{tab:led}. As expected, most broadband error
sources cancel out, resulting in an accuracy of the order of 1 permil,
matching the requirements for the StarDICE experiment.

In contrast, the sensitivity to wavelength calibration generally
increases in a way that depends on the spectra of the LEDs. The flux of
LEDs whose spectrum overlaps with the edges of the filters are more
strongly affected by wavelength calibration errors than LEDs whose
spectrum is largely contained in the filter passband. The design of
the StarDICE artificial star includes both cases, with the idea that
overlapping LEDs can provide a handle to test (or correct) for filter
front errors. Here we report sensitivity for non-overlapping LEDs
assuming no specific correction.

\begin{table}
  \centering
  \caption{Uncertainties in the flux of G191B2B after recalibration by observation of narrow-spectrum LEDs centered on the filter passband.}
  \label{tab:led}
  \begin{tabular}{@{}l@{}r@{}r@{}r@{}r@{}r@{}r@{}}
    \toprule
    Source & u & g & r & i & z & y \\
           & $[\text{\textperthousand}]$ & $[\text{\textperthousand}]$ & $[\text{\textperthousand}]$ & $[\text{\textperthousand}]$ & $[\text{\textperthousand}]$ & $[\text{\textperthousand}]$\\
    \midrule
    StarDICE & 0.2 & 0.3 & 0.2 & 0.2 & 0.5 & 1.8 \\
    $\Esolar$ & 0.4 & 0.1 & 0.1 & 0.1 & $<0.1$ & 0.1 \\
    CBP & 0.5 & 0.5 & 1.1 & $<0.1$ & $<0.1$ & $<0.1$ \\
    \midrule
    Stat (total) & 0.7 & 0.6 & 1.1 & 0.2 & 0.5 & 1.8 \\
    \midrule
    Scattered light & $<0.1$ & 0.1 & $<0.1$ & $<0.1$ & $<0.1$ & $<0.1$ \\
    Repeatability & $<0.1$ & $<0.1$ & 0.1 & $<0.1$ & $<0.1$ & $<0.1$ \\
    Linearity & $<0.1$ & $<0.1$ & $<0.1$ & $<0.1$ & $<0.1$ & $<0.1$ \\
    Contamination & 0.1 & 0.1 & $<0.1$ & $<0.1$ & $<0.1$ & $<0.1$ \\
    Mirror sampling noise & $<0.1$ & $<0.1$ & $<0.1$ & $<0.1$ & $<0.1$ & $<0.1$ \\
    NIST photodiode & 0.1 & 0.1 & $<0.1$ & $<0.1$ & $<0.1$ & 0.1 \\
    Solar cell temperature & $<0.1$ & $<0.1$ & $<0.1$ & $<0.1$ & $<0.1$ & 0.6 \\
    Wavelength calibration$^\dag$ & 0.3 & 0.5 & 0.6 & 1.1 & 0.9 & 1.4 \\
    \midrule
    Syst (total) & 0.3 & 0.5 & 0.6 & 1.1 & 0.9 & 1.5 \\
    \midrule
    Total & 0.8 & 0.8 & 1.3 & 1.1 & 1.1 & 2.4 \\
    \bottomrule
  \end{tabular}
  \tablefoot{$^\dag$The impact of wavelength calibration on
    led-calibrated fluxes depend significantly on the exact spectrum
    of the LEDs. The numbers presented come from a simulation of a
    realistic case, but the numbers for actual LEDs may differ.}
  
\end{table}

\section{Conclusion}
\label{sec:discussion}

We determined the response of the 16'' StarDICE telescope using a
collimated beam projector (CBP) powered by a tunable laser. Our
procedure involved three main steps. In the first step, we measured the
throughput of the CBP with a wavelength sampling of \SI{1}{nm} between
350 and \SI{1100}{nm} by illuminating a calibrated solar cell. We then
used the calibrated beam to map the response of the StarDICE telescope
at all wavelengths and several positions on its primary
mirror. Lastly, we interpolated between the sample points to
synthesize the equivalent response of the telescope to the illumination
of its full pupil.

A key aspect of calibrating the CBP beam is that the
sensitive area of the photodiode is large enough to cover the beam entirely. While the measurement is simple in its principle, special
care must be given to several aspects of the setup and analysis to
reach the required level of accuracy. Most notably: (1) the laser beam must be filtered to improve the purity of its wavelength
composition, even with the mitigation measures we adopted, unavoidable
contamination had to be subtracted in the analysis; (2) the setup must minimize scattered light which affects differently the calibration solar cell and the calibrated telescope; (3) achieving sufficient signal to noise ratio in the solar cell implies to fight with a high level of $1/f$ noise (pink noise).

To deal with this issue, we selected a solar cell with high
impedance, and grouped the light deposit in short bursts to reduce the
pink noise effects while illuminating the solar cell.
These bursts were separated by dark periods. Using good synchronization in time
between the bursts and the photocurrent measurements, 
we could estimate the dark current contribution and correct it.
Additionally, we increased the collected flux in the solar cell 
by selecting a larger pinhole, inducing a small but measurable change in the
chromatic throughput of the CBP, which must be accounted for.
We did not find significant non-linearities in our measurement
chain. Repeating the measurements allowed us to identify a slight
evolution in the throughput, which was easily corrected.

Most of the measurement time is spent in mapping the telescope
response which scales linearly with the number of filters,
resolution in wavelength, and number of mirror samples taken. Our main
4-positions analysis, which yields subpercent precision on the
synthesized passbands involved about \num{30000} images. The complete
dataset, including runs for the characterization of systematics,
pinhole inter-calibration, and focal plane measurements, close \num{200000}
images. With our fully automated setup, collecting this dataset required
7 days of uninterrupted operation. 

Our analysis allowed us to determine filter fronts with a precision of
about \SI{0.2}{\nano\meter}, to accurately measure out-of-band leaks at a relative
level of \num{e-3} and revealed an unexpected and strong dependency of
the mirror reflectivity with the incidence angle in the UV. As a
result, we were able to show through simulations that the uncertainties in our determination of the
telescope passbands will not contribute more than 1\textperthousand\
uncertainty in the measurement of broadband flux with StarDICE after
calibration by observations of the artificial star. 

The results presented in this paper also serve as a proof of concept for the Rubin CBP, specifically designed for measuring the LSST passbands at the Vera C. Rubin Observatory. Although scaling this setup from a 16'' to a \SI{8}{\meter} telescope will be challenging in many regards (more optical surfaces inducing ghosts, larger filters, more sensors), similar numbers are to be expected for the determination of LSST passbands if the necessary amount of calibration time is provided. This calibration will allow an accurate interpretation of the flux ratios between the established photometric standards and the SNe Ia observed to map the Universe's expansion history. On the other hand, developing a traveling version of the CBP is currently undergoing \citep{2024RASTI...3..125S} to calibrate the filter transmissions of ongoing and future SNe Ia surveys.

\bibliographystyle{aa}
\bibliography{main}

\begin{acknowledgements}
  This work received support from the Programme National Cosmology et
  Galaxies (PNCG) of CNRS/INSU with INP and IN2P3, co-funded by CEA
  and CNES and from the DIM ACAV program of the Île-de-France region.
  CWS, EU, and SB are grateful to the US Department of Energy for support under Cosmic Frontier award DE- SC0007881. 

  This paper has undergone internal review in the LSST Dark Energy Science Collaboration. The internal reviewers were Johan Bregeon and Parker Fagrelius.\\

Thierry Souverin is the primary author of the paper, and he contributed to the hardware assembly, the data taking, the data analysis and the paper edition. Jérémy Neveu is the primary author's advisor and participated to the hardware assembly, the data taking, the data analysis and the paper edition.  Marc Betoule is the StarDICE PI and participated to the hardware assembly, data taking, data reduction, lead the pupill stitching analysis and participated to the paper edition. Sébastien Bongard contributed to the logistics, the hardware assembly, the data analysis and the paper edition.  Christopher W. Stubbs is the instigator of the CBP concept and he contributed to a large part of the hardware funding, and to the solar cell calibration. Elana Urbach participated to the hardware assembly, the data taking and the solar cell calibration. 
Sasha Brownsberger participated to the hardware assembly, the data taking, and the solar cell calibration. Pierre Éric Blanc is member of the StarDICE collaboration. Johann Cohen Tanugi is member of the StarDICE collaboration. Sylvie Dagoret-Campagne is member of the StarDICE collaboration. Fabrice Feinstein is member of the StarDICE collaboration. Delphine Hardin is member of the StarDICE collaboration. Claire Juramy is member of the StarDICE collaboration. Laurent Le Guillou is member of the StarDICE collaboration and contributed with software and hardware support. Auguste Le Van Suu is member of the StarDICE collaboration. Marc Moniez is member of the StarDICE collaboration. Éric Nuss (posthumous) is member of the StarDICE collaboration. Bertrand
Plez is member of the StarDICE collaboration. Nicolas Regnault is member of the StarDICE collaboration. Eduardo Sepulveda is member of the StarDICE collaboration and contributed with software and hardware support. Kélian Sommer is member of the StarDICE collaboration. All members of the StarDICE collaboration contributed by their general support to the collaboration and to the collaboration review of the paper.

  The DESC acknowledges ongoing support from the Institut National de 
Physique Nucl\'eaire et de Physique des Particules in France; the 
Science \& Technology Facilities Council in the United Kingdom; and the
Department of Energy, the National Science Foundation, and the LSST 
Corporation in the United States.  DESC uses resources of the IN2P3 
Computing Center (CC-IN2P3--Lyon/Villeurbanne - France) funded by the 
Centre National de la Recherche Scientifique; the National Energy 
Research Scientific Computing Center, a DOE Office of Science User 
Facility supported by the Office of Science of the U.S.\ Department of
Energy under Contract No.\ DE-AC02-05CH11231; STFC DiRAC HPC Facilities, 
funded by UK BEIS National E-infrastructure capital grants; and the UK 
particle physics grid, supported by the GridPP Collaboration.  This 
work was performed in part under DOE Contract DE-AC02-76SF00515.

\end{acknowledgements}

\appendix
\appendix

\section{Ghost photometry}
\label{sec:ghost_photometry}

Let be $G_0(\lambda)$ the quantity of light collected in the main spot, and $G_\mathrm{n}(\lambda)$ the quantity of light collected in the ghost of order $n$. The ratio of the 1\up{st} order ghost $G_1(\lambda)$ over the main spot $G_0(\lambda)$ is:

\begin{equation}
    \Kghost = \frac{G_1(\lambda)}{G_0(\lambda)}.
    \label{eq:ratio_ghost}
\end{equation}

We measure $G_1(\lambda)$ with the \spinhole pinhole where it is well separated from $G_0(\lambda)$. We build a mask with the expected ghost shape like, shown in Figure~\ref{fig:ghost_contrast}, and we fit its best position on the image. To estimate the background at this position, we assume that the main spot exhibits vertical spatial symmetry and measure the flux within a symmetric mask at the vertically opposite position relative to the main spot. $G_1(\lambda)$ is the sum of the ADUs in the ghost mask after background subtraction. On the other hand, $G_0(\lambda)$ is measured with the baseline photometry detailed in Section~\ref{sec:photometry_small}. 

This study is pursued for the 3 runs of dataset No.~4 and 2 runs of dataset No.~2 from Table~\ref{tab:schedule}. The results obtained for $\Kghost$ are shown in Figure~\ref{fig:ghost_ratio}, for which the mean spline is displayed in the third panel of Figure~\ref{fig:result_params}.

\begin{figure}[h]
     \centering
     \resizebox{\hsize}{!}{\includegraphics{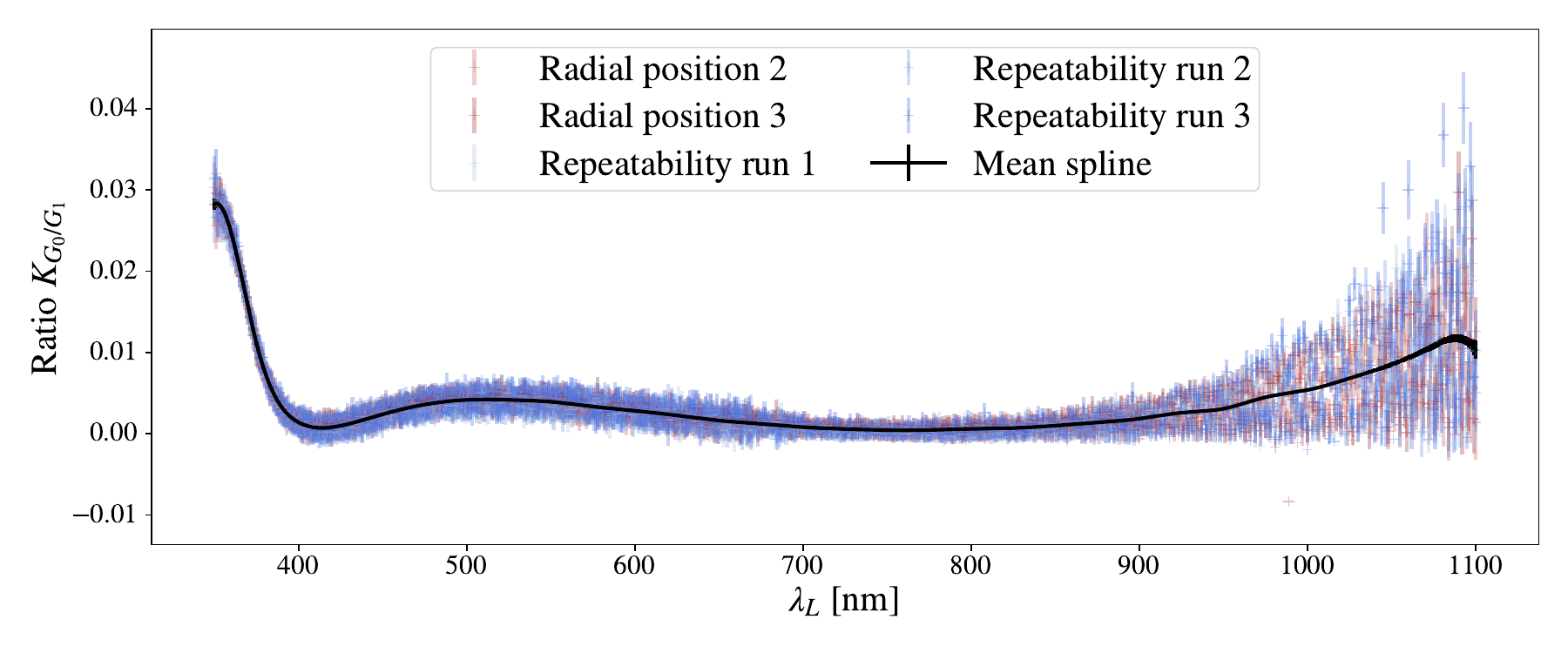}}
     \caption{Ratio $\Kghost$ with respect to $\lambda_L$. The mean spline goes through the five datasets.}
     \label{fig:ghost_ratio}
\end{figure}

This method allows for estimation of $G_1(\lambda)$, but we must investigate the contribution of ghosts of higher order $G_{n>1}$. Because of the faint intensity of these orders, the method described above cannot be performed, so we estimate their contribution from the $1^{\mathrm{st}}$ order ghost analysis. Let be $\Rwindow$ the reflection coefficient at the interface air-window, and $\Rccd$ the reflection coefficient at the interface air-CCD. We know the transmission of the window $T_\mathrm{window}(\lambda)$ from manufacturer datasheets, so we can infer $\Rwindow$:
 
\begin{equation}
    \Rwindow = 1 - T_\mathrm{window}(\lambda),
    \label{eq:rwindow}
\end{equation}
which is represented as the blue curve in Figure~\ref{fig:reflectivities}. We define $F(\lambda)$ the flux collected in the camera, $G_0(\lambda)$ the quantity of light collected in the main spot, and $G_\mathrm{n}(\lambda)$ the quantity of light collected in the ghost of order $n$. $G_0(\lambda)$ and $G_\mathrm{n}(\lambda)$ are a function of $F(\lambda)$: 

\begin{equation}
     G_0(\lambda) = (1-\Rwindow)^{2} \times (1-\Rccd) \times F(\lambda), 
     \label{eq:g0}
\end{equation}

\begin{equation}
\begin{aligned}
    G_\mathrm{n}(\lambda) & = G_{0}(\lambda) \times [\Rccd \Rwindow + \Rccd(1-\Rwindow)\Rwindow]^{n} \\
    & = G_{0}(\lambda) \times [2 \Rccd \Rwindow - \Rccd \Rwindow^{2}] ^{n}. \\
     \label{eq:gn}
\end{aligned}
\end{equation}

\noindent The sum of all the ghosts $G_{\mathrm{n>0}}(\lambda)$ is defined as
\footnote{If |q| < 1, the serie $\left( \sum_{n=0}^{m} q^n \right)_{\mathrm{m \in \mathbb{N}}}$ strictly converge and \\ $\sum_{n=0}^{\infty} q^n \equiv \lim\limits_{m \rightarrow \infty} \sum_{n=0}^{m} q^n = \frac{1}{1-q}$}:

 \begin{equation}
 \begin{aligned}
     G_{n>0}(\lambda)&=\sum_{n=0}^{\mathrm{n} \rightarrow \infty} G_\mathrm{n}(\lambda) - G_0(\lambda) \\
     & = G_0(\lambda) \times \sum_{n=0}^{n \rightarrow \infty} \left( [2 \Rccd \Rwindow - \Rccd \Rwindow^{2}] ^{n} - G_0(\lambda) \right)\\
     & = G_0(\lambda) \times\left( \frac{1}{1- [2 \Rccd \Rwindow - \Rccd \Rwindow^{2}]} - 1\right),
     \label{eq:sum_ghost}
 \end{aligned}
 \end{equation}
 
\noindent and the sum of the ghost of order higher than 1 is defined as:

 \begin{equation}
     G_{n>1}(\lambda) = G_{n>0}(\lambda) - G_1(\lambda).
     \label{eq:sum_ghost_sup_1}
 \end{equation}

\noindent As we measure the photometry of the 1\up{st} order ghost $G_1(\lambda)$, we want to verify that the ghosts of higher orders $G_{\mathrm{n}>1}(\lambda)$ are negligible. Using the Equations~\ref{eq:ratio_ghost}, \ref{eq:g0} and \ref{eq:gn}, we can compute $\Rccd$:

\begin{equation}
    \Rccd = \frac{\Kghost}{\Rwindow (2 - \Rwindow)},
    \label{eq:rccd}
\end{equation}
reported as the red curve in Figure~\ref{fig:reflectivities}.

\begin{figure}[h]
    \centering
    \includegraphics[width=\columnwidth]{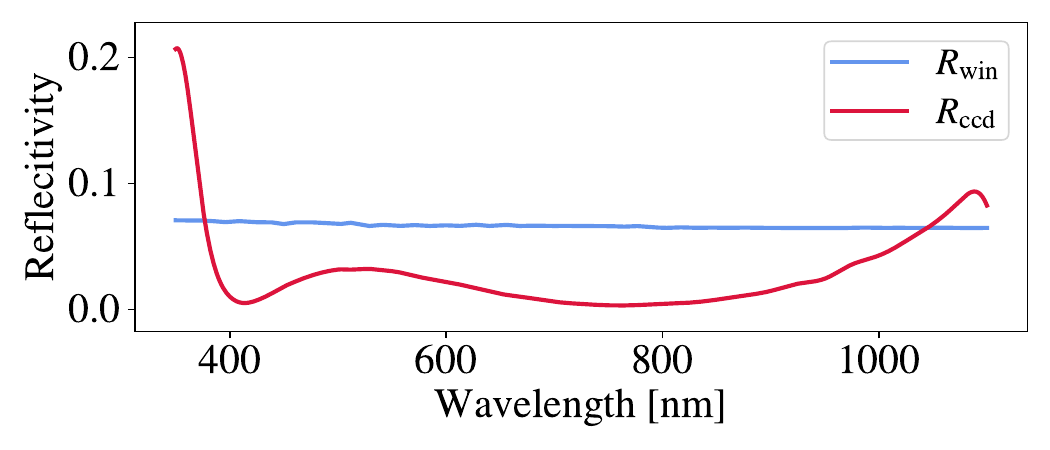}
    \caption{Reflectivities of the CCD $\Rccd$ in red, and the camera window $\Rwindow$ in blue, both against the wavelength.}
    \label{fig:reflectivities}
\end{figure}

We measured $G_1(\lambda)$ with photometry, and estimate $G_{\mathrm{n}>1}$ with the equations above. We show the ratio $K_{G_{\mathrm{n}>1}/G_0}$ in Figure~\ref{fig:ratio_ginf_g0}, and we see that $K_{G_{\mathrm{n}>1}/G_0}$ is always below the per mil level. Since $G_{\mathrm{n}>1}$ contribute for less than a per mil of $G_0(\lambda)$, we decide to neglect it.

\begin{figure}[h]
    \centering
    \includegraphics[width=\columnwidth]{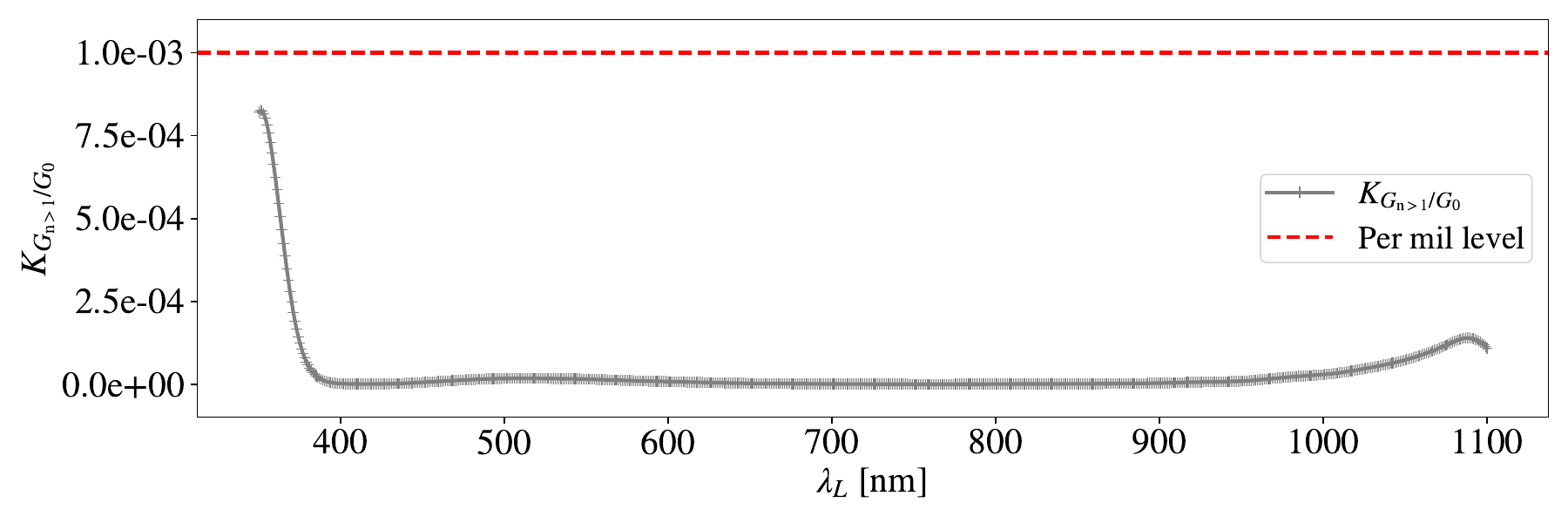}
    \caption{Ratio $K_{G_{\mathrm{n}>1}/G_0}(\lambda)$ (which corresponds to the sum of the ghosts at order higher than 1 $G_{\mathrm{n}>1}(\lambda)$ over $G_0(\lambda)$) with respect to $\lambda_L$.}
    \label{fig:ratio_ginf_g0}
\end{figure}

\end{document}